\documentclass[10pt,a4paper]{article}
\usepackage[T1]{fontenc}
\usepackage[a4paper, hmargin=3.5cm, vmargin=4cm]{geometry}
\usepackage{newtxtext}
\usepackage{newtxmath}
\usepackage{authblk}
\usepackage{graphicx}
\usepackage{amsmath}
\usepackage{caption}
\usepackage{subcaption}
\usepackage[section]{placeins}
\usepackage{float}
\usepackage{enumitem}
\usepackage[sort&compress,numbers]{natbib}
\usepackage{physunits}
\usepackage{siunitx}
\usepackage{bm}
\usepackage[symbol]{footmisc}
\usepackage[hidelinks]{hyperref}
\usepackage{orcidlink}

\newcommand{\vtheta}{\ensuremath{\bm{\theta}}}
\newcommand{\vx}{\mathbf{x}}

\newcommand{\vw}{\mathbf{w}}
\newcommand{\MET}{\ensuremath{E_{\mathrm{T}}^{\mathrm{miss}}}}
\newcommand{\pt}{\ensuremath{p_{\mathrm{T}}}}
\newcommand{\jets}{\ensuremath{N_\text{jets}}}
\newcommand{\mt}{\ensuremath{m_{\mathrm{T}}}}

\title{\Large{Neural Fake Factor Estimation Using Data-Based Inference}}

\author[1,2]{Jan~Gavranovič\,\orcidlink{0000-0002-8760-9518}\footnote{\href{mailto:jan.gavranovic@ijs.si}{\texttt{jan.gavranovic@ijs.si}}}}
\author[3]{Lara~Čalić\,\orcidlink{0000-0001-9253-9350}}
\author[1,2]{Jernej~Debevc\,\orcidlink{0000-0001-9324-719X}}
\author[3]{Else~Lytken\,\orcidlink{0000-0002-8141-3995}}
\author[1,2]{Borut~Paul~Kerševan\,\orcidlink{0000-0002-4529-452X}\footnote{\href{mailto:borut.kersevan@ijs.si}{\texttt{borut.kersevan@ijs.si}}}}
\affil[1]{Jožef Stefan Institute, Ljubljana, Slovenia}
\affil[2]{Faculty of Mathematics and Physics, University of Ljubljana, Ljubljana, Slovenia}
\affil[3]{Division of Particle and Nuclear Physics, Lund University, Lund, Sweden}

\date{\vspace{-8mm}}

\begin{document}

\maketitle

\begin{abstract}
\noindent
In a high-energy physics data analysis, the term \textit{“fake”}
backgrounds refers to events that would formally not satisfy the (signal) process selection
criteria, but are accepted nonetheless due to mis-reconstructed particles.
This can occur, e.g., when leptons from secondary
decays are incorrectly identified as originating from the hard-scatter
interaction point (known as \textit{non-prompt leptons}), or when other physics
objects, such as hadronic jets, are mistakenly reconstructed as leptons
(resulting in \textit{mis-identified leptons}). These \textit{fake leptons} are usually
estimated using data-driven techniques, one of the most common being the Fake Factor method.
This method relies on predicting the fake lepton contribution by reweighting data events,
using a scale factor (i.e.\ fake factor) function. Traditionally, fake factors
have been estimated by histogramming and computing the ratio of two
data distributions, typically as functions of a few relevant physics variables such
as the transverse momentum $\pt$ and pseudorapidity $\eta$. In this work, we introduce a novel approach of fake factor calculation,
based on density ratio estimation using neural networks trained directly on  data
in a higher-dimensional feature space. We show that our method enables the
computation of a continuous, unbinned fake factor on a per-event basis, offering
a more flexible, precise, and higher-dimensional alternative to the conventional
method, making it applicable to a wide range of analyses. A simple LHC open data analysis we implemented confirms the feasibility of the method and demonstrates that the ML-based fake factor provides smoother, more stable
estimates across the phase space than traditional methods, reducing binning
artifacts and improving extrapolation to signal regions.

\end{abstract}

\section{Introduction}

Monte Carlo~(MC) simulations are widely used in high-energy physics~(HEP) to model physics processes and the behavior of particles and their interactions within a detector. However, while this approach is very successful in general, it specifically fails to accurately describe occurrences where the detector and reconstruction algorithms \textit{failed} to perform as expected, e.g.\ to correctly reconstruct and identify leptons. The underlying cause is that the detector response is complex and consequently not understood well enough to be accurately modeled in MC simulation. Furthermore, while in the recent years there has been significant progress in the accuracy of MC simulations describing these mis-reconstructed and mis-identified objects, the other intrinsic MC limitation remains, namely the limited statistics of the simulated samples. This is especially true for processes with large cross-sections, where the mis-reconstructed objects appear only in small fractions of events, leading to very limited MC statistics in the kinematic regions of interest. A good example also comes from flavor physics, where an additional limitation comes from the large number of hadron decay modes that must be accounted for, often involving partially reconstructed final states, which makes a complete simulation-based description impractical~\cite{LHCb:2022vje}. To sum up, the \textit{data-driven} approaches are used to estimate the contributions of such types of processes directly from data.

One of the approaches commonly used for this purpose is the Fake Factor method~\cite{LEHMANN2023168376,Aad_2023}. It is used to estimate the contribution of events with mis-reconstructed and/or mis-identified leptons (collectively called \textit{fake leptons}) in the kinematic region where one tries to measure (discover) a physics process of interest, often referred to as the \textit{signal region}~(SR). The estimation is done by extrapolating the fake lepton contribution from data in a kinematically adjacent, \textit{looser} kinematic region~(SR$^{\text{L}}$)\footnote{In this paper, the label SR$^{\text{L}}$ corresponds to the region defined by the loose lepton selection and excluding events passing the nominal (tight) selection. This region corresponds to what is sometimes also referred to as the \emph{application region} (AR) in the Fake Factor method. Analogously, CR$^{\text{L}}$ refers to the control region with the loose lepton selection applied.}, where the term \textit{loose} is commonly used and refers to the implemented lepton selection requirements that are less stringent than the usual (tight) selection used in the SR. The extrapolation is done using a scale factor (i.e.\ fake factor) function.

The fake factor is thus defined as the ratio of two probability density functions, given approximately by the number of events with a fake lepton passing the nominal selection criteria to the number of events with a fake lepton passing looser selection criteria. The fake factor function itself is typically evaluated in a dedicated, fake lepton enriched, kinematic \textit{control region}~(CR) by using either nominal or looser selection criteria~(CR$^{\text{L}}$). The fake factor function, derived in a standard way,  has binned (piece-wise constant) values, since it is estimated by binning (histogramming) the data and computing the ratio of the two data distributions in CR and CR$^{\text{L}}$, typically as functions of a few relevant physics variables such as $\pt$ and $|\eta|$. Depending on the analysis, additional variables correlated with the overall event topology, such as $\MET$, may also be included. Such variables are not intrinsically related to lepton identification, but are instead sensitive to variations in event topology, background composition, and global event kinematics. In practice, this indirect (non-intrinsic) dependence is often reduced by parameterizing the fake rate in terms of variables more directly connected to the fake-lepton production mechanism, accounting for the mother parton type and kinematics, for example reconstructing the mother particle $\pt$~\cite{Khachatryan_2016}. The CR is designed to be enriched in fake leptons, while still being kinematically similar to the SR. The fake factor function is then applied to the events in SR$^{\text{L}}$ to extrapolate the background contribution coming from fakes to the SR.
This binned Fake Factor method however suffers from several challenges, such as binning selection, extrapolation uncertainties to the SR, and the limited ability to be parameterized along more than a couple of variables due to lack of statistics, to name a few. To overcome these limitations, we propose a novel approach, based on machine learning~(ML) using (real) data, which can be applied in higher-dimensional feature spaces and provides a continuous, unbinned fake factor estimate
on a per-event basis that extrapolates better to the SR.

In this paper we aim to give an introduction to the ML-based Fake Factor method and demonstrate its advantages compared to the traditional binned Fake Factor method, especially in terms of interpolation power lost due to binning artifacts, as well as its potential to extend to higher-dimensional feature spaces. The paper aims to demonstrate broad applicability of this method to diverse analyses in high energy physics, especially those involving leptons in final states.
The paper is organized as follows. Section~\ref{sec:sec2} provides an overview of the binned Fake Factor method and its connection to the related Matrix and ABCD methods, and highlights the challenges of these methods. Section~\ref{sec:sec3} introduces the ML-based Fake Factor method. In Section~\ref{sec:sec4}, we validate the ML-based approach using the ATLAS Open Data sample and present the training results. Section~\ref{sec:sec5} compares the final results of the ML-based and binned Fake Factor methods. Finally, Section~\ref{sec:sec6} summarizes our conclusions.

\section{Overview of Data-Driven Background Estimation Methods}\label{sec:sec2}

The most popular data-driven methods for estimating fake lepton backgrounds in HEP analyses are the Matrix method, the Fake Factor method, and the ABCD method~\cite{Aad_2023, LEHMANN2023168376}. All three methods rely on defining two different lepton selection criteria: a \emph{tight} selection (the nominal selection used in a physics analysis, i.e.\ the signal region) and a \emph{loose} selection (a looser selection that defines an adjacent kinematic region, excluding the tight one). The union of both categories is called the \emph{baseline}
selection. By definition, the set of leptons passing the tight selection must be
a subset of those passing the baseline selection and does not include loose leptons.

We define the real efficiency $r$ as the probability that a correctly
reconstructed (real) lepton passing the baseline selection also passes the tight
selection, and, analogously, the fake efficiency $f$ as the probability that a
fake lepton passing the baseline selection also passes the tight selection. In
the simple case of a single lepton, the relationship between the numbers of
tight and loose leptons in data and their composition in terms of real and fake
leptons can be written as
\begin{equation}
   \begin{pmatrix}
     N^{\text{T}} \\
     N^{\text{L}}
   \end{pmatrix}
   =
   \begin{pmatrix}
     r & f\\
     1 - r & 1 - f
   \end{pmatrix}
   \begin{pmatrix}
     N_{r} \\
     N_{f}
   \end{pmatrix} \>.
   \label{eq:mm}
\end{equation}
Both $r$ and $f$ are to be estimated in separate dedicated kinematic regions (control regions). Using (state-of-the-art) MC simulation to estimate the real efficiency $r$ can be considered as reliable due to a good understanding of real lepton reconstruction and identification efficiencies in the detector. Additional data-driven corrections (scale factors) can be applied to MC to further improve the modeling of real leptons. On the other hand, the fake efficiency $f$ is generally not modeled to the same precision in MC simulation, so it is typically estimated directly from data in a fake-lepton enriched control region.

\subsection{The Matrix Method}
The Matrix method estimates the number of fake leptons in the tight region by explicitly inverting Eq.~\eqref{eq:mm}, relating the numbers of tight and loose leptons in data to their composition in terms of real and fake leptons. The method can in principle be made fully data-driven, as it does not need to rely on MC simulation to estimate any contributions, but it can suffer from possible measurement bias, as it uses data information from the signal region itself\footnote{In other words, the analysis is not completely \emph{blinded} in the SR before the final statistical analysis when using this method, and it can thus impact the evaluation of possible signal presence.}.

The number of events with a fake lepton using the tight (i.e.\ nominal) selection, found by inverting the above equation, is given by
\begin{equation}
    N_{f}^{\text{T}} = f N_f = \frac{f}{f-r}\left( (1-r)N^\text{T} - rN^\text{L}\right) \>.
\end{equation}
This equation relates the number of events in the tight and loose regions to the expected number of fake leptons in the tight region.
As already stated, since $N^{\text{T}}$ (i.e.\ the measured data in the SR) appears in the equation, information from the signal region enters the estimate, so the method is not fully blinded.

\subsection{The Fake Factor Method}
The Fake Factor method estimates the contribution of events that only contain real leptons, i.e.\ $N_{r}^{\text{T}}=r N_{r}$ and $N_{r}^{\text{L}}=(1-r) N_{r}$ in Eq.~\eqref{eq:mm}, from MC simulation predictions. The method is thus no longer fully data-driven, as it relies on MC to estimate the contribution of events with real leptons. Nonetheless, as already argued, real leptons are generally well modeled in MC, so this is not a significant degradation. On the other hand, the method does not utilize any data information from the SR, which gives it a formal advantage in terms of possible measurement bias with respect to the Matrix method, as this method can be used with a fully blinded signal region.

The number of fake leptons in the loose region can be obtained from Eq.~\eqref{eq:mm} as
\begin{equation}
    \label{eq:NfL}
    N_f^\text{L} = N^\text{L} - N_r^\text{L} = (1 - f) N_f \>.
\end{equation}
Using Eq.~\eqref{eq:mm} with Eq.~\eqref{eq:NfL}, we can express the number of fake leptons in the tight region as
\begin{equation}
    \label{eq:NfT}
    N_f^\text{T} = f N_f = F N_f^\text{L} \>,
\end{equation}
where $F$ is the fake factor defined as
\begin{equation}
    F = \frac{f}{1-f} \>.
    \label{eq:ff}
\end{equation}
Alternatively, Eq.~\eqref{eq:NfT} can be derived directly from Eq.~\eqref{eq:mm}
by multiplying the left-hand side by a row vector with the entries $(1, -F)$, as
is shown in~\cite{LEHMANN2023168376}.

The fake factor in Eq.~\eqref{eq:ff} needs to be evaluated in a control region where the fake lepton contribution can be identified without looking at the data in the (`blinded') signal region. We can write the fake factor as
\begin{equation}
    \label{eq:ff_main}
    F = \frac{N_f^\text{T}}{N_f^\text{L}} = \frac{N^\text{T} - N_r^\text{T}}{N^\text{L} - N_r^\text{L}}
    = \frac{N_\text{data}^\text{T} - N_\text{MC}^\text{T}}{N_\text{data}^\text{L} - N_\text{MC}^\text{L}} \>,
\end{equation}
which can be estimated in a control region by subtracting the contribution of real leptons ($N_r$) from data using MC simulation predictions to estimate the real lepton contribution.

The Fake Factor method (and the Matrix method) can be extended to multiple leptons. For two leptons, one distinguishes four categories: \textit{{tight--tight (TT)}}, \textit{{tight--loose (TL)}}, \textit{{loose--tight (LT)}}, and \textit{{loose--loose (LL)}}.
Efficiency matrices analogous to Eq.~\eqref{eq:mm} can be constructed that relate the fake and real composition of the different categories. Using the Fake Factor method, the number of events with two fake leptons in the tight region can be expressed as
\begin{equation}
    N_{\text{data}}^\text{TT}
    = F_1\!\left(N_{\text{data}}^\text{LT}-N_{\text{MC}}^\text{LT}\right)
    + F_2\!\left(N_{\text{data}}^\text{TL}-N_{\text{MC}}^\text{TL}\right)
    - F_1F_2\!\left(N_{\text{data}}^\text{LL}-N_{\text{MC}}^\text{LL}\right) \>,
\end{equation}
where $F_1$ and $F_2$ are fake factors calculated for the first and second
lepton candidate respectively. This expression assumes that the probabilities for the two lepton candidates to satisfy the tight selection are factorizable (i.e.\ the fake rates of the two leptons are independent). This assumption may not be valid in the case where fake leptons are produced in a correlated way, for example when 2 leptons have an origin from the same heavy-flavor jet, and in which case we need to use dedicated methods (as discussed e.g.\ in the long-lived HNL search in Ref.~\cite{CMS_HNL_2024}).

\begin{figure}[H]
	\centering
	\includegraphics[width=0.55\linewidth]{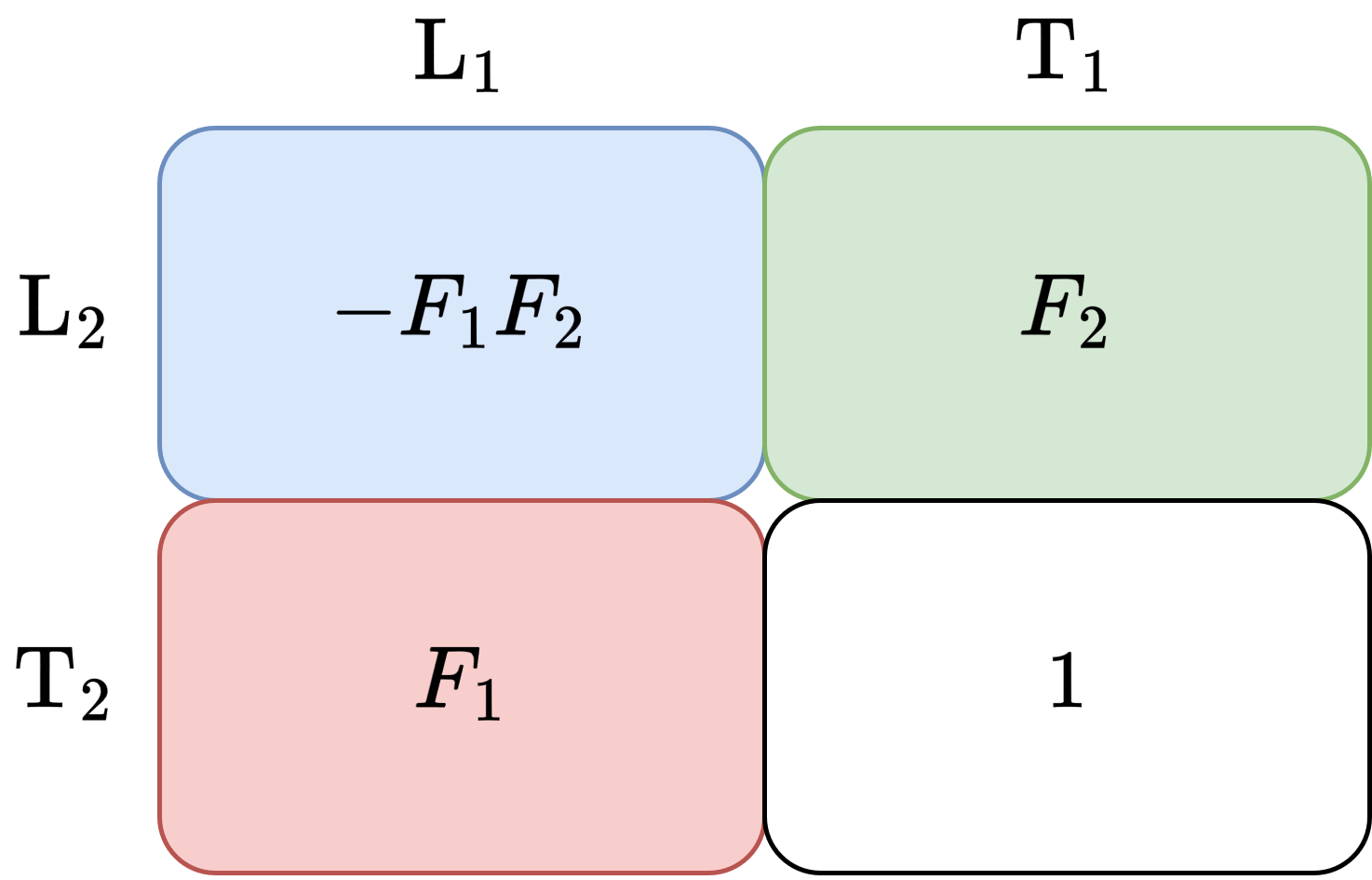}
	\caption{Fake factor method diagram in case of two leptons. In events with two leptons, the Fake Factor method is applied individually to each lepton. For LT and TL combinations, only one fake factor is assigned. In the signal region, where both leptons are tight (TT), the event weight is 1. When both leptons are loose (LL), two fake factors are applied for both leptons.}
	\label{fig:ff-2scheme}
\end{figure}

This equation is visualized in
Figure~\ref{fig:ff-2scheme} and is used in analyses with two leptons in the
final state, for example in Refs.~\cite{dch, typeIII, Khachatryan_2016, CMS_HNL_DV_2022}. Related fake-rate implementations in multi-lepton final states are used in ~\cite{CMS_HNL_2024}.

\subsection{The Binned Fake Factor Method}

\begin{figure}[H]
    \centering
    \includegraphics[width=0.5\textwidth]{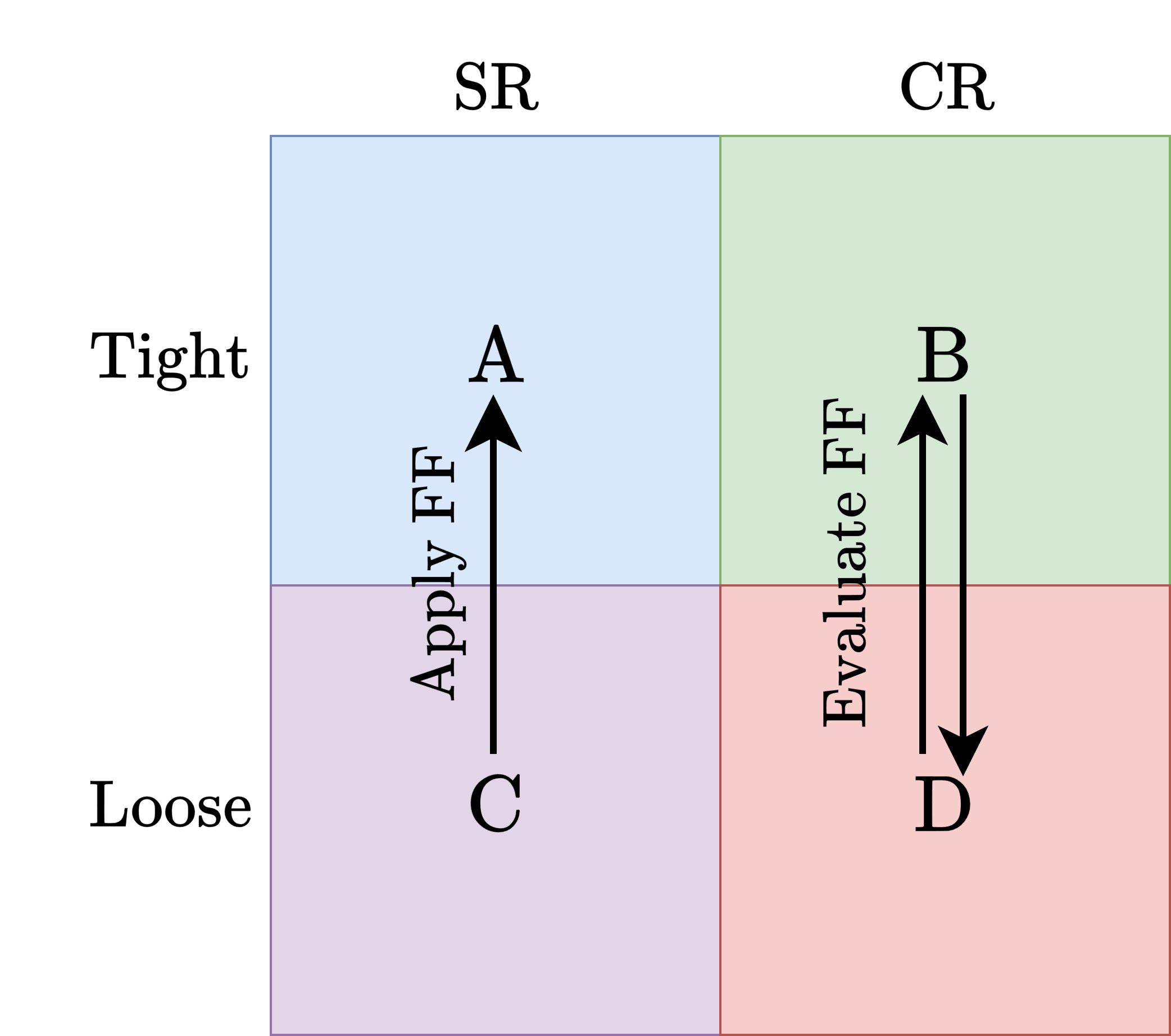}
    \caption{
        Visualization of the ABCD method, which is equivalent to the Fake Factor method when using only event counts. To estimate the number of fake leptons
        in region~A, the ratio of the number of fake leptons in kinematically
        orthogonal tight and loose regions B and D is evaluated first, giving
        the value of $F$. Since this ratio is assumed to be equal in both SR and
        CR, the number of fake leptons in region~A can be obtained by applying
        $F$ as a transfer factor to the number of fake leptons in region~C.
    }
    \label{fig:abcd2}
\end{figure}

Using simple event-counts, as the methods were in fact presented above, the fake factor is a single number. In such a case, the Matrix and Fake Factor methods are equivalent to the ABCD method~\cite{buttinger2018background,ABCDisCo}. The ABCD relation follows directly
from the same tight/loose factorization used above. In the single-lepton case, the
binned fake factor estimate (Eq.~\eqref{eq:NfT}) is equivalent to the ABCD relation:
\begin{equation}
  N_\text{A}=N_\text{C}\,\frac{N_\text{B}}{N_\text{D}}
  =\big(N_{\text{data}}^\text{L} - N_{\text{MC}}^\text{L}\big)\,
       \frac{N_\text{data}^{\prime\,\text{T}} - N_\text{MC}^{\prime\,\text{T}}}
            {N_\text{data}^{\prime\,\text{L}} - N_\text{MC}^{\prime\,\text{L}}} \>,
\end{equation}
with $N_f^\text{T}=N_\text{A}$,  $N_f^\text{L}=N_\text{C}$ and $F=N_\text{B}/N_\text{D}$. Here, regions $\text{A}$ and $\text{C}$ correspond to the tight and loose signal regions, while regions $\text{B}$ and $\text{D}$ correspond to the tight and loose control regions, respectively. The primed quantities refer to the control region. Real-lepton contamination is subtracted in both regions before building the ABCD relation. The visualization of the method is shown in Figure~\ref{fig:abcd2}.

In the actual implementations of LHC analyses, however, the fake factor cannot be considered a constant, but rather a function of some relevant kinematic variables, such as transverse momentum  $\pt$, pseudorapidity $|\eta|$, with additional topology-correlated variables (e.g.\ $\MET$) included in some analyses.
Consequently, the fake factor is evaluated as a function of these variables by binning (histogramming) the data in the control region and computing the ratio of the two data distributions in the nominal (tight) CR and loose CR$^\text{L}$.  The resulting binned fake factor function is then applied to data in the loose SR$^\text{L}$ region on a per-event basis, by assigning each event the fake factor corresponding to the bin in which the event falls.

The number of fake leptons in the SR can thus be estimated by applying the fake factor function to data in the loose SR$^\text{L} $ region, subtracting the contribution of real leptons using MC simulation. Equation~\eqref{eq:NfT} then evolves into predicting the number of fake leptons in the tight region for each kinematic bin $b$, defined by the chosen binning scheme and kinematic observables:
\begin{equation}
    \label{eq:ff_application}
    N_{f,b}^\text{T} = \sum_{i \in b} F_{\text{data},i} - \sum_{j \in b} F_{\text{MC},j} \>.
\end{equation}
In the above equation, the fake factor introduces an event weight given as a function of $\pt$, $|\eta|$ and, where applicable, $\MET$, applied as a weight to events
in the loose region. This is referred to as the \emph{binned Fake Factor method}.

The binned Fake Factor method suffers from several challenges. Most importantly, the choice of binning can significantly impact the results, as too coarse binning can lead to a loss of important kinematic features, while too fine binning can result in statistical fluctuations and empty (or even negative) bins, due to poor data and MC statistics. Furthermore,  the method is typically limited to a few dimensions due to statistical limitations, which can restrict its ability to capture complex dependencies in the data and degrade the extrapolation to the SR. Finally, the method can suffer from artifacts due to binning, producing discontinuities in the fake factor function and affecting the stability of the background estimation.

\section{The ML-Based Fake Factor Method}\label{sec:sec3}
The machine-learning based Fake Factor method, presented in this paper, is a novel approach to the traditional binned Fake Factor method. The main
idea is to use machine learning to estimate an \emph{unbinned (continuous)} fake factor function. This is achieved by estimating the probability densities, specifically the density ratio, of the real-subtracted tight and loose samples using neural networks. This is in fact related to the very active topic of simulation-based inference~(SBI) approaches for likelihood ratio estimation~\cite{KyleSBI2020}, albeit in this case based on data itself, which we can refer to as the data-based inference~(DBI) approach.

\subsection{Neural Fake Factor Estimation}

The goal is to estimate the fake factor defined in Eq.~\eqref{eq:ff_main}. By substituting the numbers of events with the corresponding probability densities, i.e.\ normalizing the numerator and denominator by the total number of baseline selection events $N_f=N_f^\text{T}+N_f^\text{L}$, we can rewrite the fake factor as a probability ratio:
\begin{equation}
    \label{eq:rr_def}
    F = \frac{N_f^\text{T}}{N_f^\text{L}} = \frac{p^\text{T}}{p^\text{L}}=\frac{f}{1-f} \>,  \quad p^{\text{T,L}} = \frac{N_f^{\text{T,L}}}{N_f} \>,
\end{equation}
reinterpreting $F$ of Eq.~\eqref{eq:ff} as the ratio of the probability densities of fake leptons in the tight and loose regions. Given event observables $\vx$, this translates to determining the functional dependence of $F$ on the observables $\vx$ as estimating the \emph{density ratio}:
\begin{equation}
    \label{eq:ff-ratio}
    F(\vx) = r_F(\vx) = \frac{p^{\text{T}}(\vx)}{p^{\text{L}}(\vx)} \>
\end{equation}
So far, this has been done using binning/slicing in some low-dimensional space, using a few observables, e.g.\ $\pt$ and $|\eta|$, where the binning was necessarily coarse due to limited statistics. Here we propose to estimate the fake factor as the density ratio in a higher-dimensional feature space using neural networks, thus avoiding binning artifacts and enabling a more precise and flexible estimation of the fake factor.

The density ratio $r(\vx)$ of two probability densities can be estimated by training a binary classifier to distinguish
two hypotheses (loose and tight) using the \emph{likelihood ratio trick}~\cite{Nachman2023ratios,cranmer2016lr,sugiyama2012density}. There are many choices for the loss function for training
the classifier~\cite{pmlr-v48-menon16}. We use the binary cross-entropy and
a squared regularization term to prevent `exploding' densities. The loss function
is given by:
\begin{equation}
    \label{eq:loss}
    \mathcal{L}(\vtheta)
    = \sum_{i=1}^{N} w_i \left[
        - y_i \log \sigma\!\left(q(\vx_i; \vtheta)\right)
        - (1 - y_i) \log \!\left(1 - \sigma\!\left(q(\vx_i; \vtheta)\right)\right)
    \right]
    - \sum_{i=1}^{N} \lambda q(\vx_i; \vtheta)^2 \>,
\end{equation}
where $\sigma$ is the sigmoid function, $y_i \in \{0, 1\}$ are the binary labels corresponding to the two hypotheses, $w_i$ are the event weights, $q(\vx; \vtheta)$ is the classifier output (in this case used as the \emph{logit} value), $\vtheta$ are the model parameters, $\lambda$ is a regularization parameter, and $N$ is the batch size. After the classifier has been trained, the density ratio can be estimated as $r(\vx) = \exp(q(\vx; \vtheta))$. The training dataset this contains events from both the tight and loose regions, with labels  $y_i$ assigned accordingly.
We can then use the ratio as the fake factor function on a per-event basis, reweighting loose events by assigning each event the fake factor corresponding to its features $\vx$.

A special consideration is needed to account for the real lepton contamination in both the tight and loose regions, which needs to be subtracted before estimating the density ratio, i.e.\ $N_f^\text{T,L}=N^\text{T,L}_{\text{data}}-N^\text{T,L}_{\text{MC}}$ as given by Eq.~\eqref{eq:ff_main}. Simulated MC events (and sometimes also data events) are in practice  weighted events, to account for scaling to the data luminosity and process cross-section, simulation corrections (scale factors) and similar. The probabilities are thus given as (cf.\ Eq.~\eqref{eq:rr_def}):
\begin{equation}
  p^\text{T,L}(\vx) \simeq \frac{1}{N_f}\bigg[ \sum_{i\in \Omega(\vx)} w^\text{T,L}_\text{data,i} - \sum_{j \in \Omega(\vx)} w^\text{T,L}_\text{MC,j}\bigg]\>,
\end{equation}
where $\Omega(\vx)$ denotes the set of events with features close to the value of $\vx$. The number of data and MC events in the tight and loose regions needs to be high enough to avoid too sparsely populated kinematic regions, resulting in locally negative probability estimation, but the formula is valid and only requires careful checks of possible
negative probability values in low-statistics regions.

Subsequently, to incorporate this necessary subtraction, the ML training procedure needs to account for weighted events, including events contributing negative weights due to subtraction.
Using negative weights in the ML loss function definition has been an open topic of discussion in the HEP ML community~(advanced MC generators in fact produce a fraction of negative-weighted events, see e.g.\ Ref.~\cite{Nachman2020negweights}), but we note that in this case it is a well-defined problem, because what we are doing in ML training is estimating the model density using cross-entropy, as described in~\cite{goodfellow2016deep}:
\begin{equation}
    \begin{aligned}
        \mathcal{L(\vtheta)}&=-\mathbb{E}_{\vx\sim p^\text{T,L}}\!\big[\log{p^\text{T,L}_\text{model}(\vx; \vtheta)}\big] \\
        &\simeq \frac{1}{N_f} \bigg[\sum_{i \in \Omega(\vx)} w^\text{T,L}_\text{data,i} \log{p^\text{T,L}_\text{model}(\vx_i; \vtheta)} -\sum_{j \in \Omega(\vx)} w^\text{T,L}_\text{MC,j} \log{p^\text{T,L}_\text{model}(\vx_j; \vtheta)} \bigg] \>.
    \end{aligned}
\end{equation}
As long as the data and simulation statistics are sufficiently high, the model density $p_\text{model}(x)$ can be learned.

We can improve the method further by again introducing the density ratio trick to estimate the (real-)subtracted densities in both the tight and loose regions separately, determining a ratio that would reweight either data or MC events to the subtracted density. This is similar to the approach used in Ref.~\cite{Nachman2020reweighting}, where the authors use classifiers to reweight simulated samples to match data distributions. Here, we apply this idea to reweight either data or MC samples to obtain the subtracted densities $p^\text{T,L}(\vx)$ in  tight and loose regions. We introduce a ratio of  data to MC (it works symmetrically for the opposite MC to data ratio as well), using Eq.~\eqref{eq:ff_main}:
\begin{equation}
    \label{eq:rtl}
    r^{\text{T,L}} = \frac{N_\text{data}^{\text{T,L}}}{N_\text{MC}^{\text{T,L}}} \>, ~~~~ N_f^{\text{T,L}}= N_\text{data}^{\text{T,L}}-N_\text{MC}^{\text{T,L}}=N_\text{data}^{\text{T,L}} \left(1-\frac{1}{r^{\text{T,L}}}\right) = N_\text{MC}^{\text{T,L}}\left(r^{\text{T,L}}-1 \right)  \>.
\end{equation}
By learning this ratio, we can effectively reweight either data or MC to obtain the subtracted densities in both tight and loose regions.
Furthermore, this approach has to ensure that the derived correction factors are positive by construction, which means we need to ensure that the  ratio $r^{\text{T,L}}$ is always greater than one.  The ratio approach also greatly improves the stability of the learning process, which we find to be an important achievement in itself and crucial in practice. The loss function remains the same as in Eq.~\eqref{eq:loss}, with the labels assigned accordingly to distinguish the data and MC events in this case.

\begin{figure}[H]
    \centering
    \includegraphics[width=0.67\linewidth]{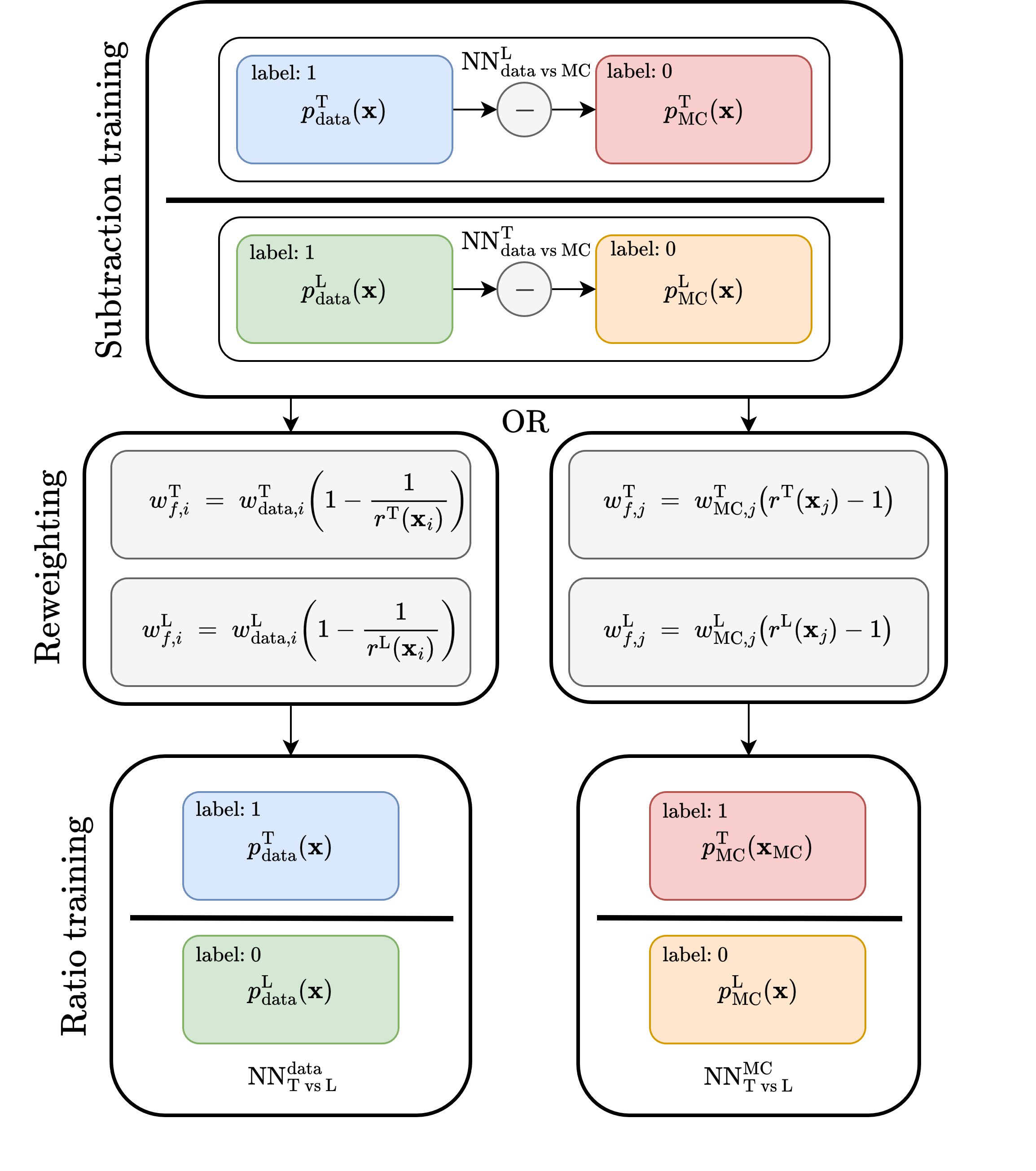}
    \caption{
        Flow diagram of the ML-based method to obtain the fake factor $F$ as a
        density ratio $r_F(\vx)$. Firstly, two independent classifiers are
        trained in the tight and loose regions to model the ratios
        $r^\text{T,L}$ between data and MC. These can then be used as correction
        factors to obtain prompt-subtracted densities by reweighting either data
        or MC events, giving the two branches of the diagram. Lastly, a third
        classifier is trained on reweighted events to separate loose and tight
        prompt-subtracted distributions, which gives the final density ratio
        $r_F(\vx)$.
    }
    \label{fig:mlff-scheme}
\end{figure}

The complete ML fake factor estimation method consists of two steps: modeling the \emph{subtraction} and \emph{ratio}. The subtraction step is applied separately in the tight and loose regions to obtain real-subtracted densities. The ratio step then estimates the fake factor as the ratio of the real-subtracted tight and loose samples. The overall procedure is shown in Figure~\ref{fig:mlff-scheme} and detailed below.

\paragraph{Subtraction Step}
We construct the numerator and denominator terms of the fake factor density ratio $r_F$ of Eq.~\eqref{eq:ff-ratio} by
applying the subtraction procedure separately in the tight and loose regions, as defined in Eq.~\eqref{eq:rtl}. We train two independent (\textit{subtraction}) classifiers and use them to derive correction factors $r^{\text{T,L}}$. These correction factors are then used to reweight either data or MC samples on an event-by-event basis. Since the data and MC events are weighted, we include the event weights and write the  reweighting procedure on an event-by-event basis as:
\begin{equation}
    \label{eq:subtraction}
    w_{f,i}^{\text{T,L}} = w_\text{data,i}^{\text{T,L}} \left( 1 - \frac{1}{r^\text{T,L}(\vx_i)} \right)
    \quad \text{or} \quad w_{f,j}^{\text{T,L}}  =
    w_\text{MC,j}^{\text{T,L}} \left( r^{\text{T,L}}(\vx_j) - 1 \right) \>.
\end{equation}
To obtain the functions $r^\text{T,L}(\vx_i)$, we train two classifiers to estimate these two ratios separately on the tight (T) and loose (L) datasets, constructed as a union of data and MC samples, setting MC labels to 0 and data labels to 1:

\begin{equation}
    \mathcal{D}^{\text{T,L}} = \{ (\vx^\text{T,L}_\text{data}, \vw^\text{T,L}_\text{data}, \bm{1}) \}
    \cup \{ (\vx^\text{T,L}_\text{MC}, \vw^\text{T,L}_\text{MC}, \bm{0}) \} \>.
\end{equation}
In our implementation, we chose to reweight the data rather than the MC to stay more data-driven (left formula of Eq.~\eqref{eq:subtraction}), although either approach is valid. We need to ensure that the ratio condition $r^{\text{T,L}}(\vx) > 1$ is met, to keep the weights positive.
This is straightforward to achieve by requiring the classifier logit outputs $q^{\text{T,L}}(\vx;\vtheta)$ to be positive, since the ratio function is then $r^{\text{T,L}}(\vx) =\exp(q^{\text{T,L}}(\vx;\vtheta))$, as defined by the likelihood ratio trick and the loss function of Eq.~\eqref{eq:loss}.
The actual implementation of this constraint is given in Section~\ref{sec:model-arch}.

\paragraph{Ratio Step}
After subtraction, we train a third (\textit{ratio}) classifier to distinguish the
numerator (tight subtraction) from the denominator (loose subtraction),
using the new weights derived in the subtraction steps described  above in the new loss function, constructed again as given by Eq.~\eqref{eq:loss}. This will give us the final ratio $r_F(\vx)$ to be used as the fake factor function $F(\vx)$.
According to the reweighting choice made in the subtraction step, we train an ML model on data
\begin{equation}
    \mathcal{D}_\text{data} = \{ (\vx^\text{T}, \vw_{f}^{\text{T}}, \bm{1}) \}
    \cup \{ (\vx^\text{L},\vw_{f}^{\text{L}}, \bm{0}) \} \>.
\end{equation}
The end result is a parametric model $r_F: \vx \in \mathbb{R}^d \rightarrow \mathbb{R}^+$ that estimates the fake factor given an event that has features $\vx$  of dimension $d$.

\subsection{Model Architecture and Training Setup}
\label{sec:model-arch}

We begin by preprocessing the data. Numerical features are standardized to have
zero mean and unit variance, while categorical features are label-encoded. The
dataset is then split into training (\qty{50}{\percent}), validation
(\qty{25}{\percent}), and test (\qty{25}{\percent}) sets. During training, we
monitor the validation loss to prevent over-fitting and select the best neural network classifier model
based on validation performance. If the validation loss does not decrease for five consecutive epochs, the learning rate is reduced by a factor of \num{0.5}.

Before feeding the data into the model, we apply an embedding layer to both
categorical and numerical features~\cite{gorishniy2021revisiting,
gorishniy2022embeddings}. This maps each feature into a higher-dimensional
space, enabling the model to learn richer representations of the input data. All
embeddings are concatenated into a single feature vector, and mean pooling is
applied across the feature dimension to obtain the final input representation
for the model.

We train three binary classifiers with the same architecture for density ratio
estimation using the loss function defined in Eq.~\eqref{eq:loss}. Training is
performed with the AdamW
optimizer~\cite{loshchilov2019decoupledweightdecayregularization}, using a
learning rate of \num{3e-4}, a weight decay of \num{e-5}, a batch size of
\num{1024} epochs, and training taking up to \num{100} epochs.
Early stopping is applied based on validation loss. The regularization parameter $\lambda$ in
the loss function is set to \num{0.01} for the ratio classifier and \num{0} for the subtraction classifiers. We notice that
the higher regularization translates into better extrapolation to the signal region, but worse performance
in the control region, so this is a hyperparameter that can be tuned based on the specific analysis needs.

\begin{figure}
    \centering
    \hspace*{1.5cm}
    \includegraphics[width=0.31\linewidth]{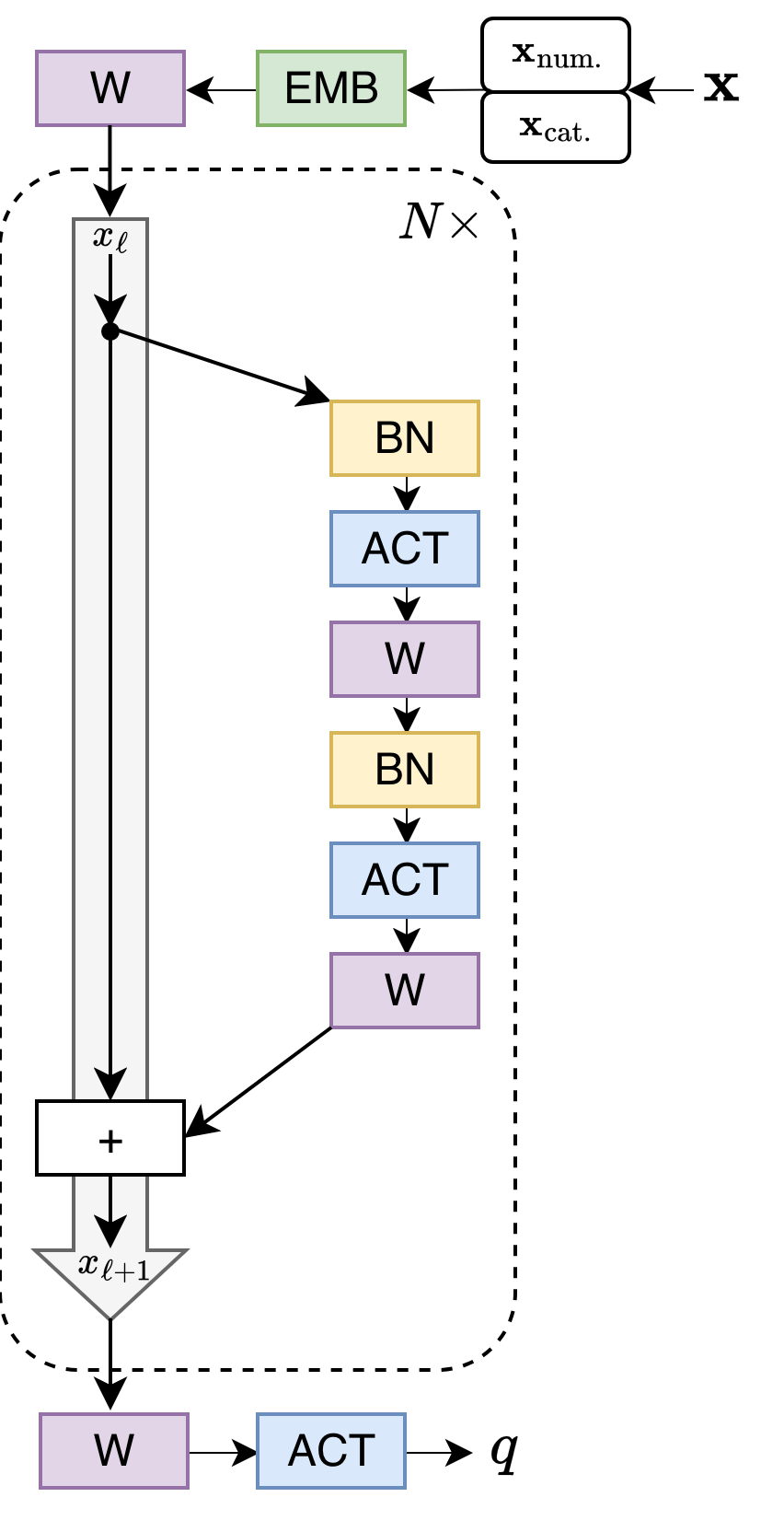}
    \caption{Schematic illustration of the classifier model architecture used in this work. Our classifiers use a pre-activation residual network~(ResNet) architecture. The numerical features $\vx_\text{num.}$, concatenated with the categorical features $\vx_\text{cat.}$, are embedded through an embedding layer~(EMB) and then passed through a projection layer (if needed) before being fed into the ResNet. The ResNet is schematically shown as a stack of batch normalization~(BN), weight multiplication~(W), and activation function~(ACT) layers, with residual connections between them. The output layer uses either a soft absolute or linear activation function to produce the final output (logit) value $q$, as described in the text.}
    \label{fig:resnet}
\end{figure}

For the classifier model architecture, schematically shown in Figure~\ref{fig:resnet}, we
adopt a \textit{pre-activation ResNet}~\cite{he2016identity} with four residual
blocks, each containing two layers of \num{128} neurons. We found that batch
normalization and dropout did not improve performance, so they are omitted in
the final model, although they can be easily added if needed. ReLU activations
are used after each layer except for the output layer. The network input
consists of the concatenated feature embeddings, with an optional projection
layer applied if the embedding dimension does not match the network input
dimension. The output layer is a single projection layer with one neuron that
produces the result $q$ (interpreted as the logit value) for the binary classification task.

We use the ResNet architecture because residual connections mitigate the problem of vanishing gradients, enabling deeper networks to be trained effectively. This design allows the classifier to learn more expressive transformations without suffering from performance degradation as depth increases. In practice, we found that using residual connections leads to more stable training dynamics and consistently better density ratio estimate compared to equivalent plain feed-forward networks, where we found that a classifier can over-fit easily.

The output activation function depends on the specific task. For the subtraction classifiers, we use a \textit{soft absolute} activation function to ensure non-negative outputs $q$ (interpreted as the logit value) for both subtraction classifiers, which is required to keep the data correction weights $(1-1/r)$ of Equation  \eqref{eq:subtraction} positive, since $r=\exp(q)$. In the ratio classifier, we use a linear activation function to allow for unbounded outputs.

\begin{figure}
    \centering
    \includegraphics[width=0.5\linewidth]{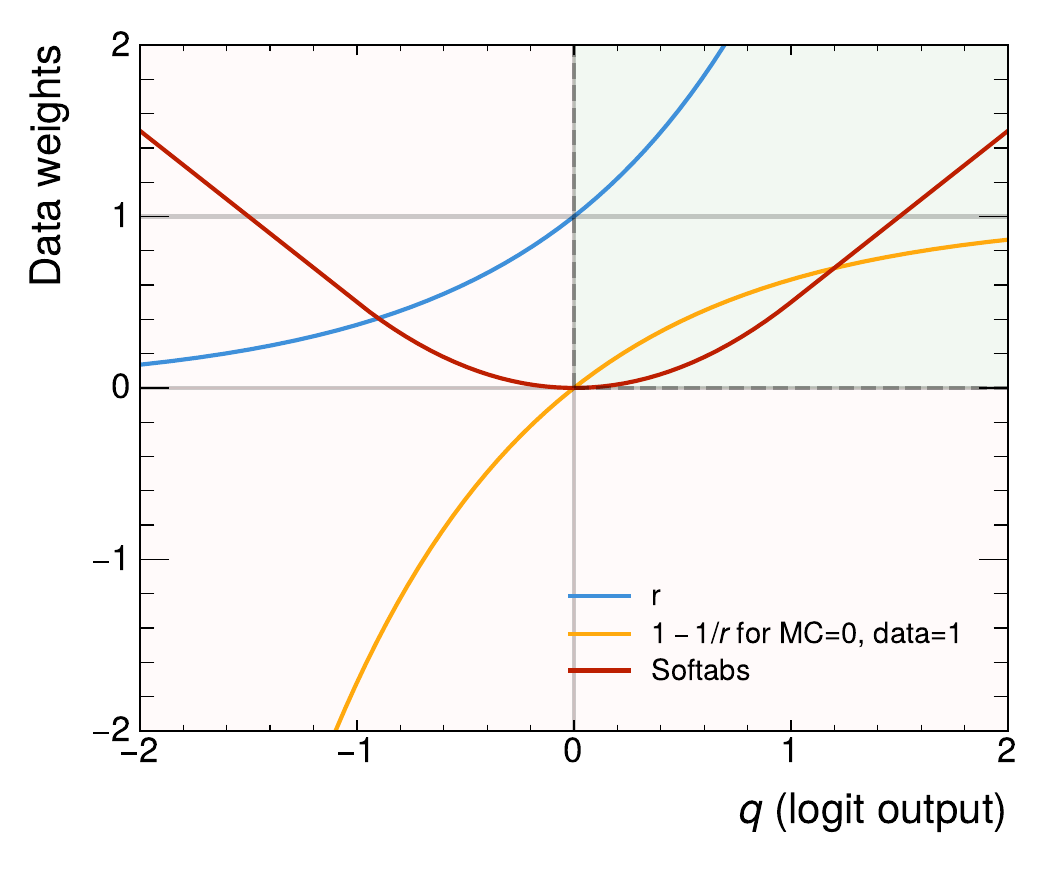}
    \caption{The soft absolute activation function  (red) constrains the (logit) outputs $q$ of both subtraction classifier networks to be non-negative, which is required to keep the data correction weights positive. The exponential of the output $r=\exp(q)$,  is used to obtain the density ratio estimate  (blue), which will be $r>1$ when using the proposed activation function. Data reweighting function (orange) is given in Eq.~\eqref{eq:subtraction}. In our implementation, we reweight data with labels 0 for MC and labels 1 for data. Using the soft absolute activation ensures that the reweighting function remains non-negative and bounded within $[0, 1]$, as required.}
    \label{fig:soft-abs}
\end{figure}

The soft absolute activation function is defined as
\begin{equation}
    f(x) =
    \begin{cases}
        \frac{1}{2} x^2, & \text{if } |x| < 1 \>, \\[6pt]
        |x| - \frac{1}{2}, & \text{otherwise} \>,
    \end{cases}
\end{equation}
which is inspired by the Huber loss~\cite{huber1964robust}. The function is
smooth and differentiable everywhere, and behaves like the absolute value
function for large inputs, while being quadratic near zero. This ensures that
the network outputs are always non-negative as depicted in
Figure~\ref{fig:soft-abs}, which is important for the subtraction step where we
need to avoid negative weights.

\section{Method Validation Using a Representative LHC Analysis}
\label{sec:sec4}
To validate the ML-based Fake Factor method, we define a straightforward data analysis to measure the $W$ boson transverse mass $\mt$ from the $W$ boson decaying into an electron and a corresponding neutrino ($W \to e \nu$). We use the ATLAS Open Data sample
from Run~2 collected in the years \num{2015} and \num{2016}~\cite{ATLAS_Open_Data}.
Events are required to pass single-electron triggers, with the reconstructed electron
required to match the triggering object. Additionally, the requirements are:
$\pt$ of the electron above \qty{25}{\GeV}, missing transverse energy in the event $\MET>\qty{30}{\GeV}$, at least one
jet present and the $b$-jet veto implemented. The main background contributions in this channel after
these selection cuts are  $W$+jets and $t\bar{t}$ events. To define our control and
signal kinematic regions, we use transverse mass $\mt$ defined as
\begin{equation}
    \mt = \sqrt{\,2\,\pt\,\MET\,\left(1 - \cos\left(\Delta \phi\right)\right)} \>,
\end{equation}
where $\Delta \phi$ is the azimuthal angle between the electron and the missing
transverse energy directions. The control region (CR) is defined with the requirement
$\mt<\qty{60}{\GeV}$, while the signal region (SR) is defined with
$\mt>\qty{60}{\GeV}$, to ensure orthogonality. This selection is loosely inspired
by the measurement of the $W$-boson mass in Ref.~\cite{Wboson}.  We deliberately set up  our benchmark analysis to be quite simple in selection, with a minimal set of cuts to demonstrate that it works well for describing the fake contribution across a wide range of kinematics and for substantial real lepton contamination (subtraction) as well.

The events are further
divided into two categories, \textit{tight} and \textit{loose}, referring to the electron reconstruction requirements, that can be used in the Fake Factor method. In these two categories, tight leptons pass the tight identification and loose
isolation criteria, while loose leptons fail both.

\begin{figure}
    \centering
    \begin{subfigure}{0.42\linewidth}
        \includegraphics[width=\linewidth]{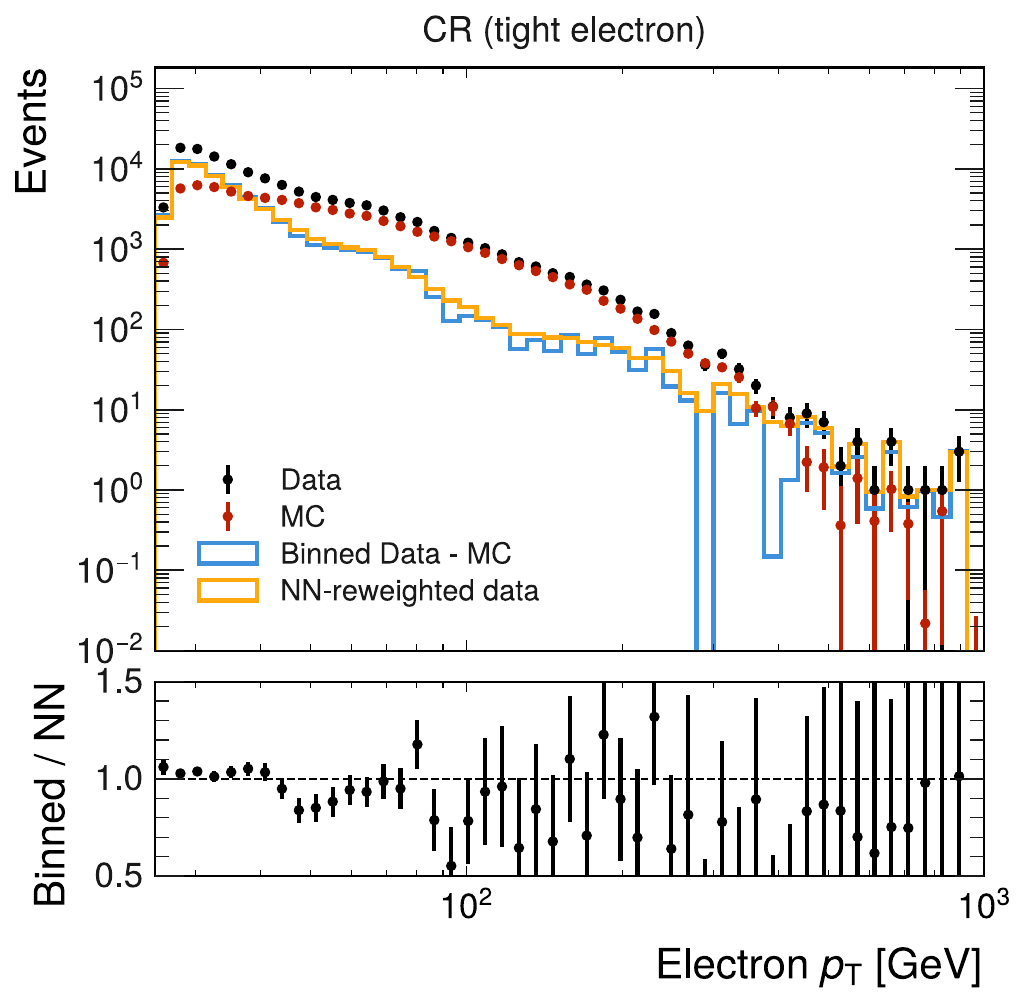}
    \end{subfigure}
    \hspace{4mm}
    \begin{subfigure}{0.42\linewidth}
        \includegraphics[width=\linewidth]{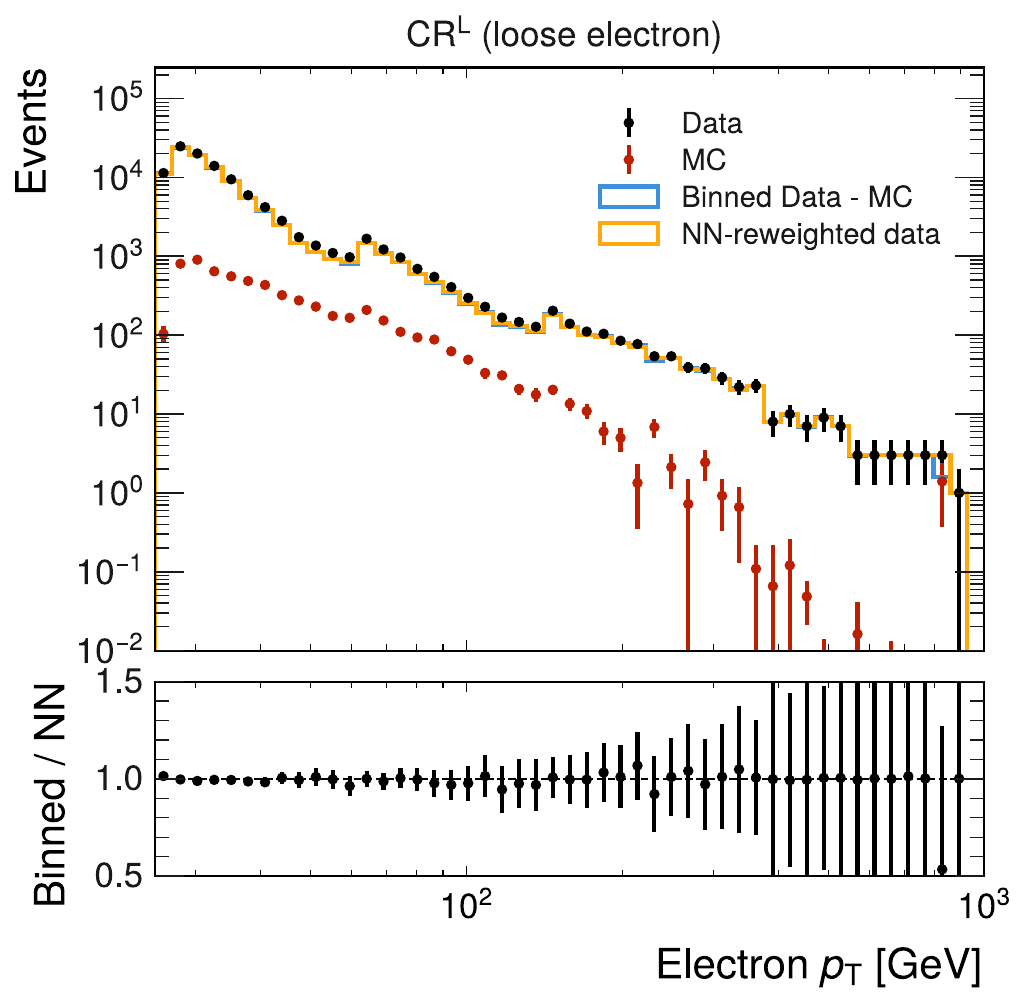}
    \end{subfigure}

    \begin{subfigure}{0.42\linewidth}
        \includegraphics[width=\linewidth]{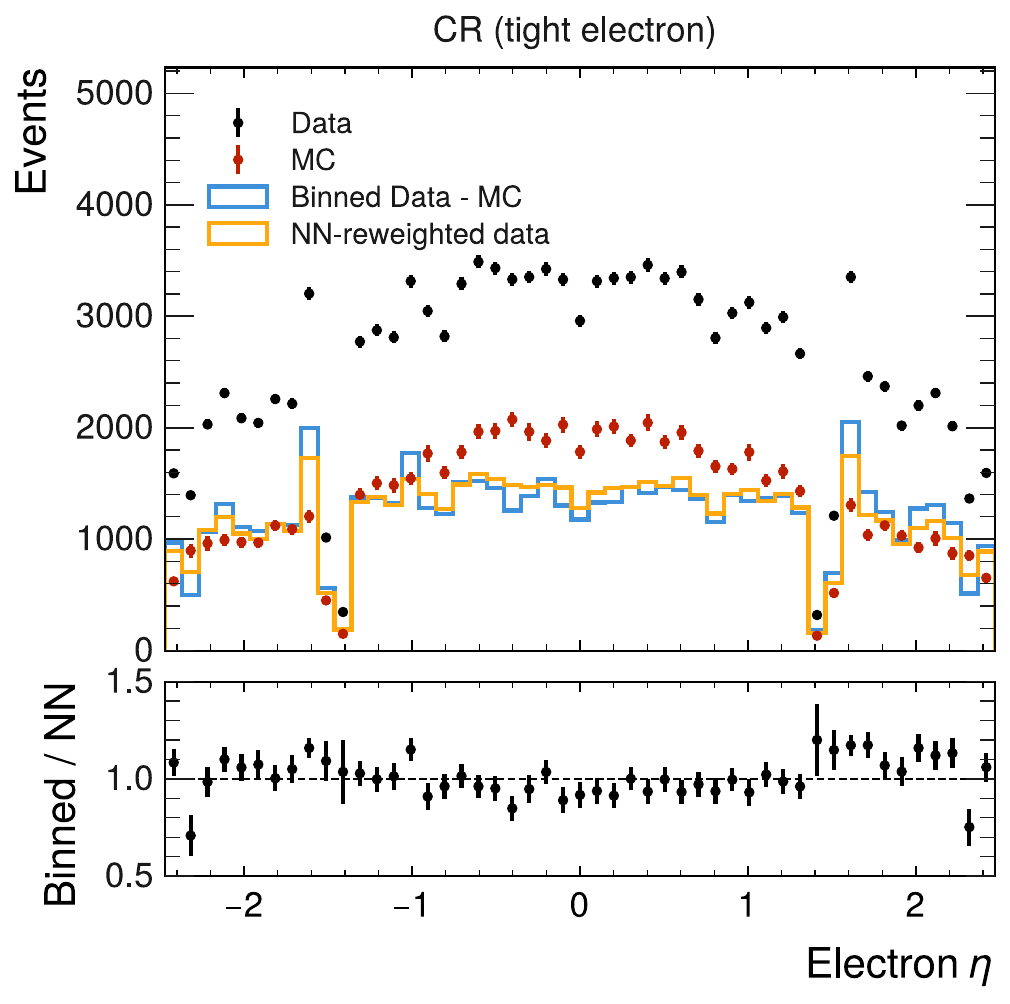}
    \end{subfigure}
    \hspace{4mm}
    \begin{subfigure}{0.42\linewidth}
        \includegraphics[width=\linewidth]{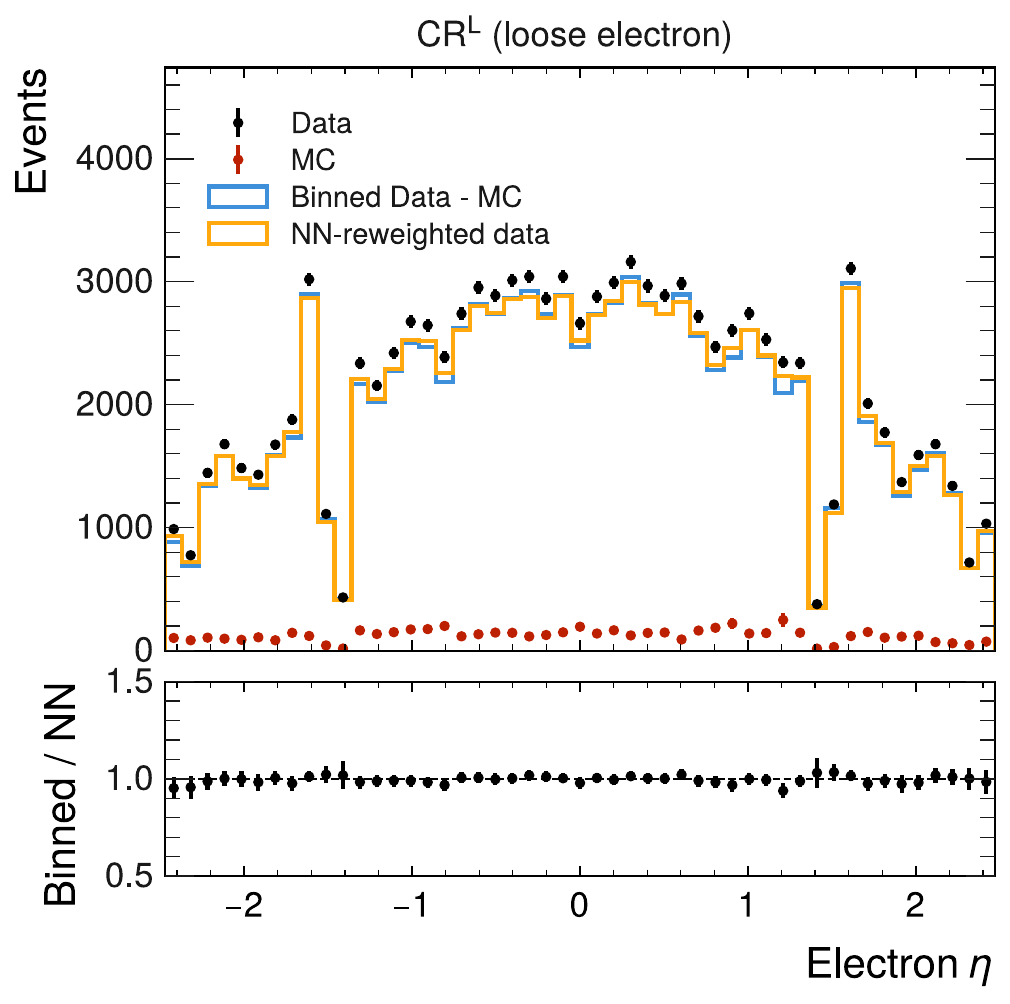}
    \end{subfigure}

    \begin{subfigure}{0.42\linewidth}
        \includegraphics[width=\linewidth]{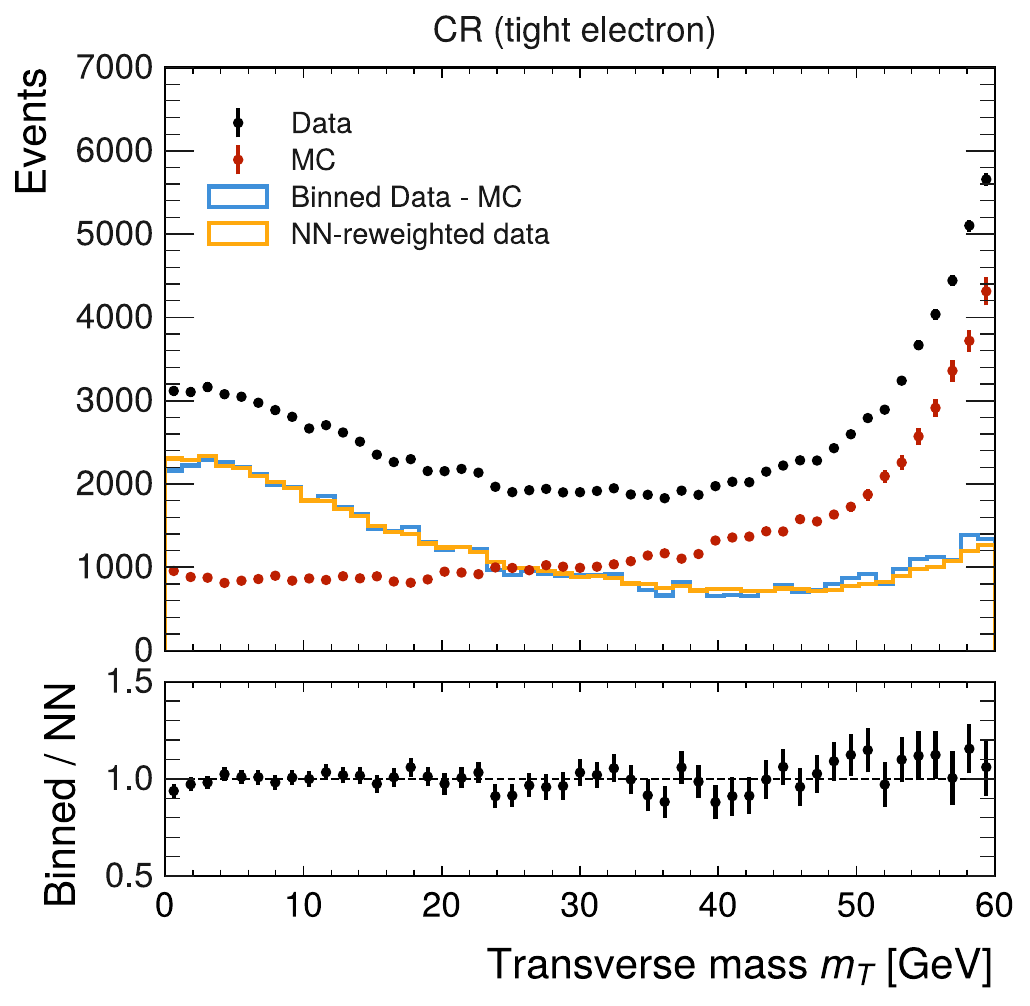}
    \end{subfigure}
    \hspace{4mm}
    \begin{subfigure}{0.42\linewidth}
        \includegraphics[width=\linewidth]{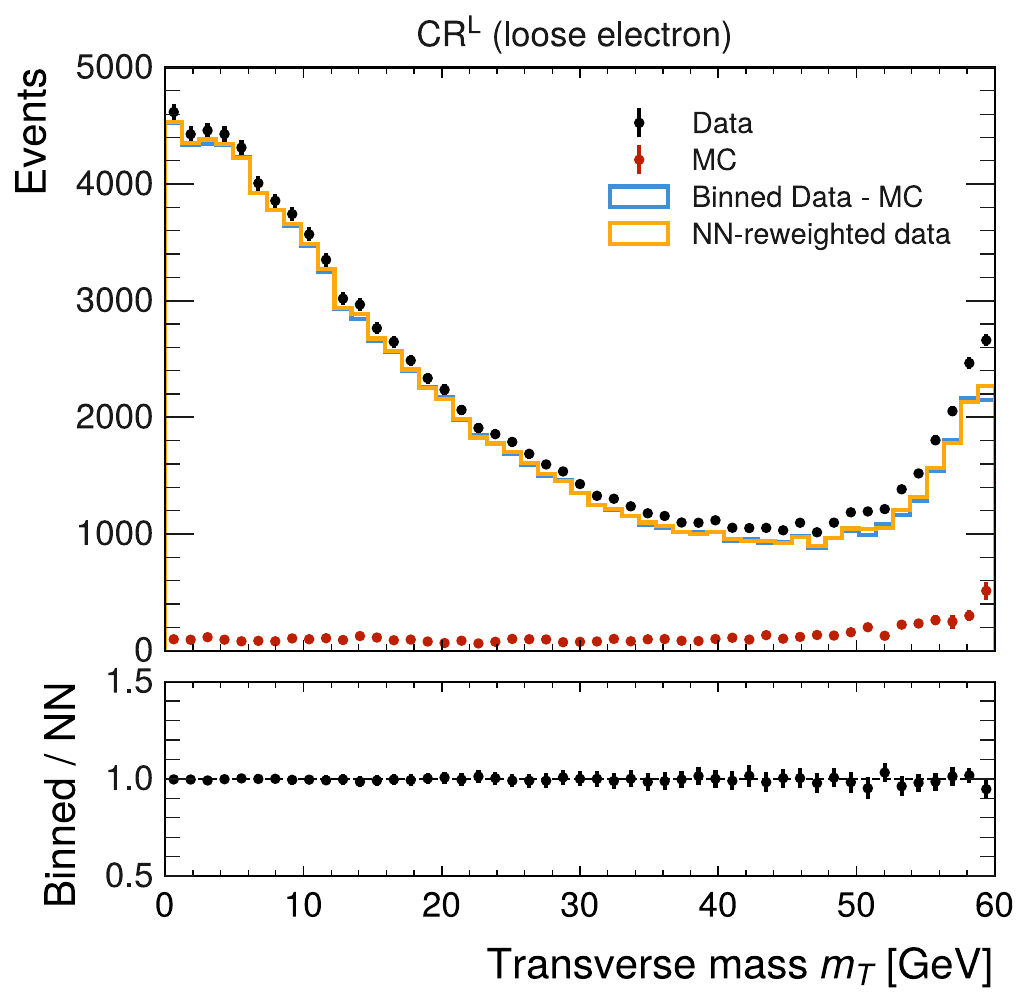}
    \end{subfigure}

    \caption{
        Distributions of $\pt$, $\eta$ and $\mt$ in the control region of the
        implemented analysis for both the tight (left column) and loose (right
        column) selections. Data and MC distributions are shown by the black and
        red error bars, respectively. The error bars show the statistical uncertainties of data or MC only.  For reference, the blue histogram shows the values
        of the subtraction of MC from data for each bin individually.
        Distributions derived by using the subtraction step of the ML-based approach, i.e.\ obtained by reweighting data events by the correction factor
        $r^\text{T,L}$, are shown in orange. One can observe that the main objective of the ML-based subtraction, which is to obtain per-event weights that reweight the data, resulting in adequate agreement with the reference, per-bin subtracted distributions, is achieved to a good precision.
    }
    \label{fig:CR_distributions}
\end{figure}

We have chosen five different kinematic observables for the ML training: $\pt$, $\MET$, $\eta$, number of jets ($\jets$), and $\mt$.
The control region distributions of a subset of these observables for both tight and loose selections are shown in Figure~\ref{fig:CR_distributions}. The machine learning model is trained using these observables (four continuous and one discrete) as input features in the control region to estimate the fake factor.

\subsection{ML Subtraction Step in the Analysis Control Region}

To validate the ML-based Fake Factor method, we first examine the performance of the subtraction step in the CR. In order to validate the subtraction, we apply the weights derived from the subtraction step in Eq.~\eqref{eq:subtraction} to reweight the data distribution in the control region. We then bin the reweighted data and compare it to the binned subtraction of data and MC in the control region. This allows us to assess the effectiveness of the subtraction step.
The results are shown in Figure~\ref{fig:CR_distributions}. We observe a good agreement between ML and binned subtraction, indicating that the ML subtraction step is working as intended.

In more detail, the difference in the total number of events between the two methods is within \qty{1}{\percent} for both numerator and denominator, which is acceptable given the statistical uncertainties in the data and MC samples. This demonstrates that the ML-based subtraction step is effective and can be used as part of the overall fake factor estimation process.

In the loose region (CR$^{\text{L}}$), the real lepton contamination is relatively small, so the subtraction step has a limited impact on the final fake factor estimate. However, it is still important to perform the subtraction to ensure that the derived fake factor is not biased by any residual real contributions in the loose region.
\clearpage

\subsection{ML Ratio Step in the Analysis Control Region}

\begin{figure}[hb]
    \centering
    \begin{subfigure}{0.35\linewidth}
        \includegraphics[width=\linewidth]{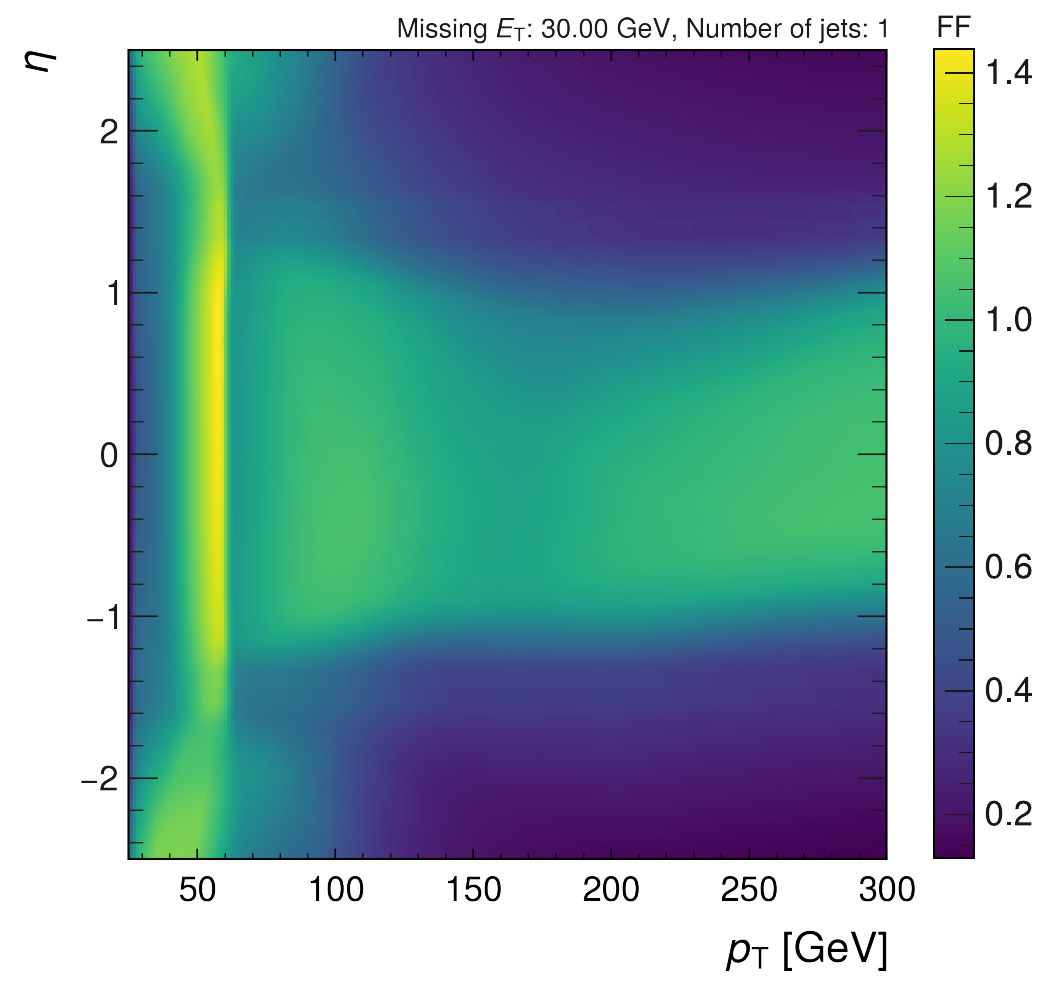}
    \end{subfigure}
    \hspace{4mm}
    \begin{subfigure}{0.35\linewidth}
        \includegraphics[width=\linewidth]{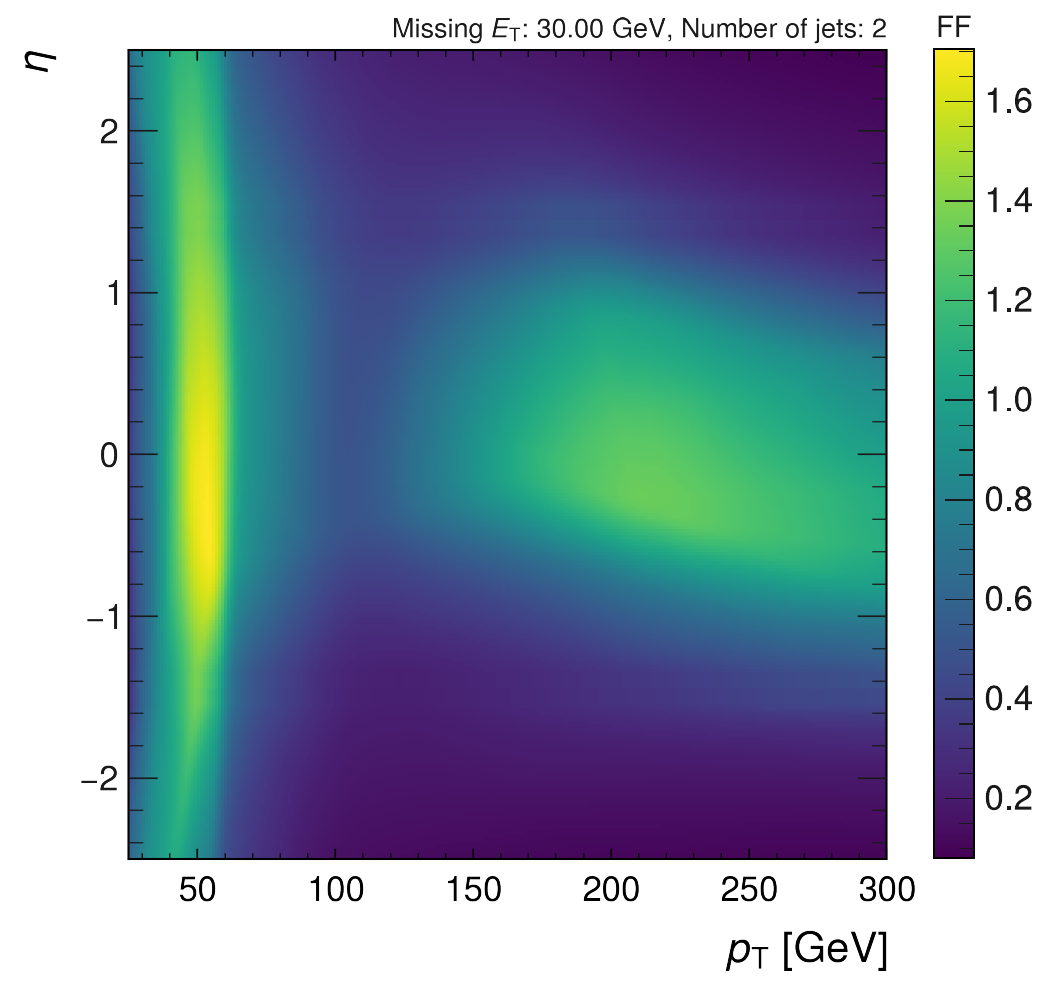}
    \end{subfigure}

    \begin{subfigure}{0.35\linewidth}
        \includegraphics[width=\linewidth]{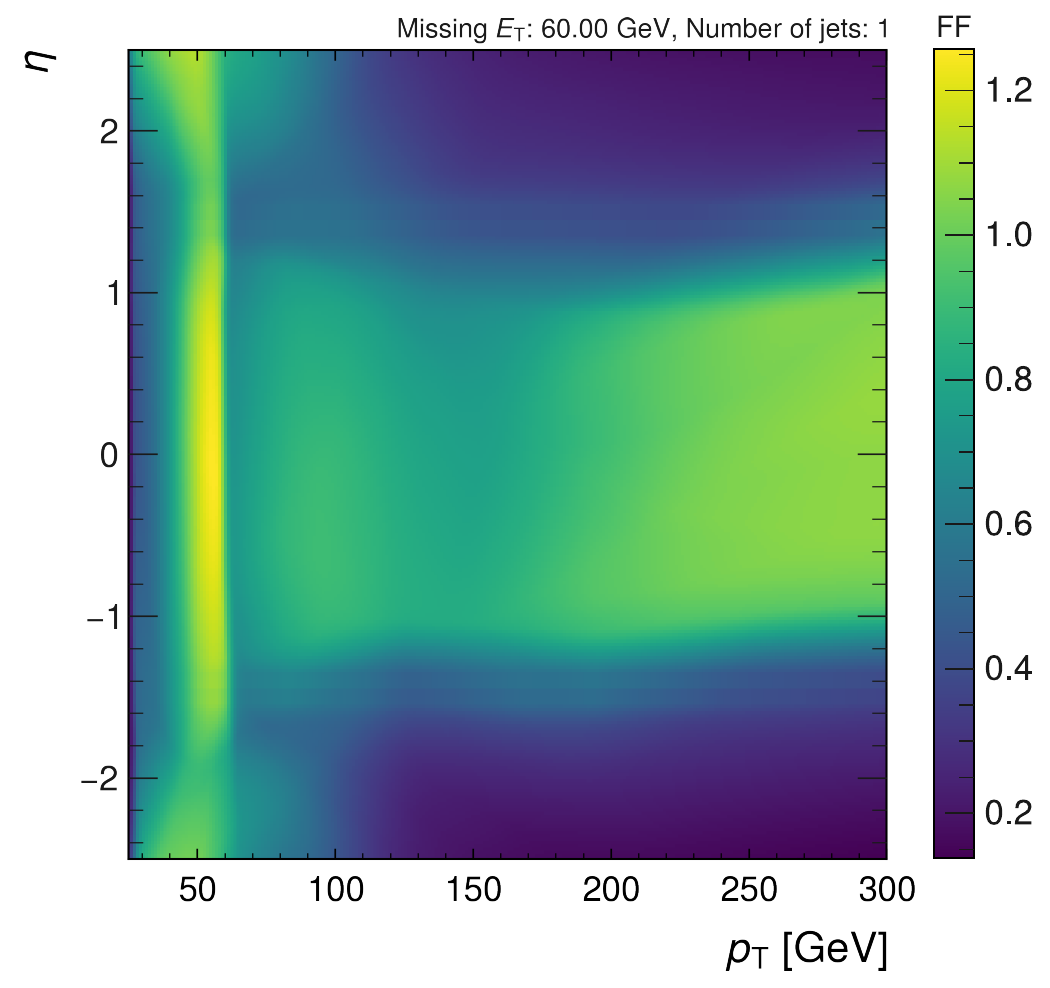}
    \end{subfigure}
    \hspace{4mm}
    \begin{subfigure}{0.35\linewidth}
        \includegraphics[width=\linewidth]{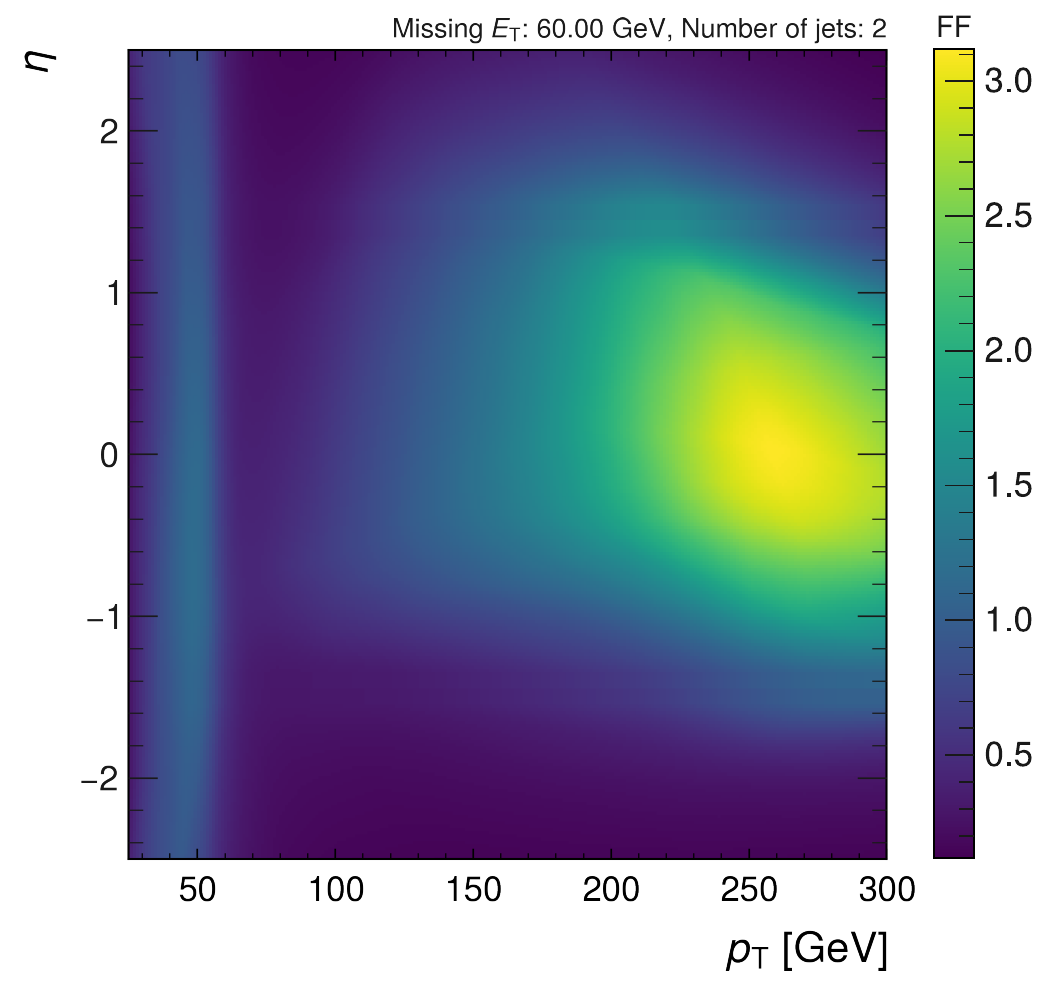}
    \end{subfigure}
    \caption{
       A 2D~projection of the fake factor obtained with the ML-based method in
        the $\pt$-$\eta$ plane for different fixed values of the other features.
        The values are approximately symmetrical in $\eta$, as would be
        expected, but have an otherwise non-trivial dependence across all input
        features. Note that all values are strictly positive by construction.
    }
    \label{fig:ml-projection_eta}
    \end{figure}

\begin{figure}
    \centering
    \begin{subfigure}{0.35\linewidth}
        \includegraphics[width=\linewidth]{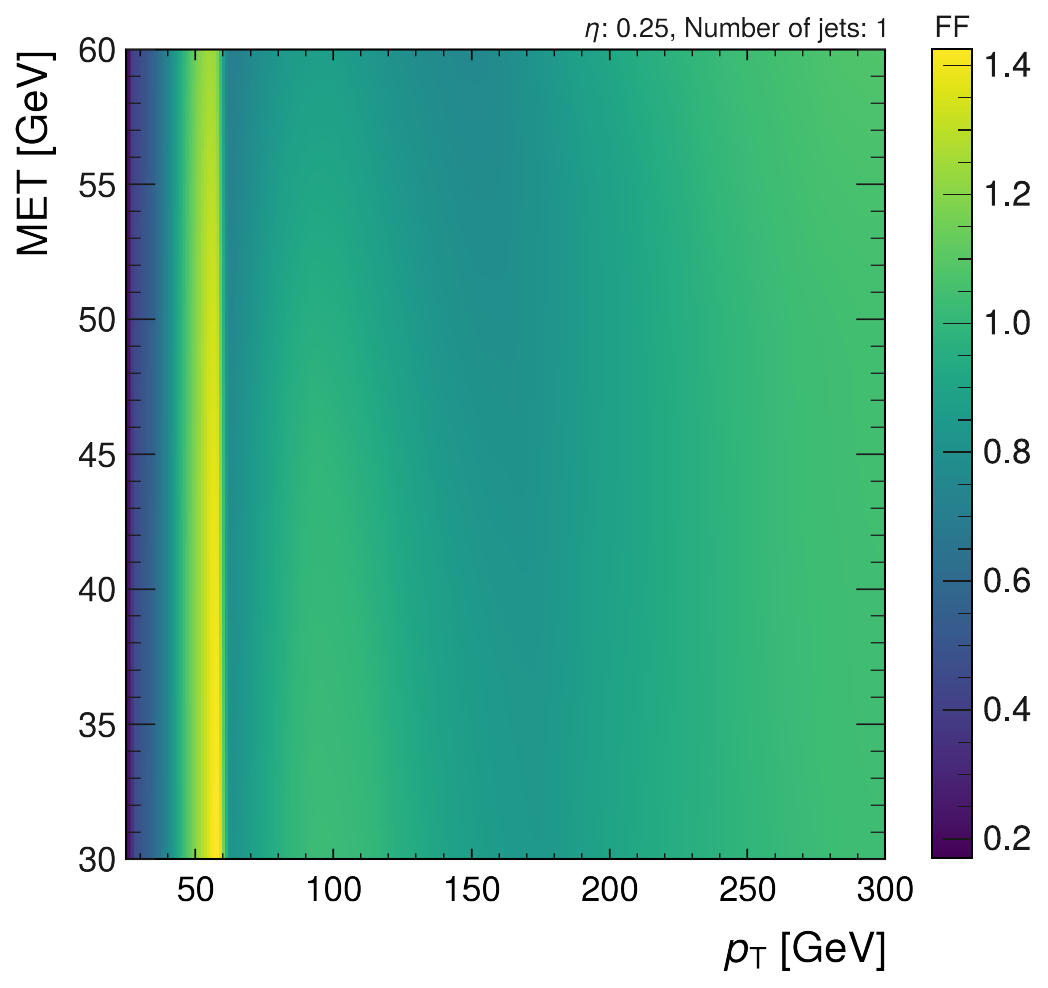}
    \end{subfigure}
    \hspace{4mm}
    \begin{subfigure}{0.35\linewidth}
        \includegraphics[width=\linewidth]{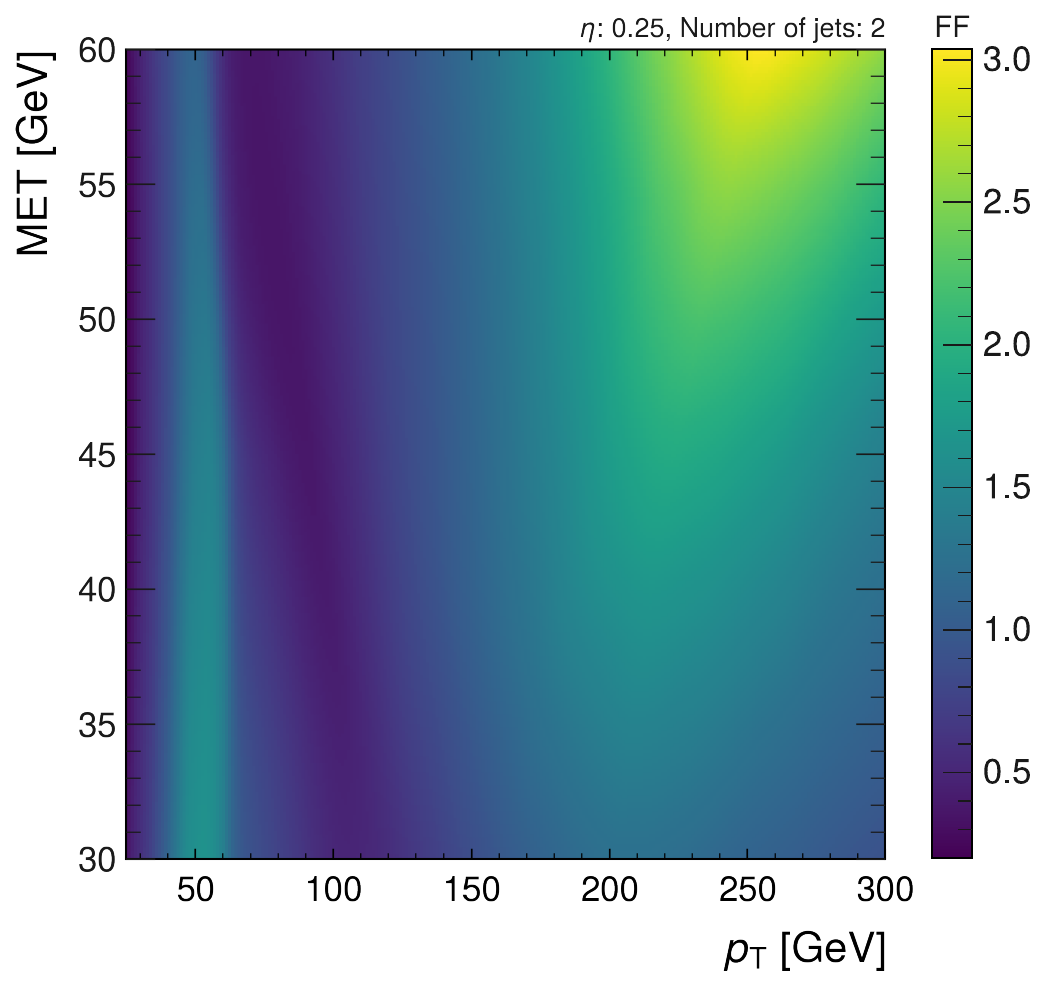}
    \end{subfigure}

    \begin{subfigure}{0.35\linewidth}
        \includegraphics[width=\linewidth]{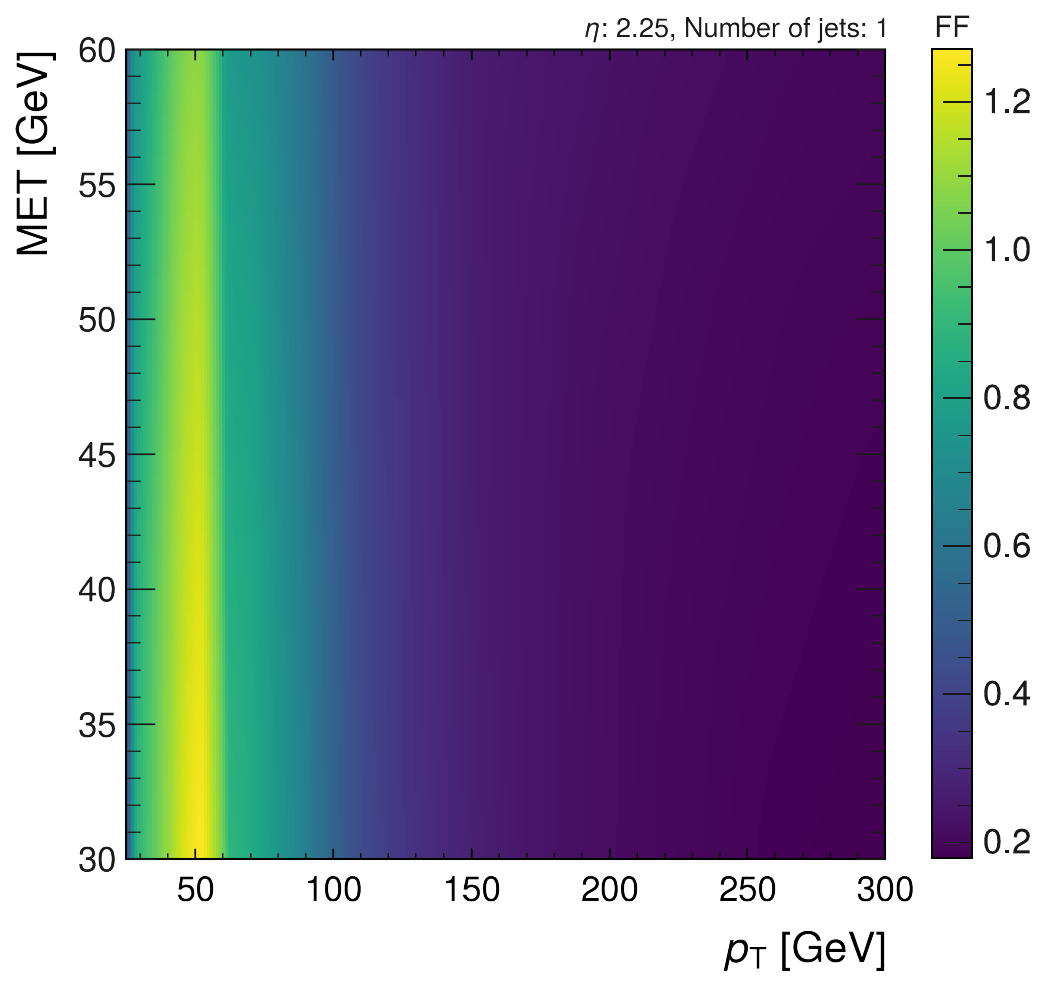}
    \end{subfigure}
    \hspace{4mm}
    \begin{subfigure}{0.35\linewidth}
        \includegraphics[width=\linewidth]{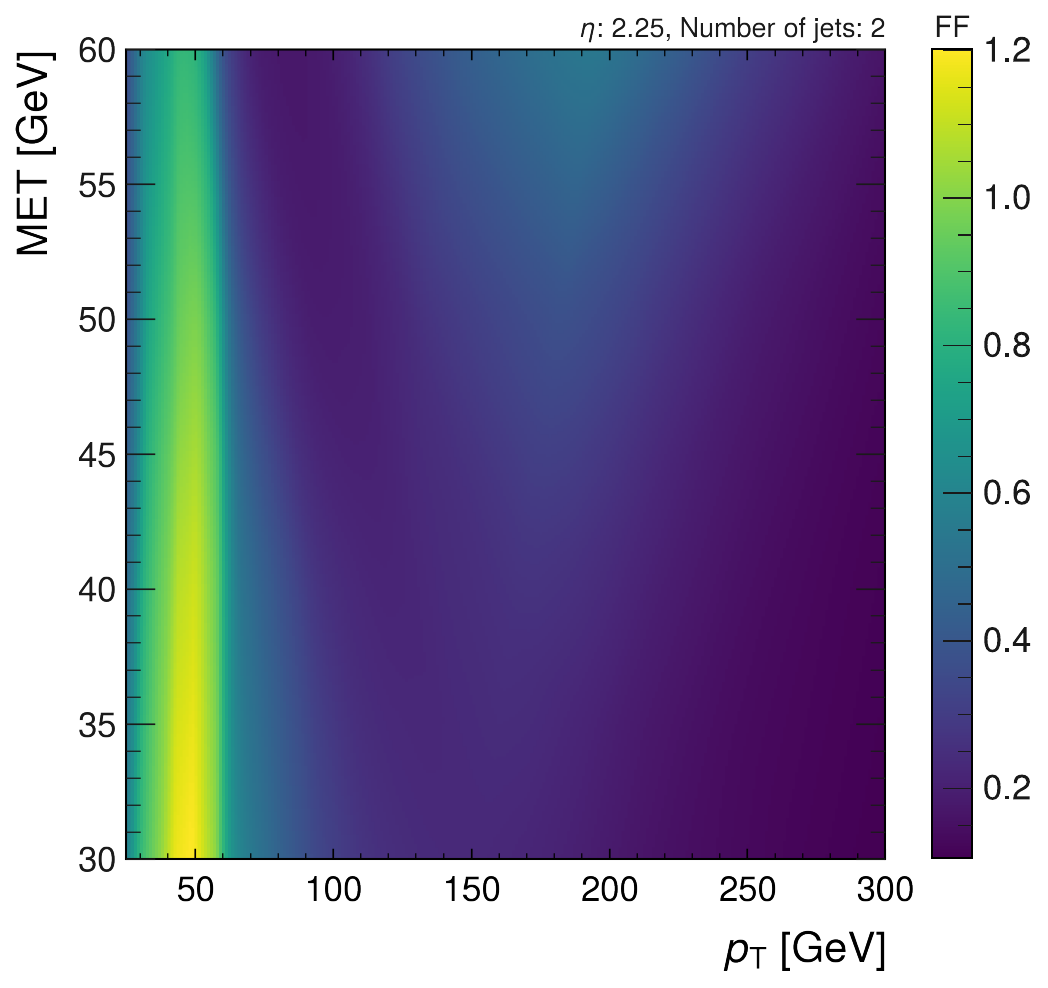}
    \end{subfigure}
    \caption{
        A 2D~projection of the fake factor obtained with the ML-based method in
        the $\pt$-$\MET$ plane for different fixed values of the other features.
    }
    \label{fig:ml-projection_met}
    \end{figure}

After validating the subtraction step, we proceed to validate the ratio step in the ML-based Fake Factor method, as denoted in Eq.~\eqref{eq:ff-ratio}. Once the final ratio classifier has been trained, we can visualize the resulting predicted value of the fake factor as a continuous function of the input features, which allows us to see how the fake factor varies across different regions of the feature space.
We create 2D~projections of the fake factor function as slices in the $\pt$-$\eta$ plane (Figure~\ref{fig:ml-projection_eta}) and $\pt$-$\MET$ plane (Figure~\ref{fig:ml-projection_met}).
We observe that the fake factor varies smoothly across the feature space, indicating that the classifier model has learned a meaningful representation of the data, and that the fake factor values increase with the jet multiplicity $\jets$, which is consistently incorporated in the ML-based fake factor parameterization\footnote{It is worth emphasizing that using a more optimal parameterization (e.g.\ mother parton $\pt$~\cite{Khachatryan_2016}) such dependence would be absent or at least greatly reduced, which is, however, beyond the scope of this paper. In this study, the successful reproduction of the kinematics in a multidimensional representation of a fake factor parameterization was the main emphasis.}.

With increasing $\jets$ and $\MET$, we also encounter lower data and MC  statistics, which can lead to larger fluctuations in the derived fake factor values. This is particularly evident in the high $\MET$ regions and with more than one jet, where the fake factor values do fluctuate significantly. Overall, the 2D~projections provide a useful way to visualize the learned fake factor and assess its behavior across different regions of the feature space.

\begin{figure}
    \centering
    \includegraphics[width=0.52\linewidth]{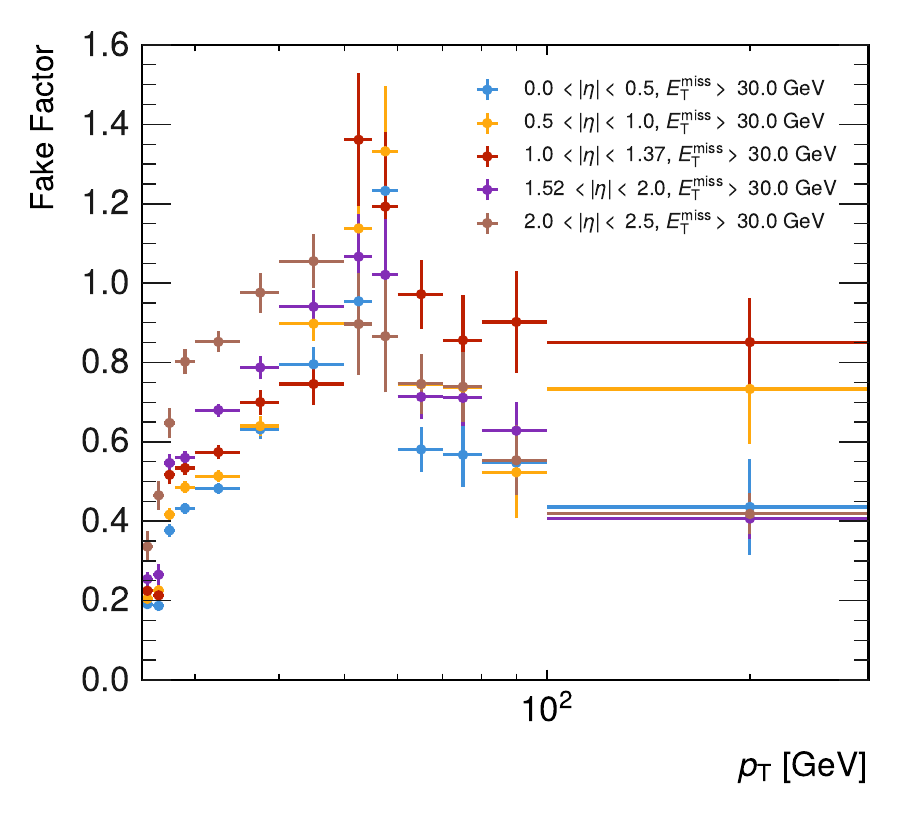}
    \caption{
        Values of the fake factor obtained using the binned Fake Factor method
        as a function of $\pt$ for different bins in $|\eta|$. Uncertainties are
        driven by the finite statistics of data and MC samples, limiting the
        possibility to further segment the binning or add an additional
        dimension.
    }
    \label{fig:ff-binned}
\end{figure}

In order to validate the newly derived ML procedure, we compare the ML-based fake factor to the results the (customary) binned fake factor approach would give. For the binned fake factor calculation, we use the same control region as for the ML-based method, and apply the same event selection criteria. The binned fake factor is calculated by taking the ratio of the number of tight events to the number of loose events in bins of $\pt$ and $|\eta|$ after the data and real-lepton MC subtraction, i.e.\ subtracting and dividing the required histograms. This provides the comparison baseline to assess the performance of the ML-based method. The calculated binned fake factors are shown in Figure~\ref{fig:ff-binned}.

%\vspace{-1cm}

\begin{figure}
    \centering
    \begin{subfigure}{0.35\linewidth}
        \includegraphics[width=\linewidth]{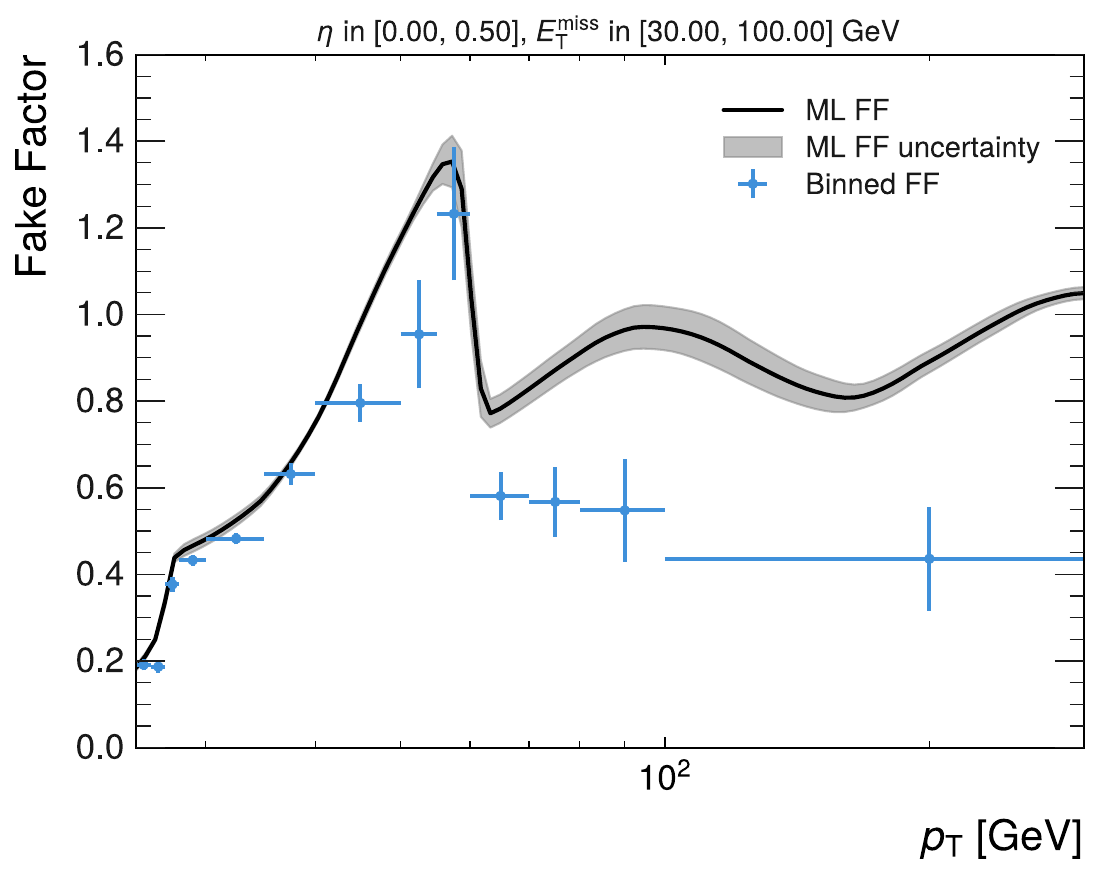}
        \label{fig:plot1-a}
    \end{subfigure}
    \hspace{4mm}
    \begin{subfigure}{0.35\linewidth}
        \includegraphics[width=\linewidth]{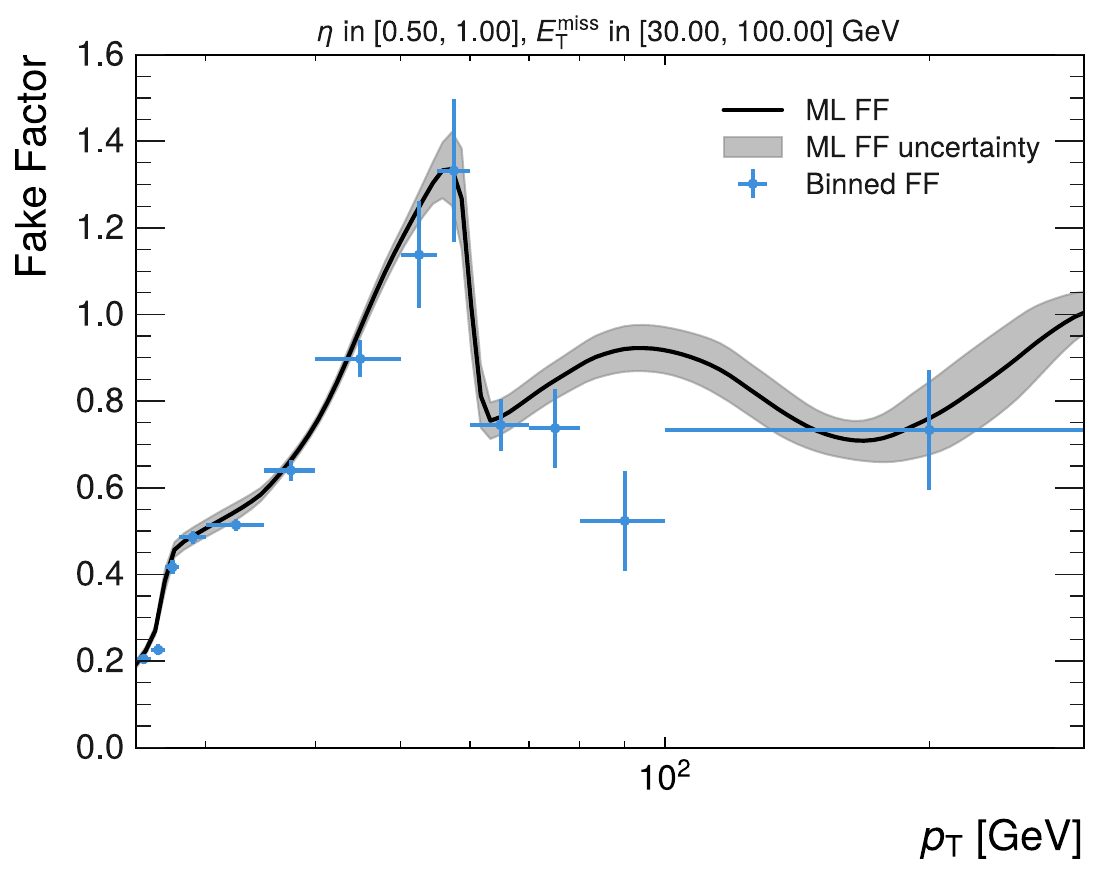}
        \label{fig:plot2-b}
    \end{subfigure}

    \begin{subfigure}{0.35\linewidth}
        \includegraphics[width=\linewidth]{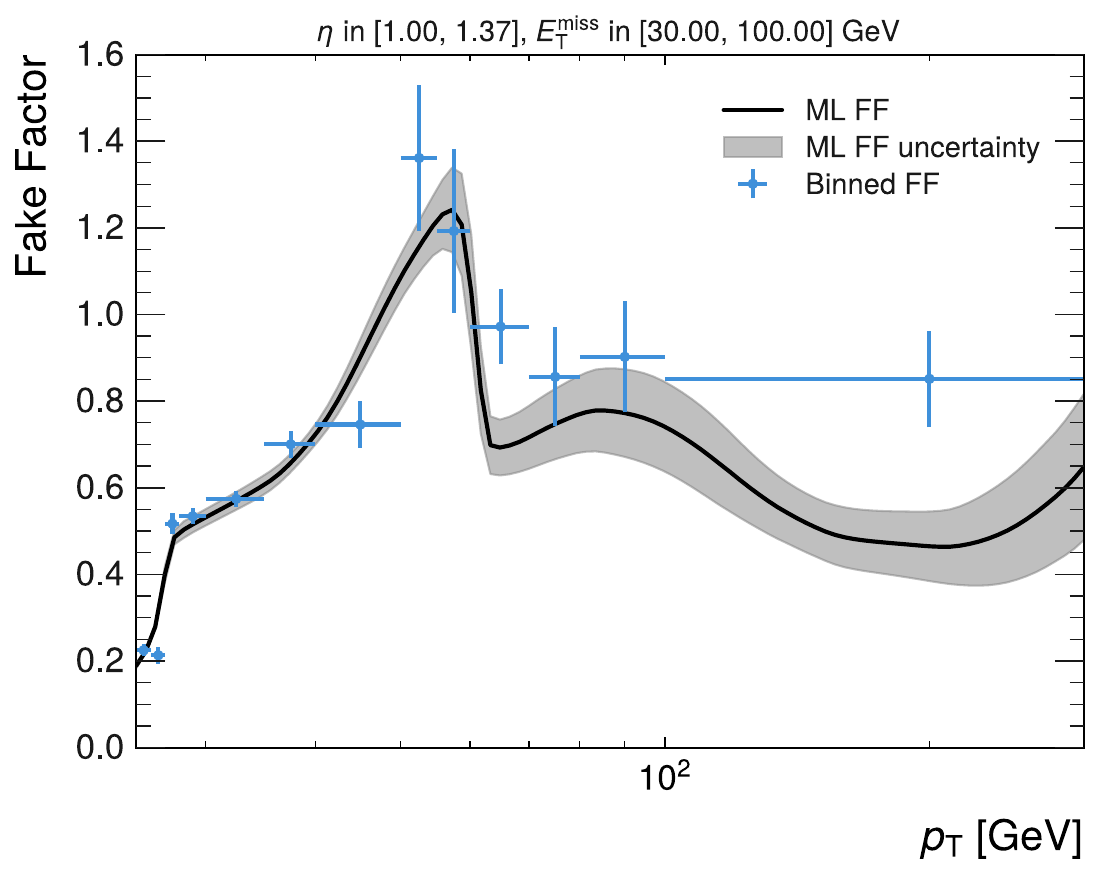}
        \label{fig:plot3-c}
    \end{subfigure}
    \hspace{4mm}
    \begin{subfigure}{0.35\linewidth}
        \includegraphics[width=\linewidth]{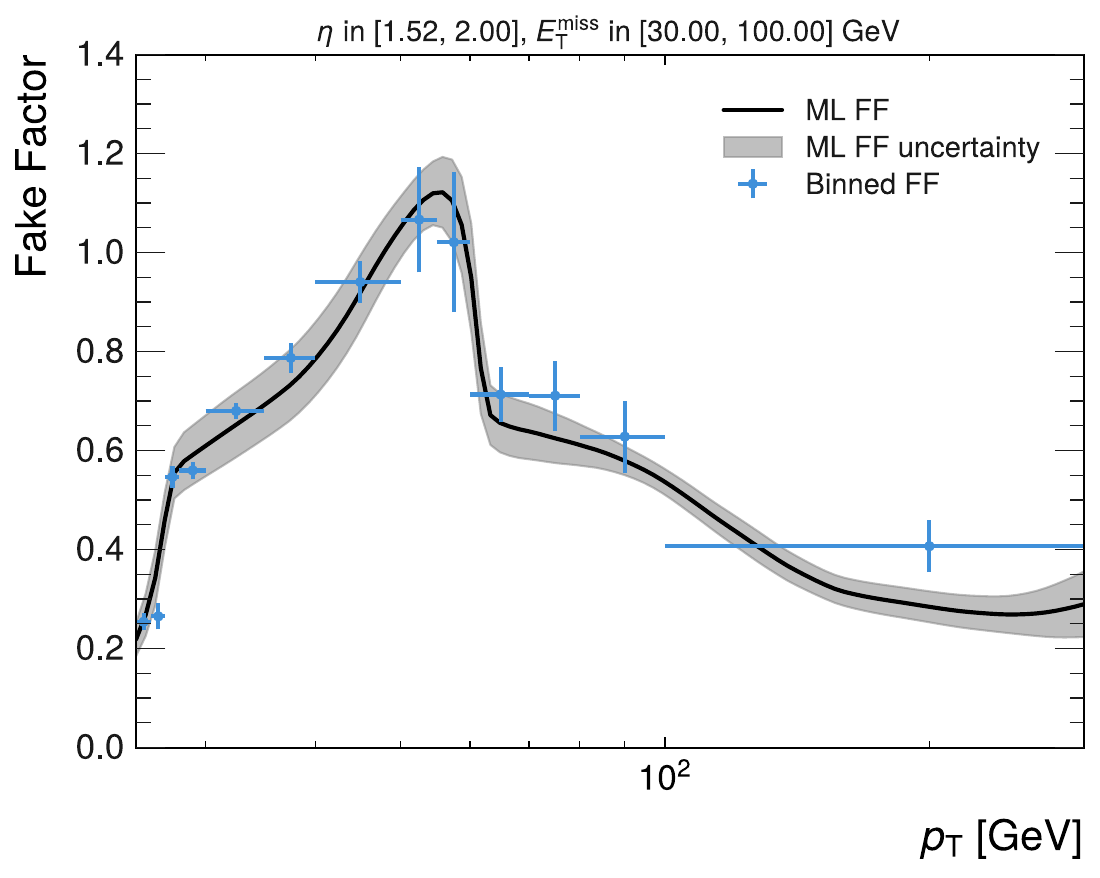}
        \label{fig:plot4-d}
    \end{subfigure}

    \begin{subfigure}{0.35\linewidth}
        \includegraphics[width=\linewidth]{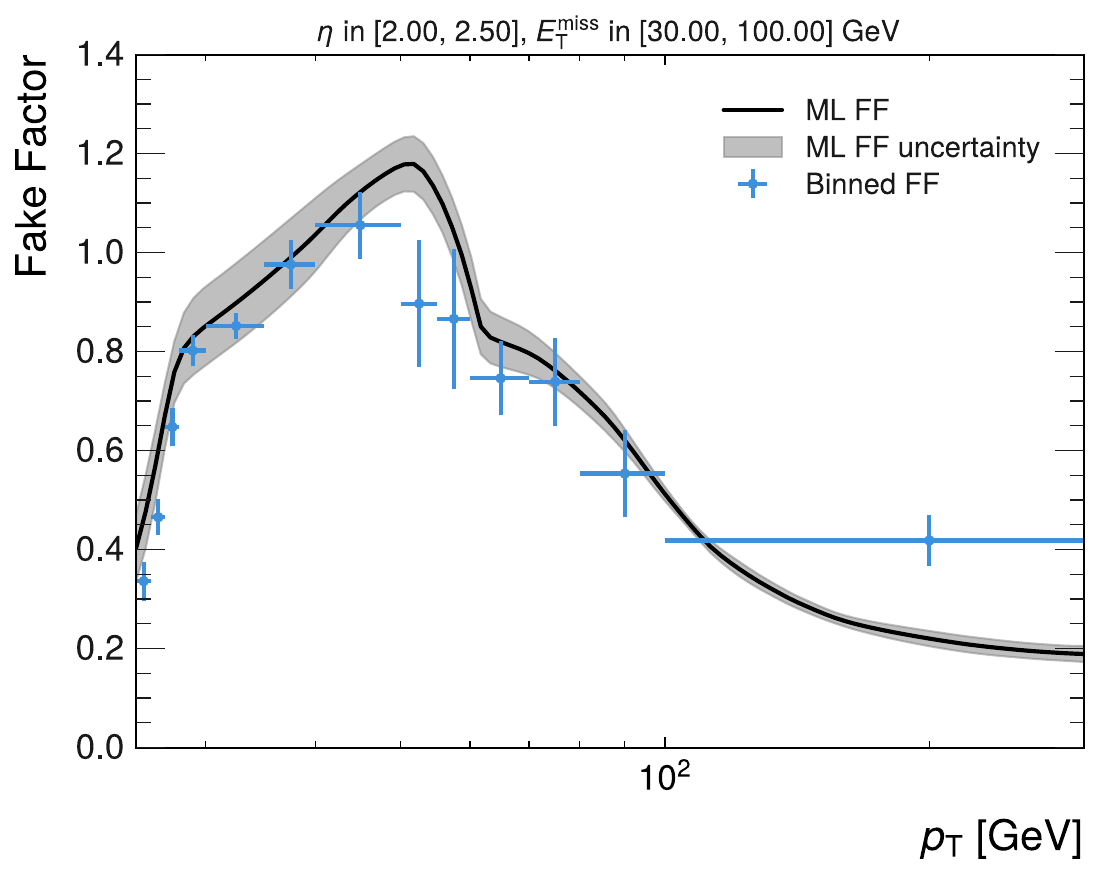}
        \label{fig:plot5-e}
    \end{subfigure}
    \caption{
        Comparison of binned fake factors and 1D~projections of the fake factors
        obtained with the ML-based method as a function of $\pt$. Individual
        plots are showing different bins (ranges) in $|\eta|$. 1D~projections of the
        ML-based method are obtained by integrating out (averaging over) all
        additional non-relevant dimensions ($\MET$, $\jets$ and $\mt$) as well
        as the applicable range in $|\eta|$. The uncertainty band represents the
        standard deviation of the fake factor values in the integration range.
    }
    \label{fig:5plots_cont_pt}
\end{figure}

\begin{figure}
    \centering
    \begin{subfigure}{0.35\linewidth}
        \includegraphics[width=\linewidth]{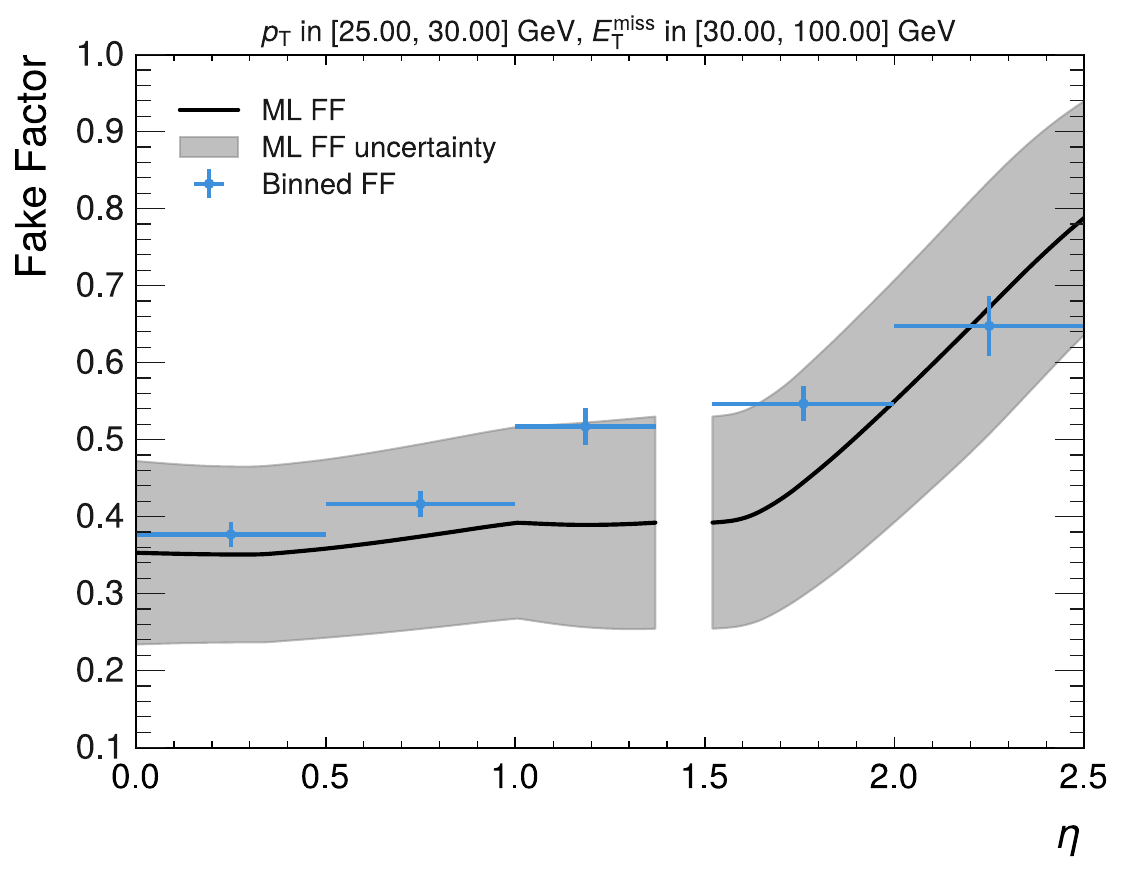}
    \end{subfigure}
    \hspace{4mm}
    \begin{subfigure}{0.35\linewidth}
        \includegraphics[width=\linewidth]{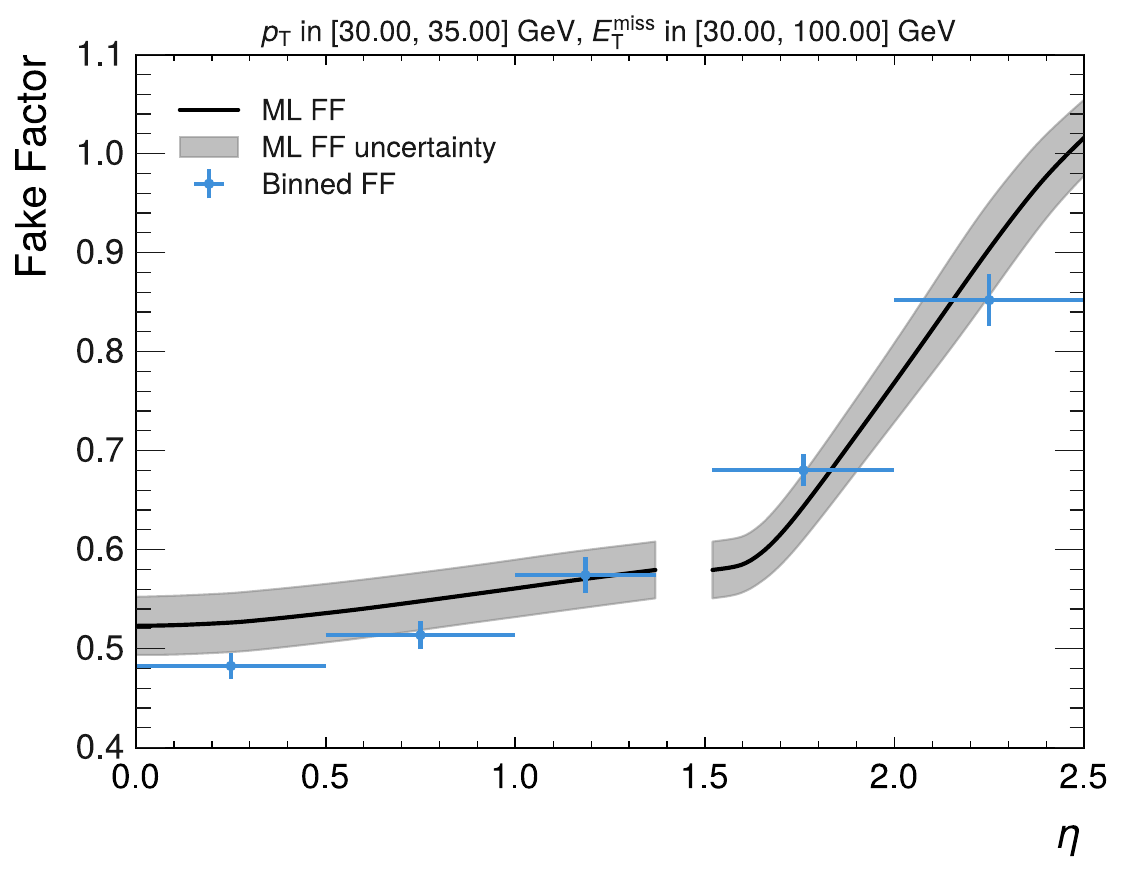}
    \end{subfigure}

    \begin{subfigure}{0.35\linewidth}
        \includegraphics[width=\linewidth]{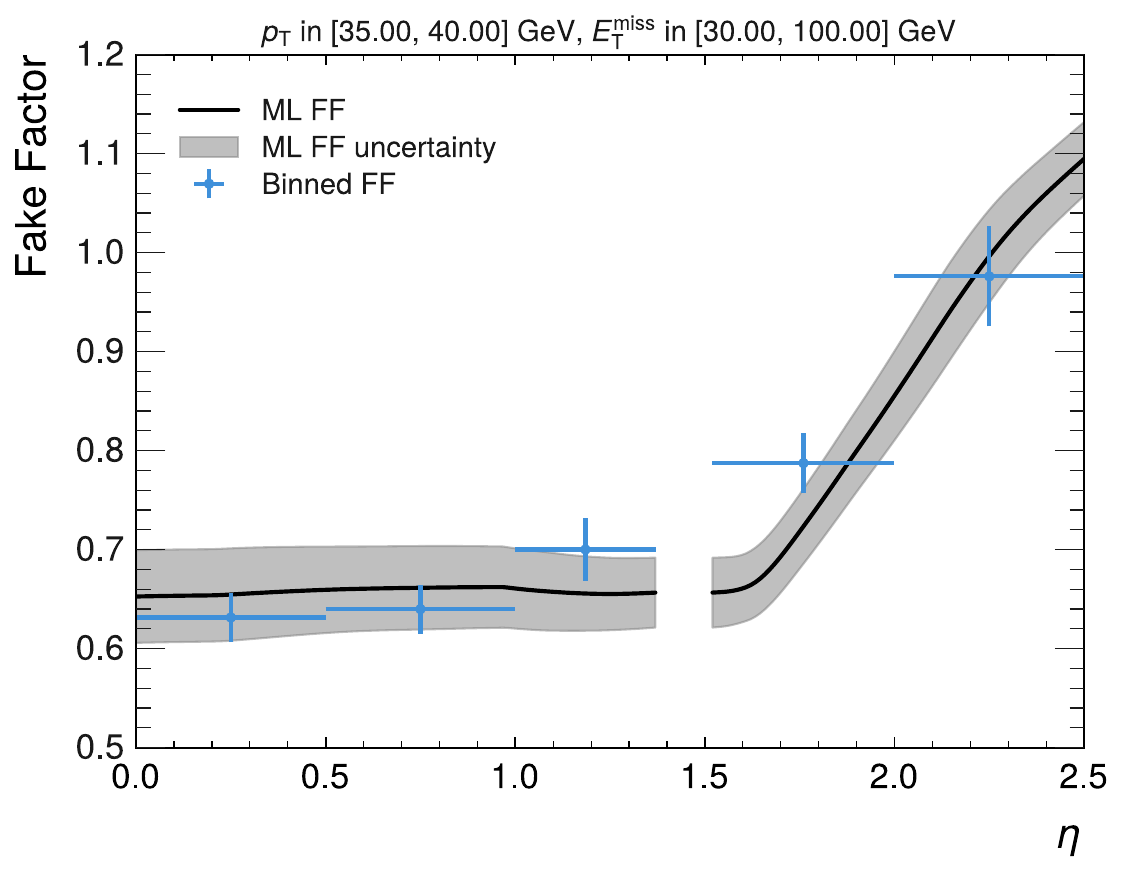}
    \end{subfigure}
    \hspace{4mm}
    \begin{subfigure}{0.35\linewidth}
        \includegraphics[width=\linewidth]{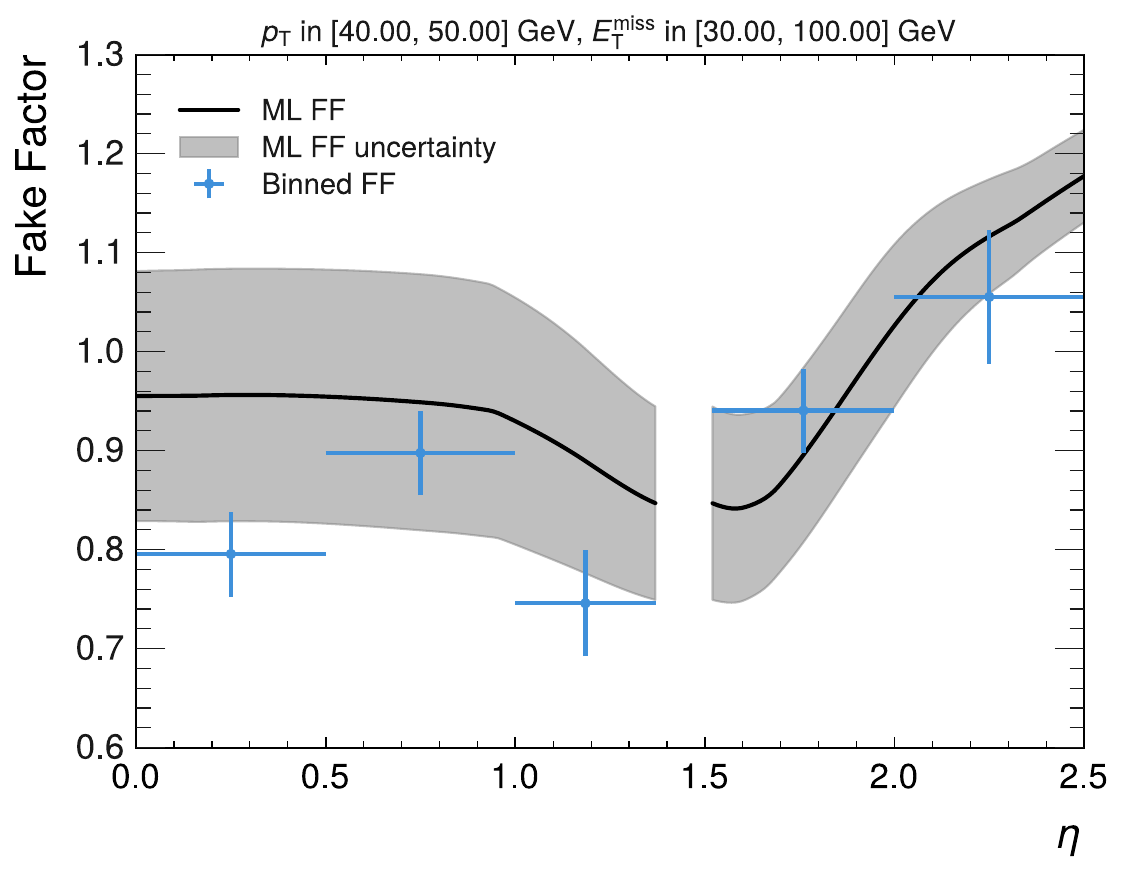}
    \end{subfigure}

    \begin{subfigure}{0.35\linewidth}
        \includegraphics[width=\linewidth]{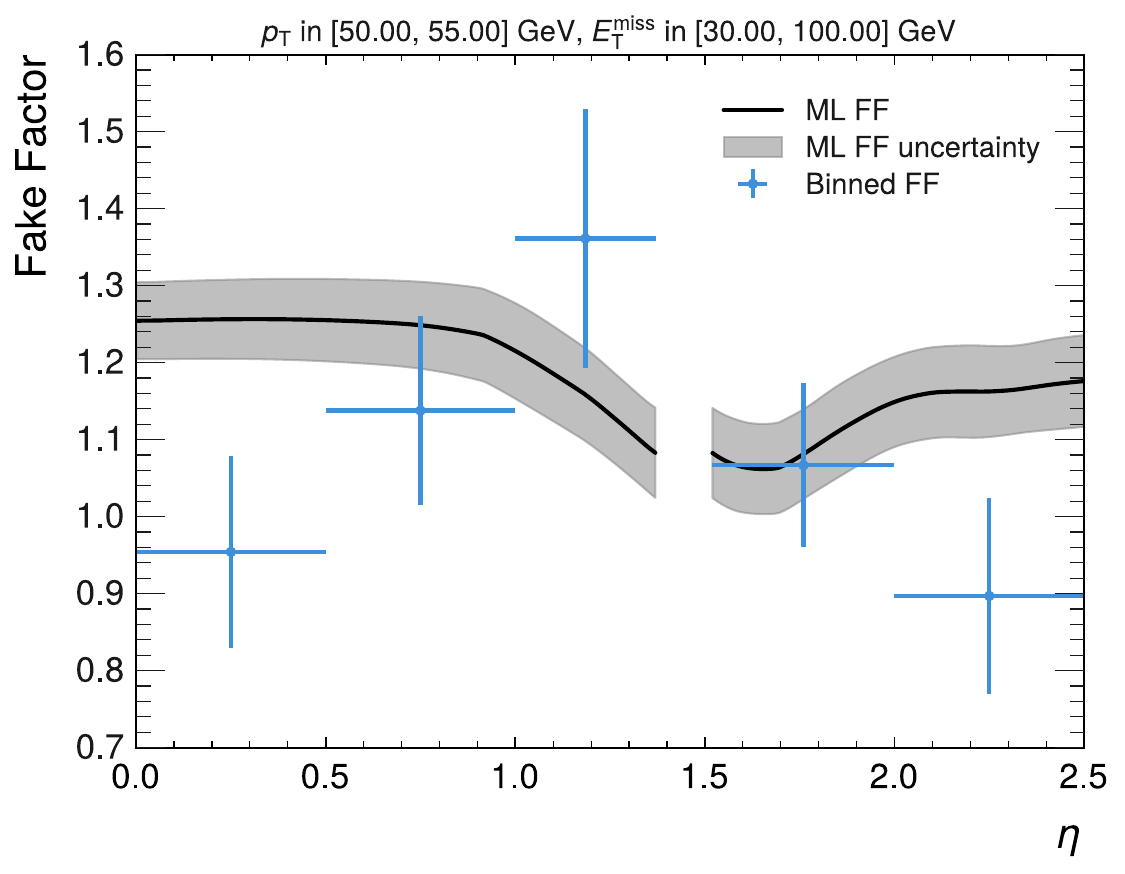}
    \end{subfigure}
    \caption{
        Comparison of binned fake factors and 1D~projections of the fake factors
        obtained with the ML-based method as a function of $|\eta|$. Individual
        plots are showing different bins (ranges) in $\pt$. 1D~projections  of
        the ML-based method are obtained by integrating out (averaging over) all
        additional non-relevant dimensions ($\MET$, $\jets$ and $\mt$) as well
        as the applicable range in $\pt$. The uncertainty band represents the
        standard deviation of the fake factor values in the integration range.
    }
    \label{fig:5plots_cont_eta}
\end{figure}

To compare the ML-based fake factor with the binned fake factor, we produce one-dimensional projections of the ML fake factor across different $\pt$ and $|\eta|$ ranges (bins). For all other variables, we perform an averaging (integration) over the ranges where the binned fake factor is defined. This procedure ensures a consistent and direct comparison between the two methods. Additionally, uncertainty bands are included for the ML-based fake factor, reflecting the variation introduced by the averaging process. The results of this comparison are presented in Figures~\ref{fig:5plots_cont_pt} and~\ref{fig:5plots_cont_eta}.

We observe good agreement in both shape and magnitude between the two methods across all bins given the uncertainties, indicating that the ML-based approach successfully captures the same underlying physics as the traditional binned method. In poor-statistics regions the ML-based method is expected to produce a more robust result, since it uses more event information in a properly correlated way, which explains the occasional deviations of the two, e.g.\ in high-$\pt$ regions. This demonstrates that the ML-based method is a valid and robust alternative for estimating fake factors, offering greater flexibility and the potential for improved accuracy.

\section{Predicting the Fake Lepton Background in the Analysis}\label{sec:sec5}

In this section, we present the results of applying both the binned and ML-based Fake Factor methods to estimate the fake electron background in the implemented analysis setup using the ATLAS Open Data sample. We validate the performance of both methods in the control region (CR) and signal region (SR) defined by transverse mass ($\mt$) cuts, as described in Section~\ref{sec:sec4}.
Performance is evaluated by comparing the estimated fake contributions to the observed data and MC predictions across several key kinematic variables, including lepton transverse momentum ($\pt$), pseudorapidity ($\eta$), missing transverse energy ($\MET$), jet multiplicity ($\jets$), and transverse mass ($\mt$). In order to ensure a fair comparison, we evaluated both methods using the same events and applied identical selection criteria.

We assess the agreement between data and fake process predictions, after applying the fake factor corrections, focusing on both the shape and normalization of the distributions.

\subsection{Control Region Performance}
\begin{figure}
    \centering
    \begin{subfigure}{0.465\linewidth}
        \includegraphics[width=\linewidth]{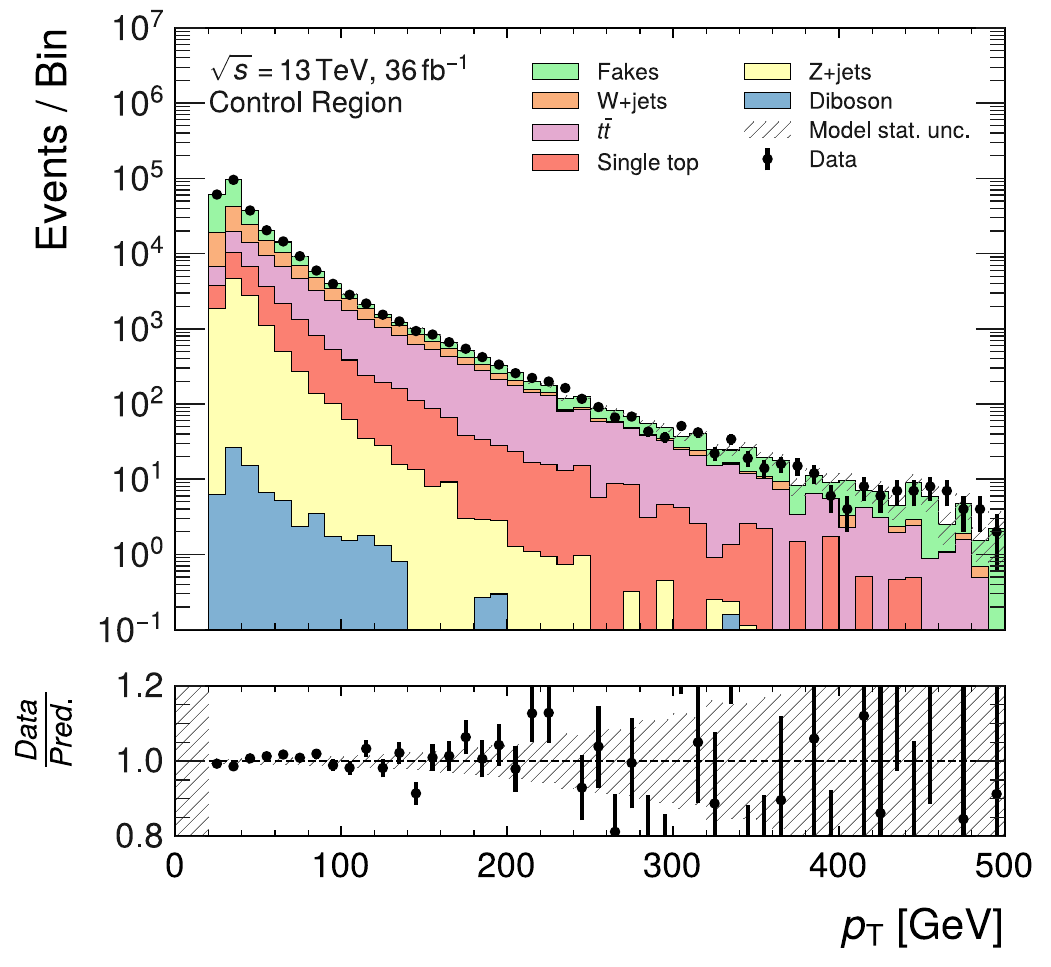}
    \end{subfigure}
    ~
    \begin{subfigure}{0.465\linewidth}
        \includegraphics[width=\linewidth]{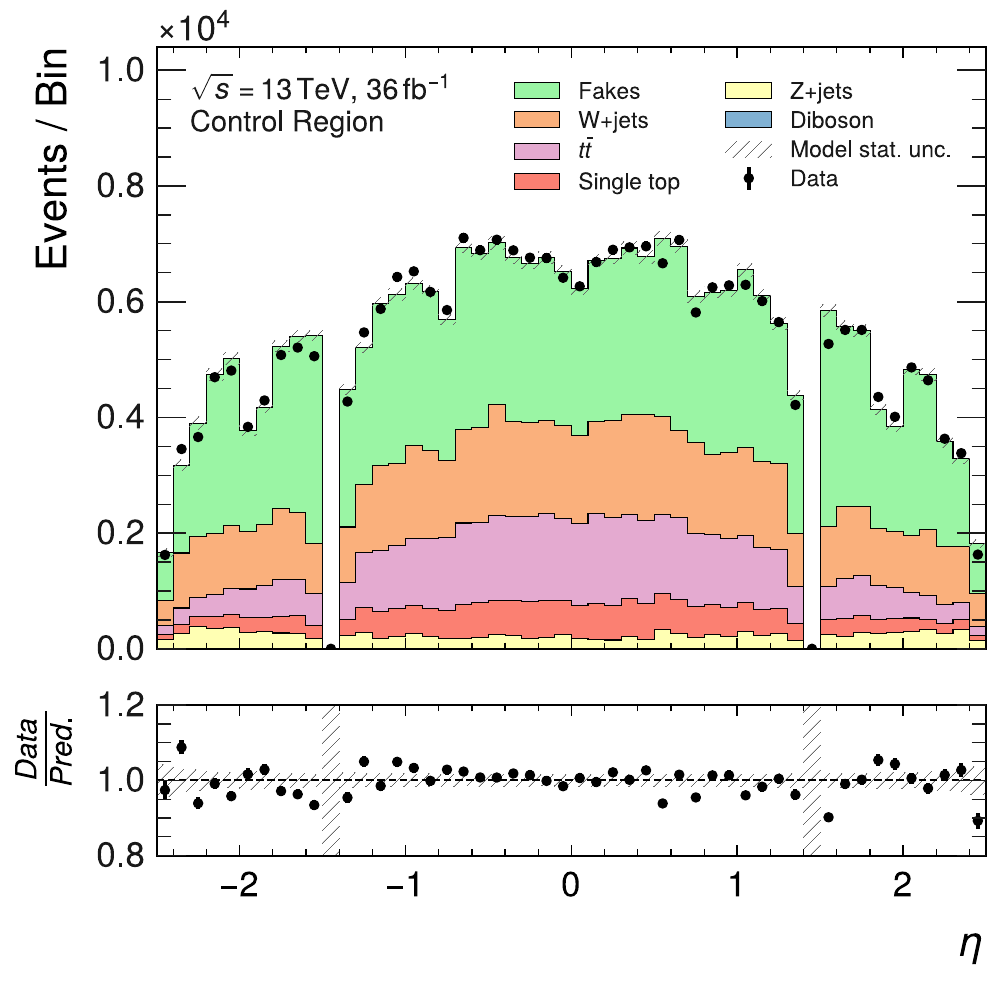}
    \end{subfigure}

    \begin{subfigure}{0.465\linewidth}
        \includegraphics[width=\linewidth]{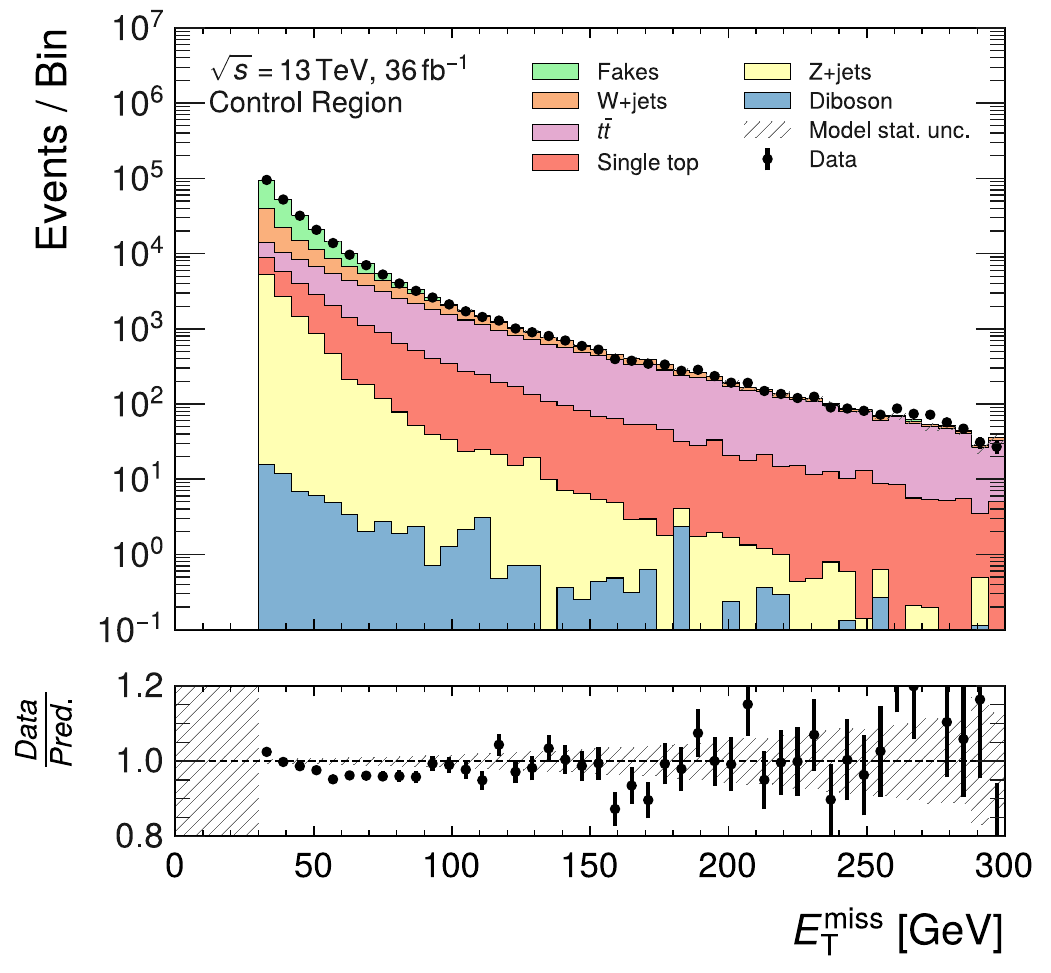}
    \end{subfigure}
    ~
    \begin{subfigure}{0.465\linewidth}
        \includegraphics[width=\linewidth]{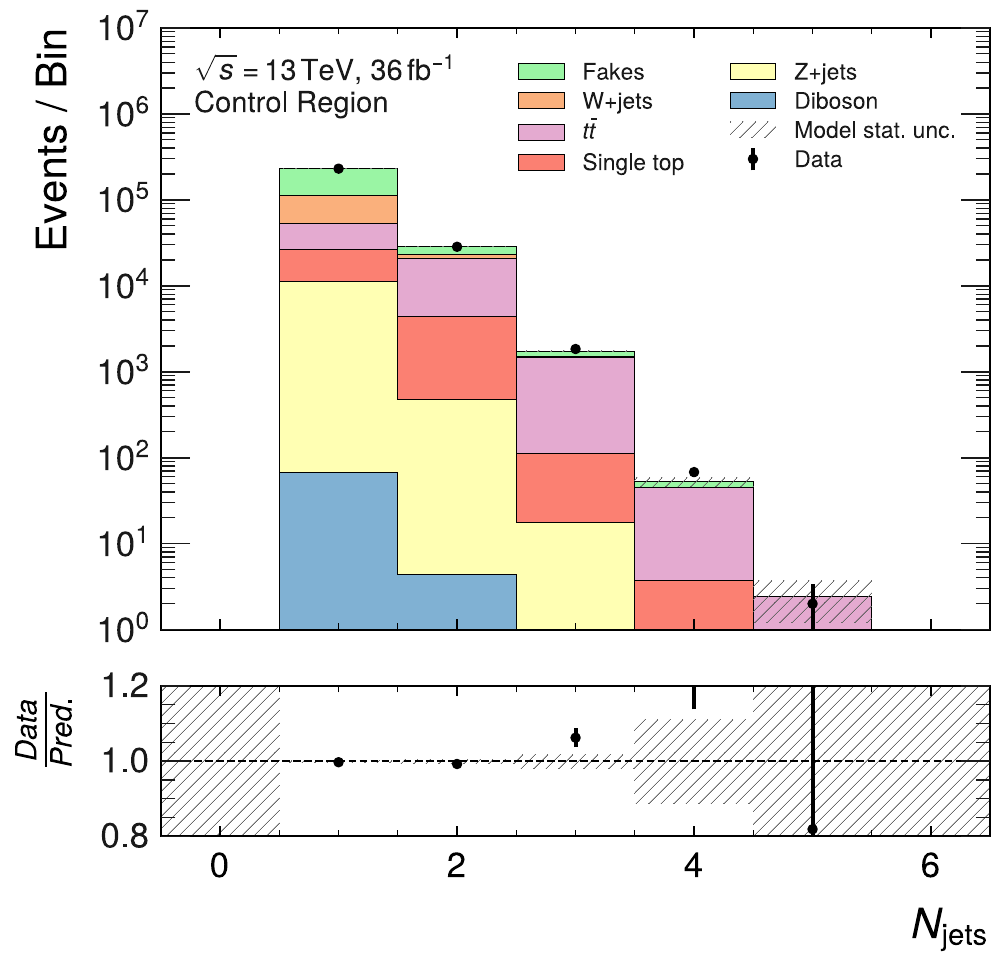}
    \end{subfigure}

    \begin{subfigure}{0.465\linewidth}
        \includegraphics[width=\linewidth]{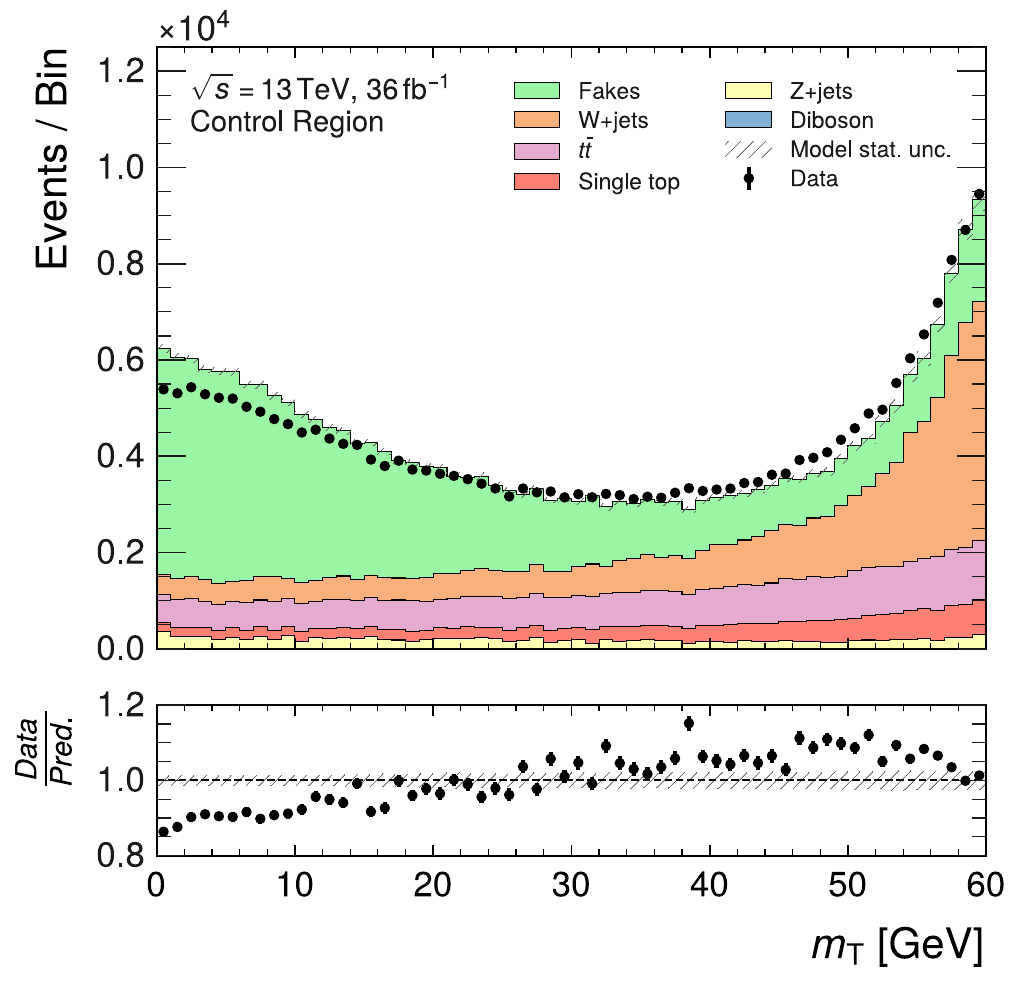}
    \end{subfigure}
    \caption{
        Closure test in the control region~(CR) of the implemented analysis using
        fake factors obtained from the binned method. As expected, distributions
        in $\pt$ and $\eta$ show good agreement since the binned method was
        parameterized in these variables and optimized to match the input data (i.e.\ `close' well).
        Distributions with respect to other variables that were not included in the
        parameterization all show a greater systematic mis-modeling, which is
        especially noticeable in the $\mt$ distribution.
    }
    \label{fig:closure_test}
\end{figure}

\begin{figure}
    \centering
    \begin{subfigure}{0.465\linewidth}
        \includegraphics[width=\linewidth]{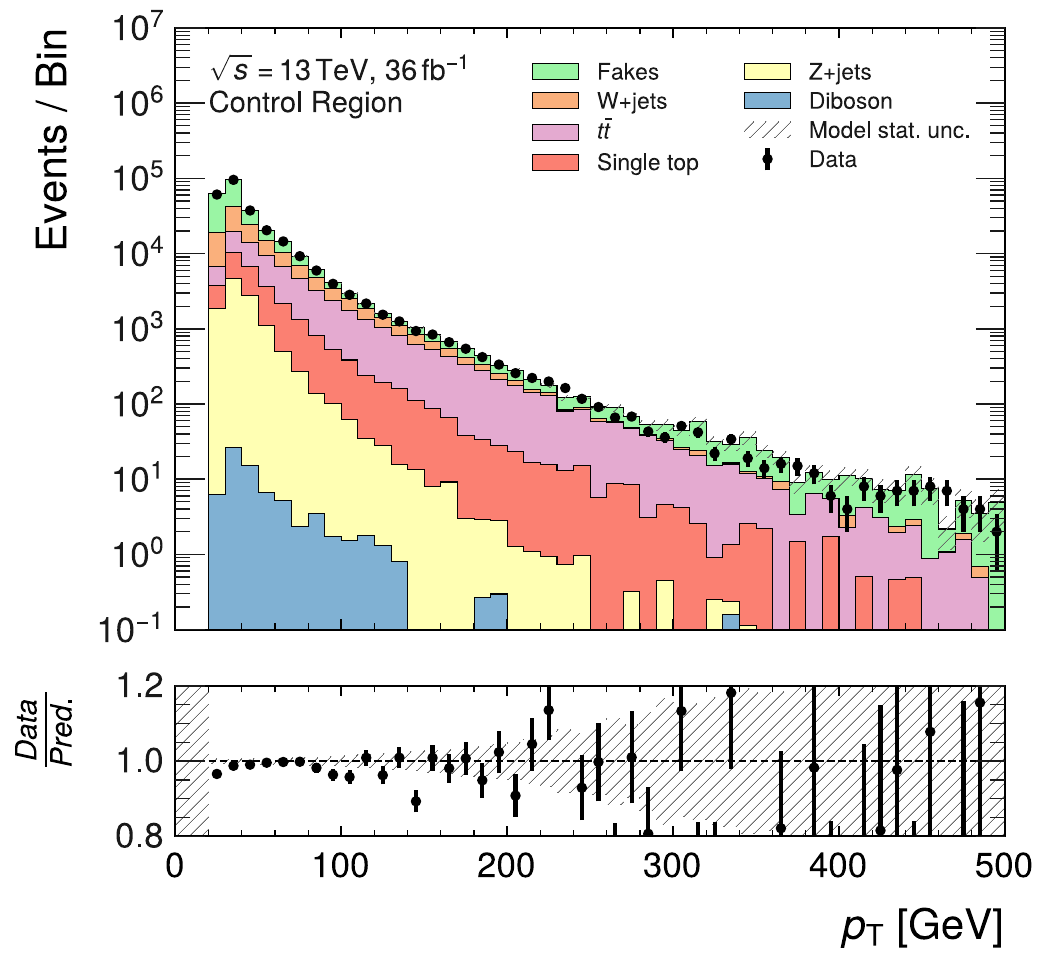}
    \end{subfigure}
    ~
    \begin{subfigure}{0.465\linewidth}
        \includegraphics[width=\linewidth]{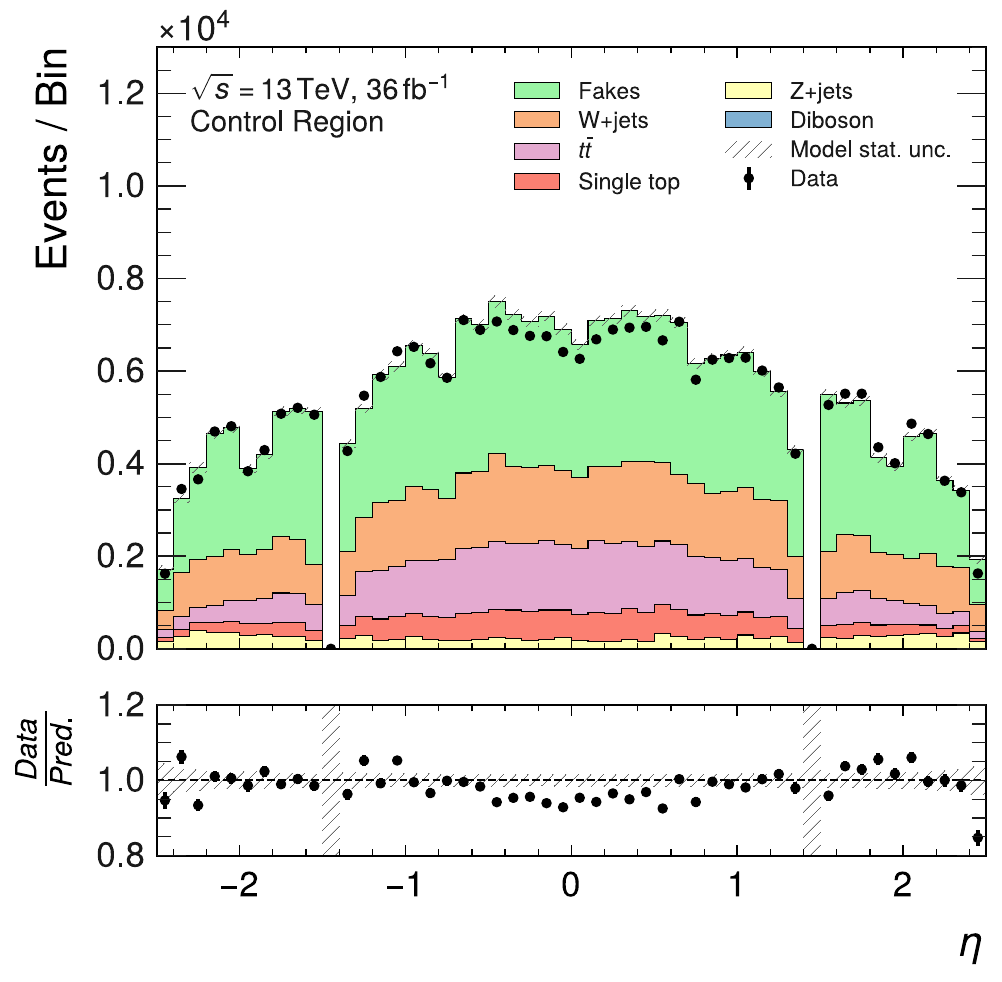}
    \end{subfigure}

    \begin{subfigure}{0.465\linewidth}
        \includegraphics[width=\linewidth]{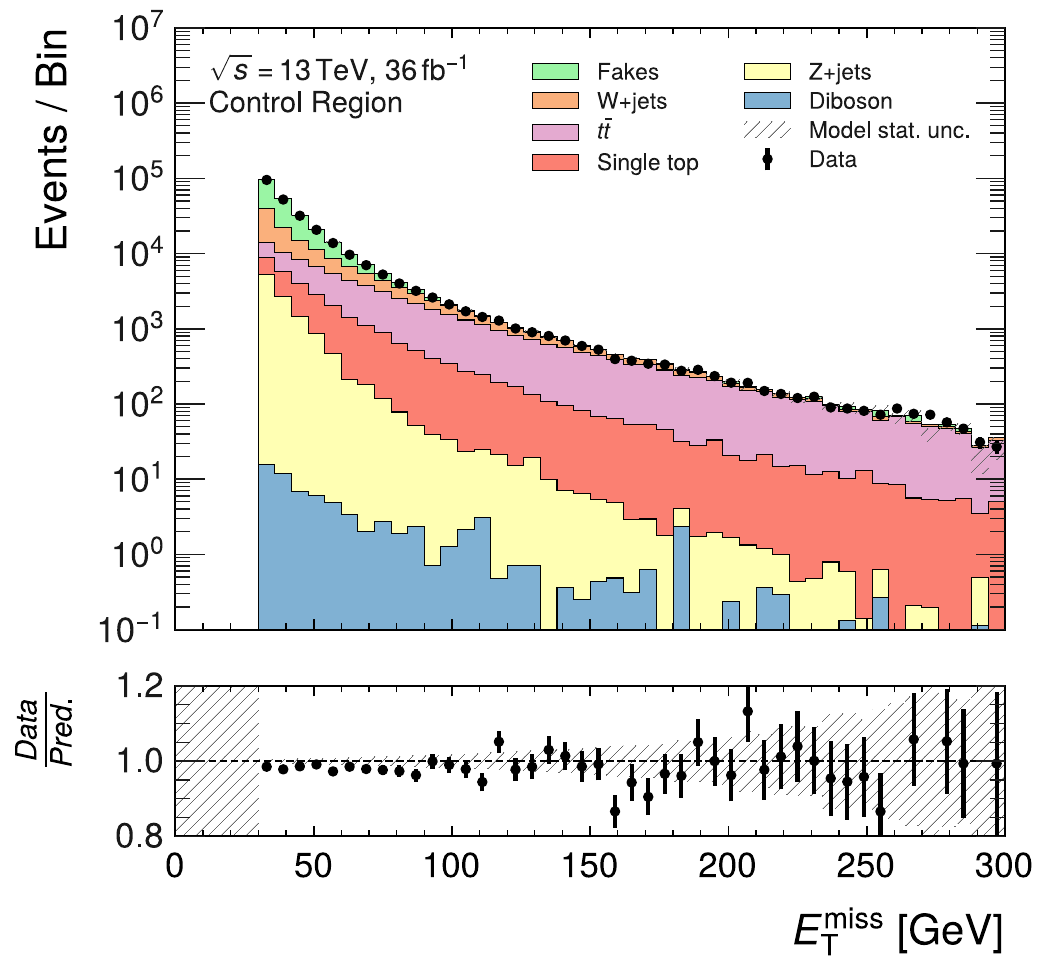}
    \end{subfigure}
    ~
    \begin{subfigure}{0.465\linewidth}
        \includegraphics[width=\linewidth]{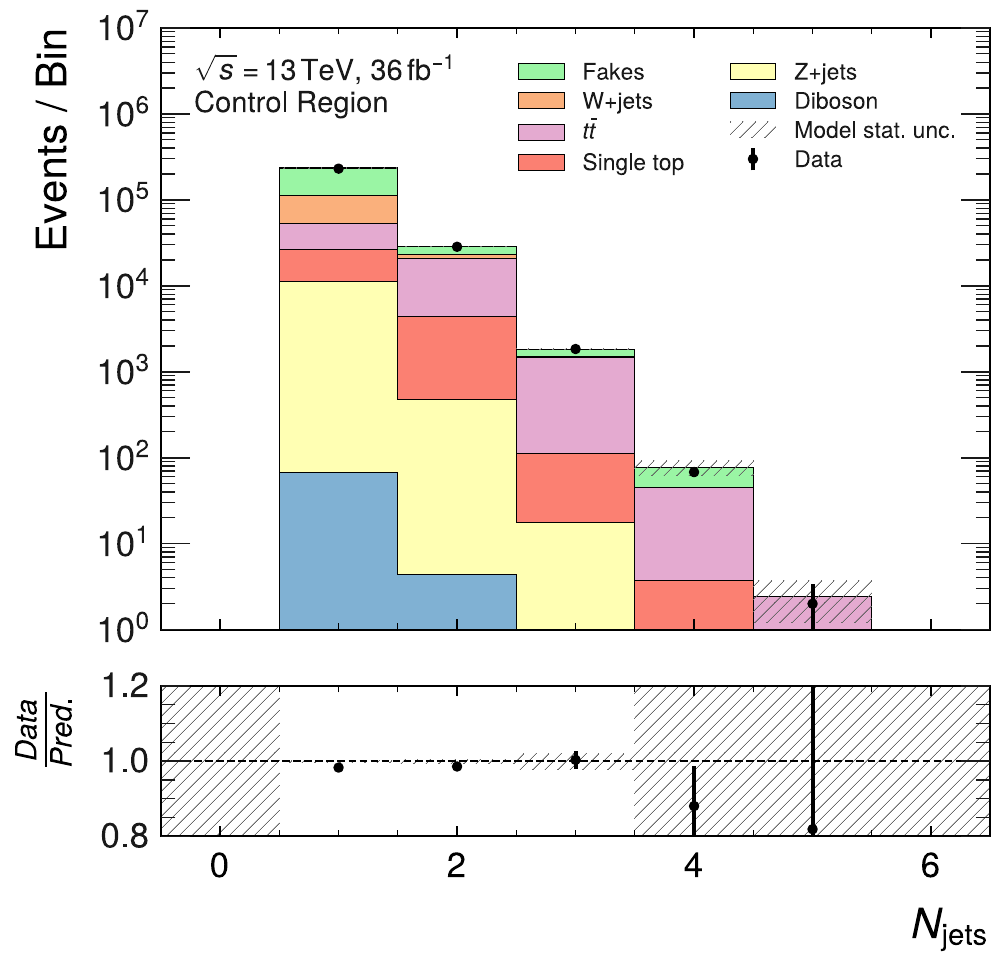}
    \end{subfigure}

    \begin{subfigure}{0.465\linewidth}
        \includegraphics[width=\linewidth]{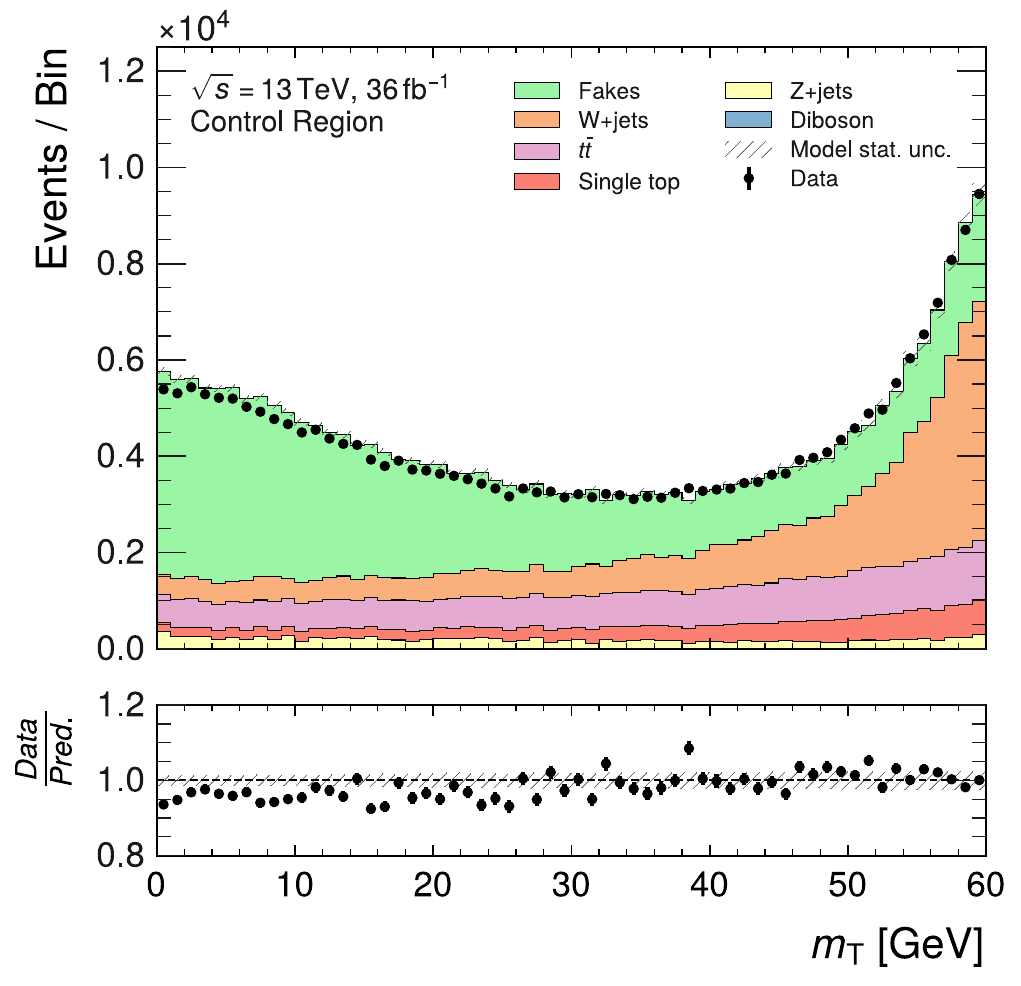}
    \end{subfigure}
    \caption{
        Closure test in the control region~(CR) of the implemented analysis using
        fake factors obtained from the ML-based method. Comparing to the results
        of the binned method in Figure~\ref{fig:closure_test}, with $\pt$ and
        $\eta$ distributions giving a slightly degraded agreement, an
        overall improvement in the closure is seen over all considered kinematic
        variables. This especially noticeable in the $\MET$ and $\mt$
        distributions, where the ML-based method captures the non-trivial
        dependence of the fake factor.
    }
    \label{fig:closure_test_ml}
\end{figure}

Firstly, we validate both the binned and ML-based Fake Factor methods in the
control region defined by $\mt<\qty{60}{\GeV}$, corresponding to the
fake-enriched region described in more detail in Section~\ref{sec:sec4}. In
Fig.~\ref{fig:closure_test} and Fig.~\ref{fig:closure_test_ml}, we show the
closure tests for both methods in $\pt$, $\eta$, $\MET$, $\jets$, and $\mt$.
Both methods show good agreement between data and MC across all variables, which
is expected since the fake factors are derived in this region. This demonstrates
that the estimated fake contributions accurately account for the shape and
normalization of the fake-lepton background within the fake-enriched control region for both
methods.

Focusing on the $\pt$ distribution, the binned method performs better at lower
$\pt$ ($\pt<\qty{100}{\GeV}$), where we have sufficient statistics in each bin
and where the binning was optimized to capture the finer features of the
distribution. The ML-based method, however, provides better modeling in $\MET$
and $\mt$ since these variables were also used in the ML training, while they cannot be used in the binned method due to statistical constraints, as already described. In the $\jets$
distribution, where we expect the jet multiplicity to be correlated with the
fake contribution (since there is a higher probability of non-real leptons in
multi-jet events), both methods capture this trend.

\subsection{Signal Region Performance}

\begin{figure}
    \centering
    \begin{subfigure}{0.465\linewidth}
        \includegraphics[width=\linewidth]{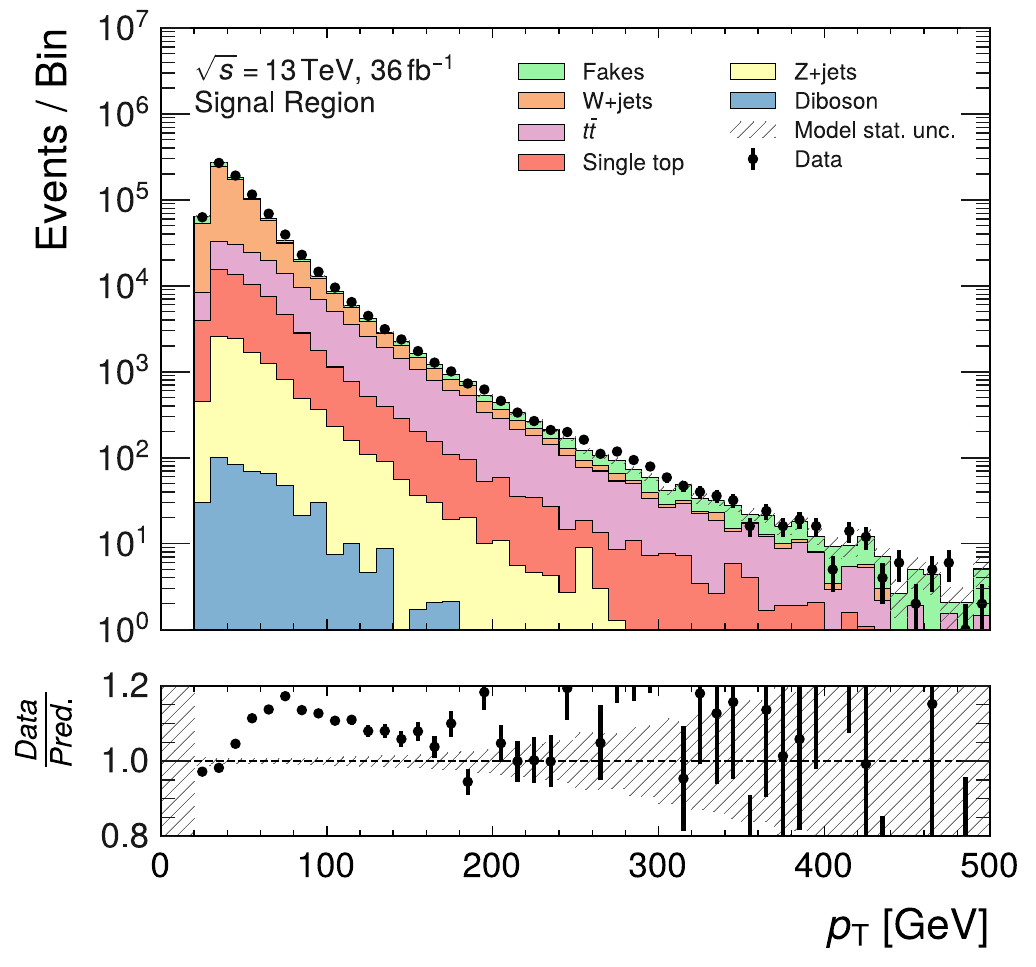}
    \end{subfigure}
    ~
    \begin{subfigure}{0.465\linewidth}
        \includegraphics[width=\linewidth]{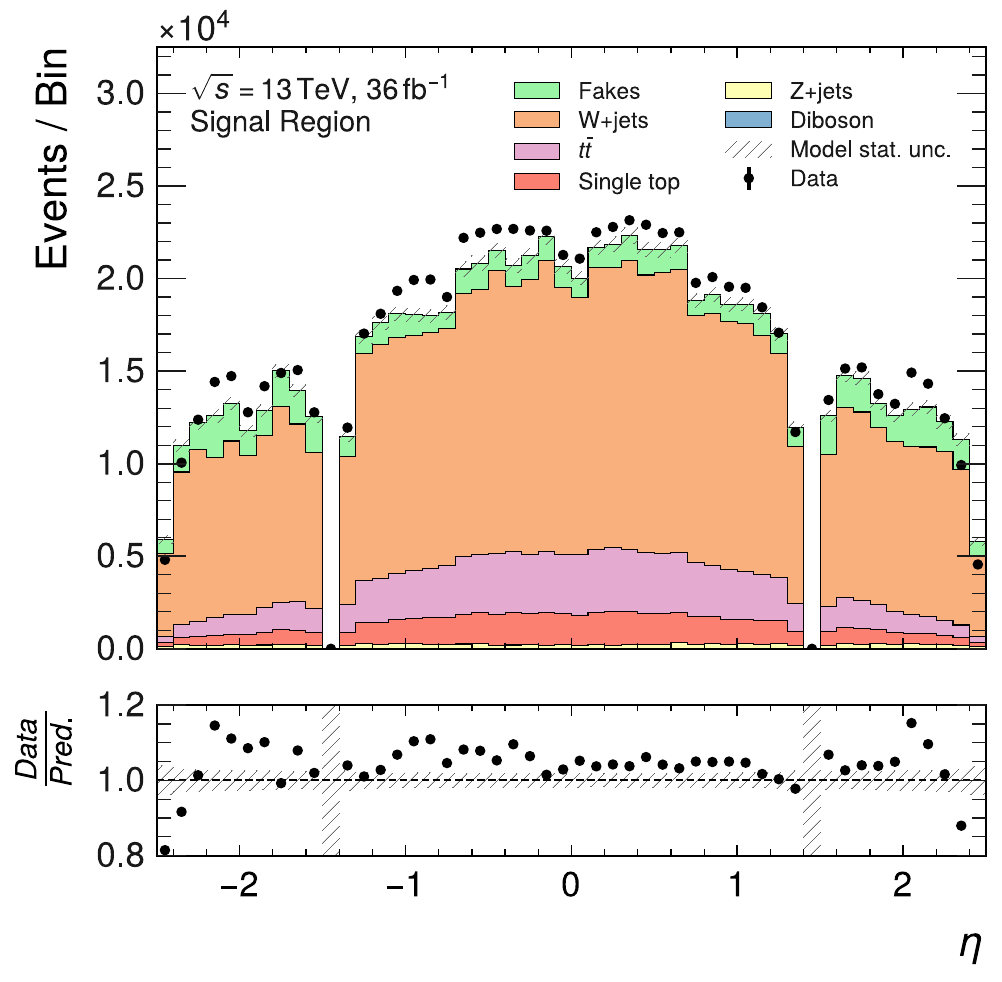}
    \end{subfigure}

    \begin{subfigure}{0.465\linewidth}
        \includegraphics[width=\linewidth]{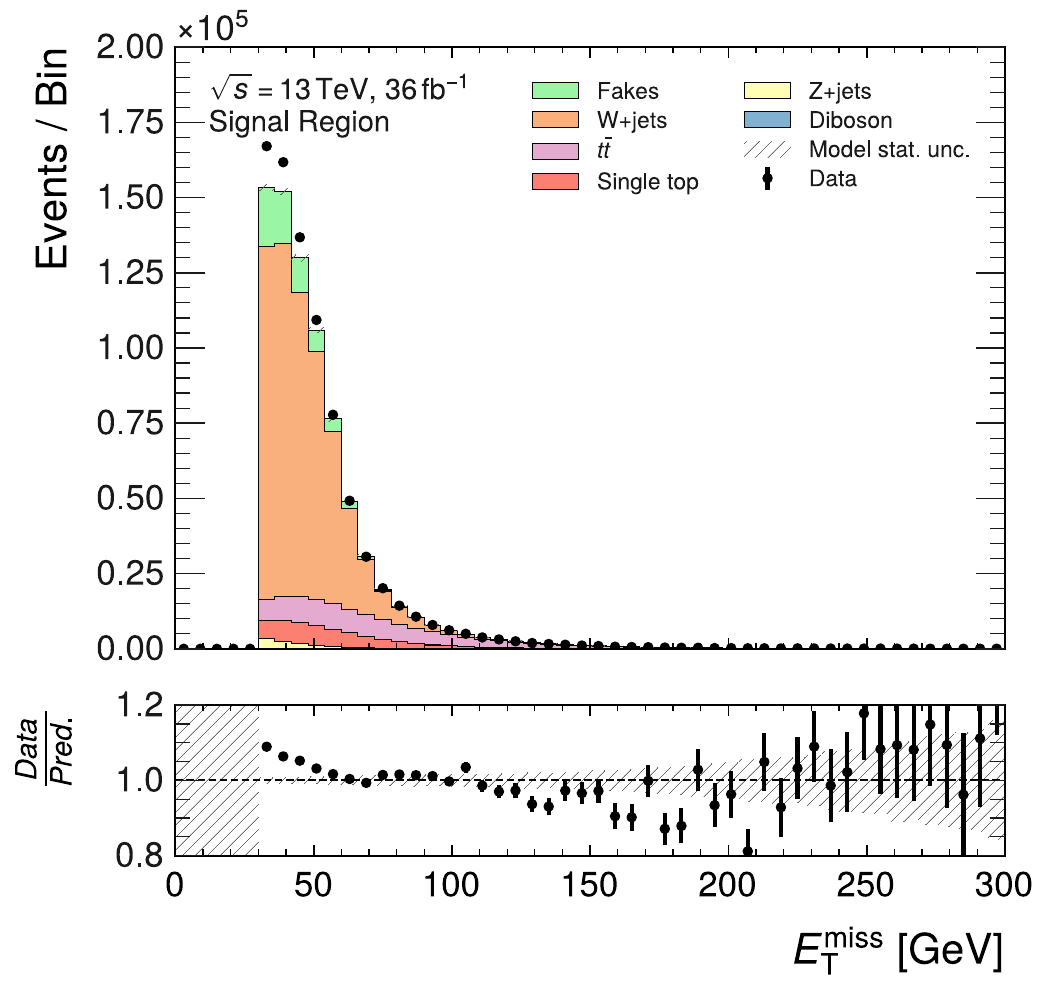}
    \end{subfigure}
    ~
    \begin{subfigure}{0.465\linewidth}
        \includegraphics[width=\linewidth]{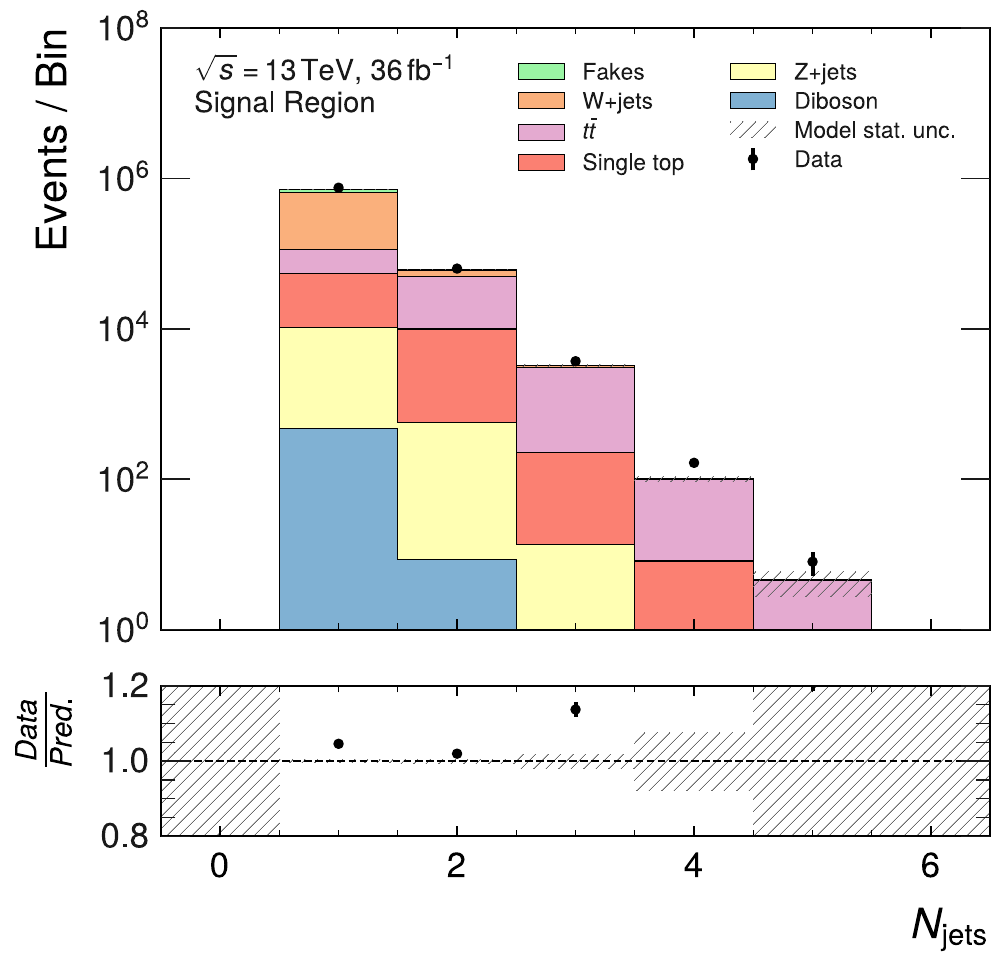}
    \end{subfigure}

    \begin{subfigure}{0.465\linewidth}
        \includegraphics[width=\linewidth]{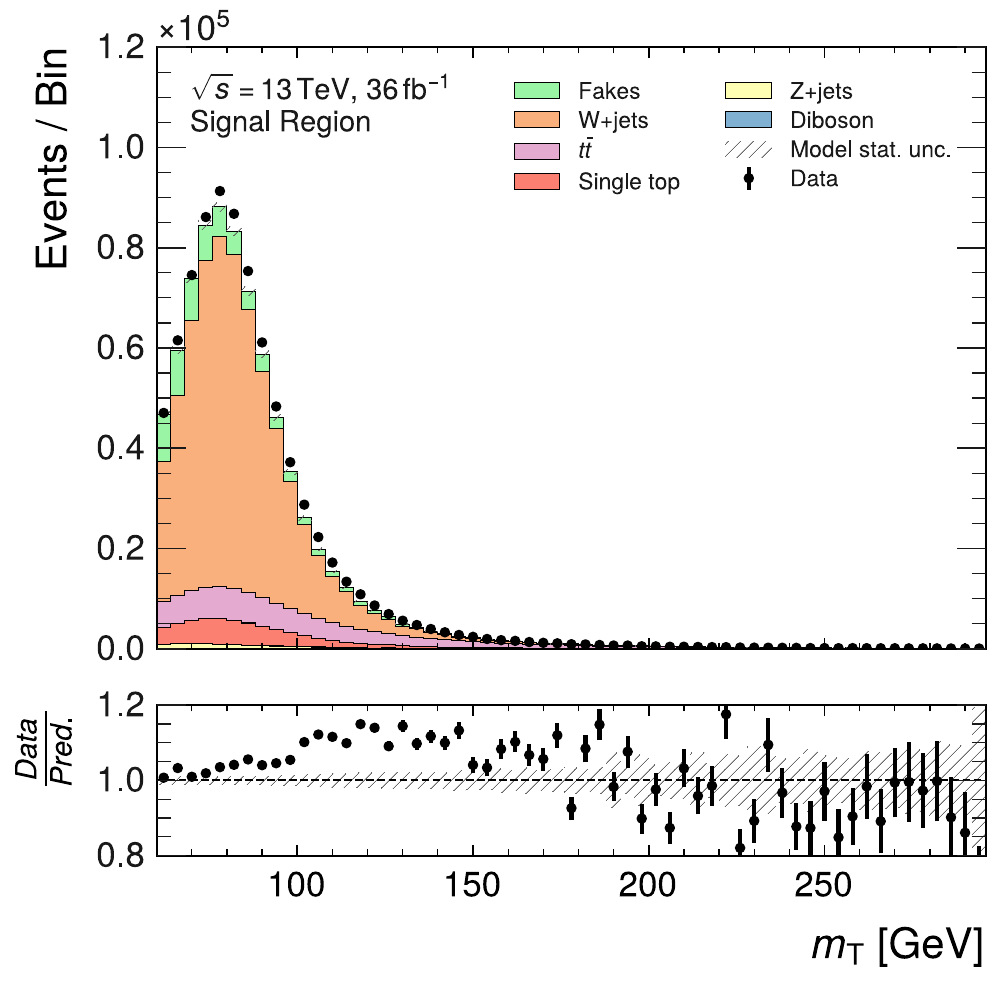}
    \end{subfigure}
    \caption{
        Signal region~(SR) of the implemented analysis with fake events modeled
        using binned fake factors. Compared to the closure study in the CR, the $\pt$ and $\eta$
        distributions show a significant systematic mis-modeling when the binned fake
        factors are extrapolated to the signal region. The same is reflected in
        the distributions of other variables as well, showing that the
        contributions from fake events are consistently underestimated.
    }
    \label{fig:sig-closure-binned}
\end{figure}

\begin{figure}
     \centering
    \begin{subfigure}{0.465\linewidth}
        \includegraphics[width=\linewidth]{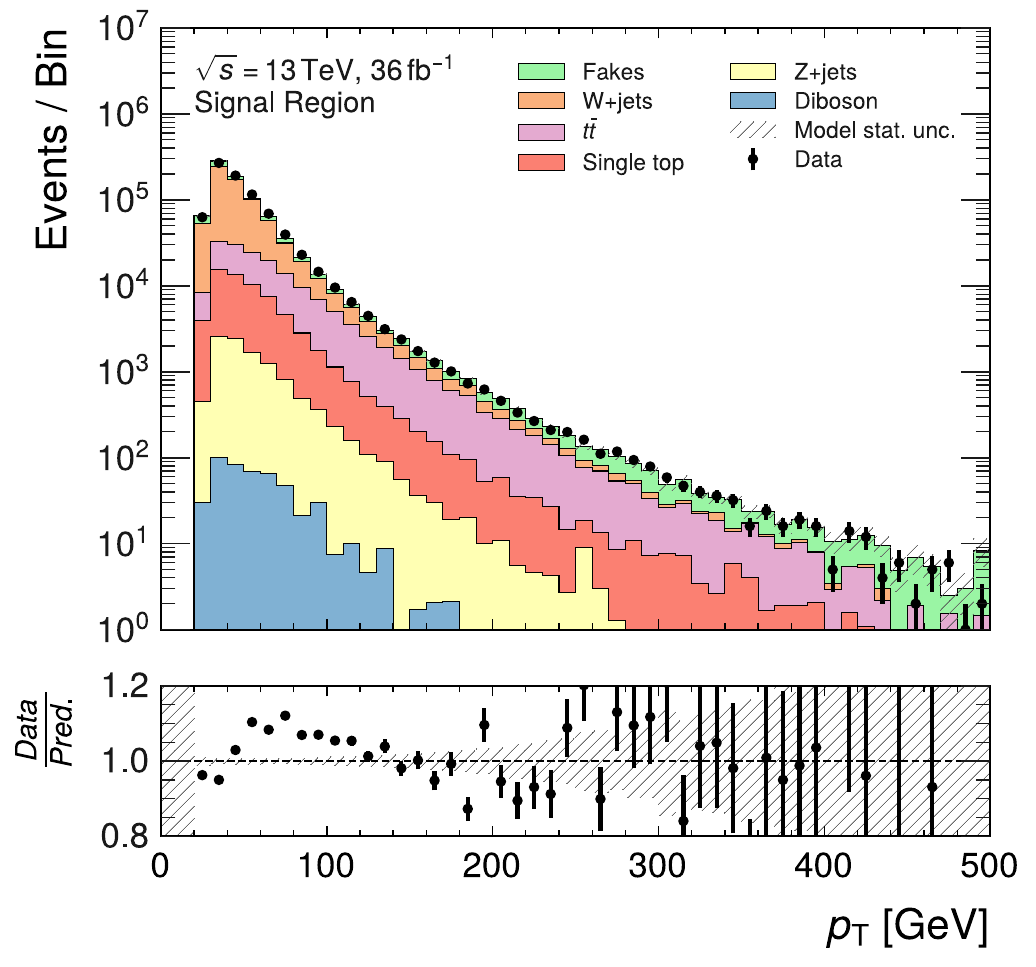}
    \end{subfigure}
    ~
    \begin{subfigure}{0.465\linewidth}
        \includegraphics[width=\linewidth]{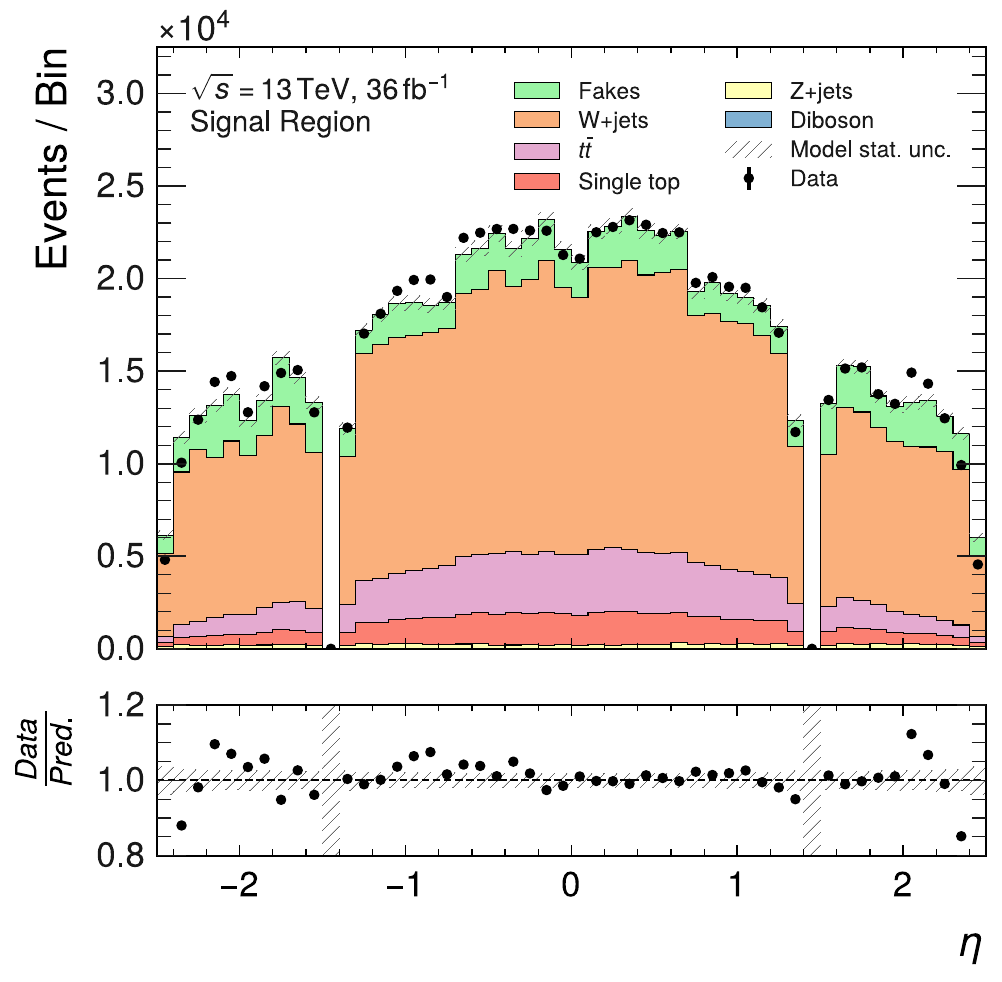}
    \end{subfigure}

    \begin{subfigure}{0.465\linewidth}
        \includegraphics[width=\linewidth]{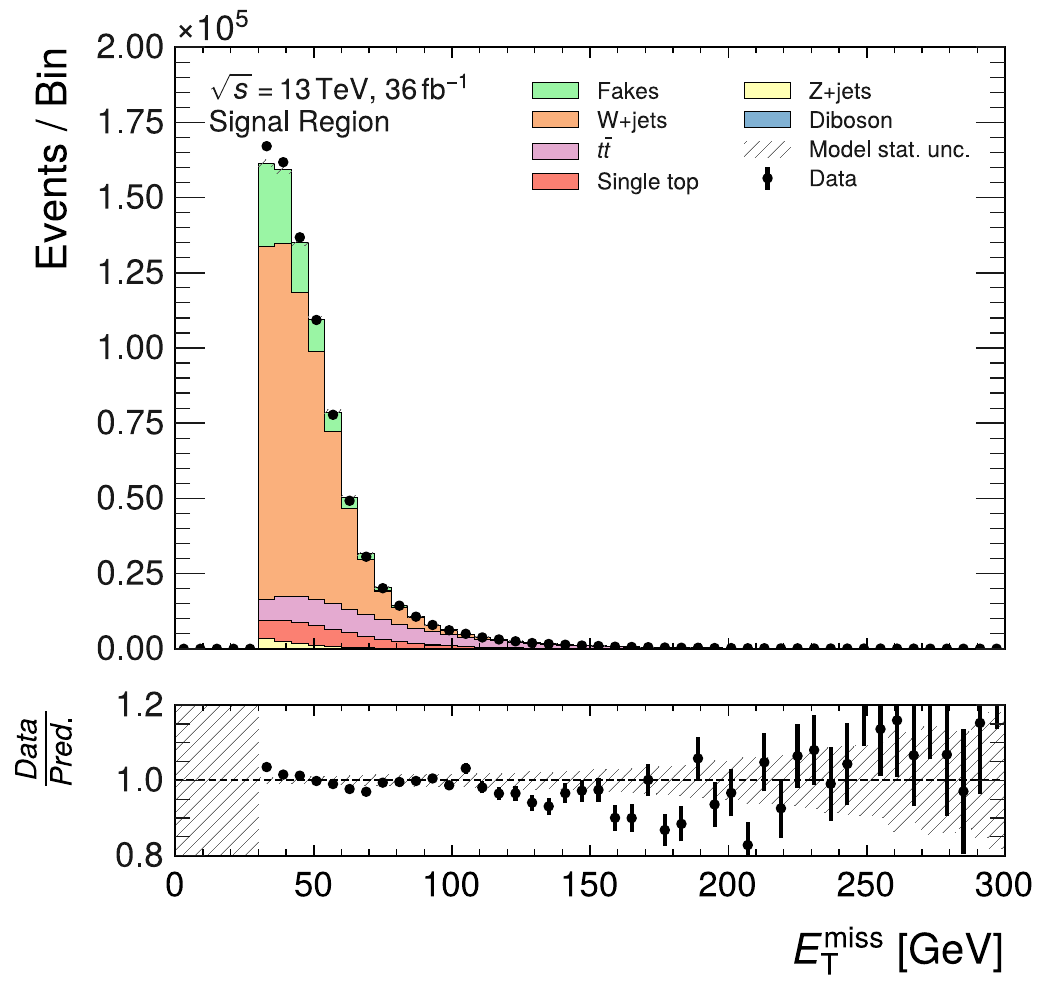}
    \end{subfigure}
    ~
    \begin{subfigure}{0.465\linewidth}
        \includegraphics[width=\linewidth]{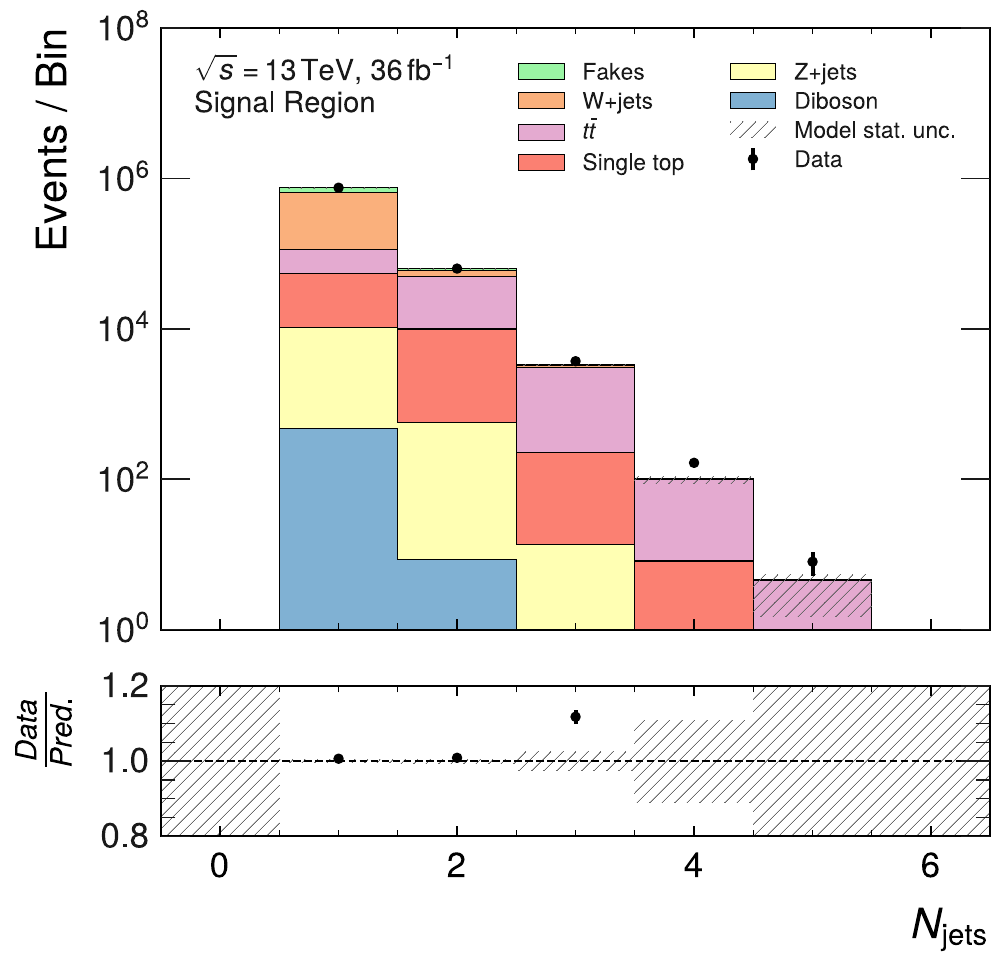}
    \end{subfigure}

    \begin{subfigure}{0.465\linewidth}
        \includegraphics[width=\linewidth]{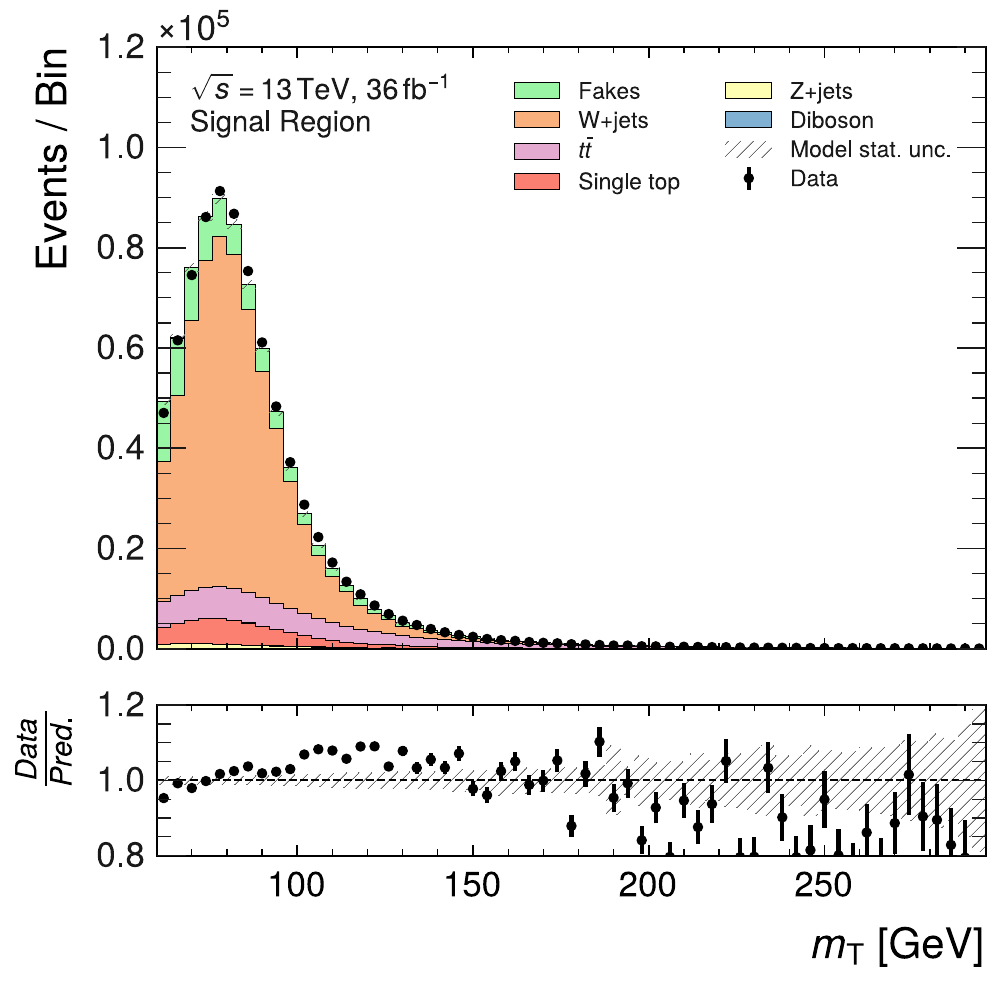}
    \end{subfigure}
    \caption{
        Signal region~(SR) of the implemented analysis with fake events modeled
        using fake factors from the ML-based method. In comparison with the
        binned method on Figure~\ref{fig:sig-closure-binned}, a significant
        improvement in the modeling is seen in the distributions of all
        variables -- both in the normalization and non-trivial shape effects.
        This indicates an improved ability of the ML-based method to accurately
        extrapolate the fake factor to the signal region, capturing a more
        complex and accurate kinematic dependence.
    }
    \label{fig:sig-closure-ml}
\end{figure}

In the signal region (SR), defined by $\mt>\qty{60}{\GeV}$, we validated both
Fake Factor methods by extrapolating from the results derived in the  control region to a kinematic regime in the SR,  where the fake contribution is smaller and real backgrounds dominate. In Fig.~\ref{fig:sig-closure-binned} and Fig.~\ref{fig:sig-closure-ml}, we show the distributions across the main observables.

In all the distributions, we observe that the ML-based method provides a better prediction in both shape and normalization with reduced fluctuations. This can be seen particularly in the $\MET$ and $\mt$ distributions where the ML-based method captures the shape more accurately.
The binned method shows larger discrepancies due to its statistical constraints on coarse binning, which can lead to inaccuracies when extrapolating to regions with limited statistics or complex correlations between variables.

With these validations, we demonstrate that the ML-based Fake Factor method can effectively extrapolate from the control region to the signal region, providing a reliable estimate of the fake lepton background in a kinematic regime where real backgrounds dominate. This is particularly important for analyses searching for rare signals, where accurate background estimation is crucial for maximizing sensitivity and avoiding potential appearance of a spurious (accidental) signal due to mis-modeling.

\section{Conclusion}\label{sec:sec6}

In this study, we have introduced a novel data-based inference method using  machine learning for estimating fake lepton backgrounds in high-energy physics analyses, demonstrating its advantages over traditional binned fake factor approaches. By leveraging neural density ratio estimation, our approach enables the calculation of continuous, unbinned fake factors on a per-event basis, allowing for precise modeling in high-dimensional feature spaces. This flexibility mitigates common limitations of conventional methods, such as coarse binning, extrapolation uncertainties, and the inability to capture complex correlations between variables.

Applying the method to a straightforward $W \to e \nu$ transverse mass analysis using ATLAS Open Data, we validated the method's performance through a two-step procedure consisting of subtraction and ratio calculations. The subtraction step effectively removes contamination from real leptons, ensuring that the derived fake factors reflect the true background contribution. The ratio step then generates smooth and physically consistent estimates across the feature space. Comparison with traditional binned fake factor results shows good agreement in both normalization and distribution shapes, while also highlighting the ML approach's strength in sparsely populated or high-dimensional regions where conventional methods often struggle.

Beyond demonstrating methodological improvements, this work highlights the broader potential of machine learning for data-driven background estimation in high-energy physics. The proposed method is inherently adaptable, capable of being extended to multi-lepton final states or other types of mis-identified objects.
Future work will focus on incorporating an advanced approach to systematic uncertainty estimation, exploring probabilistic models to quantify the uncertainty in ML-based fake factor predictions beyond simple statistical variations available through data resampling techniques (bootstrapping, Jackknife, etc.), which are already available.

Additionally, integrating this method into more complex analyses of the LHC experiments, including those searching for  new physics phenomena, will further showcase its utility and robustness.

In conclusion, the machine learning based Fake Factor method represents a significant step toward more precise, flexible, and robust data-driven background estimation in particle physics. By combining data-driven techniques with advanced computational tools, it enables better exploitation of the rich datasets produced by modern experiments, ultimately improving the sensitivity of searches for rare processes and new phenomena. As high-energy physics continues to explore more intricate final states and higher-dimensional datasets, methods like this will play a crucial role in ensuring that background modeling keeps pace with experimental and simulation capabilities, enabling utilization of the full physics potential of the data collected.

The code used in this work is available at~\cite{NeuralFakeFactor_GitHub}.

\section*{Acknowledgments}
We would like to thank the ATLAS Collaboration for providing the Open Data
used in this paper, as well as the developers of the various software tools
and libraries that facilitated this work. The work of B.P.~Kerševan, J.~Debevc and J.~Gavranovič is supported by the research grant J1-60028 and
research program P1-0135, funded by the Slovenian Research Agency~(ARIS). L.~Čalić and E.~Lytken wish to acknowledge financial support from the Swedish Research Council~(VR).

\bibliographystyle{apsrev4-2-fmf-eng}
\bibliography{references}

\end{document}